\documentclass[a4paper,fleqn,usenatbib]{mnras}

\usepackage[T1]{fontenc}
\usepackage{ae,aecompl}

\usepackage{amssymb}
\usepackage{array}
\usepackage{etoolbox}
\usepackage{threeparttable}
\usepackage{booktabs}
\usepackage{bm}
\usepackage{amsmath}
\usepackage{graphicx}
\usepackage[super]{nth}
\usepackage{xcolor}

\makeatletter
\patchcmd{\NAT@citex}
  {\@citea\NAT@hyper@{%
     \NAT@nmfmt{\NAT@nm}%
     \hyper@natlinkbreak{\NAT@aysep\NAT@spacechar}{\@citeb\@extra@b@citeb}%
     \NAT@date}}
  {\@citea\NAT@nmfmt{\NAT@nm}%
   \NAT@aysep\NAT@spacechar\NAT@hyper@{\NAT@date}}{}{}
   
\patchcmd{\NAT@citex}
  {\@citea\NAT@hyper@{%
     \NAT@nmfmt{\NAT@nm}%
     \hyper@natlinkbreak{\NAT@spacechar\NAT@@open\if*#1*\else#1\NAT@spacechar\fi}%
       {\@citeb\@extra@b@citeb}%
     \NAT@date}}
  {\@citea\NAT@nmfmt{\NAT@nm}%
   \NAT@spacechar\NAT@@open\if*#1*\else#1\NAT@spacechar\fi\NAT@hyper@{\NAT@date}}
  {}{}
\makeatother

\newcommand{\email}[1]{\mbox{\href{mailto:#1}{#1}}}

\newcolumntype{C}{>{$}c<{$}} 

\title[BayeSN]{A Hierarchical Bayesian SED Model for \\Type Ia Supernovae in the Optical to Near-Infrared }
\author[K.\ Mandel et al.]{Kaisey S. Mandel$^{1,2}$\thanks{E-mail: \email{kmandel@ast.cam.ac.uk}}, Stephen Thorp$^{1}$, Gautham Narayan$^{3,4}$, \newauthor Andrew S. Friedman$^{5}$, and Arturo Avelino$^{6,7}$\\
$^1$Institute of Astronomy and Kavli Institute for Cosmology, Madingley Road, Cambridge, CB3 0HA, UK\\
$^2$Statistical Laboratory, DPMMS, University of Cambridge, Wilberforce Road, Cambridge, CB3 0WB, UK\\
$^3$University of Illinois at Urbana-Champaign, 1003 W. Green St., IL 61801, USA\\
$^4$Center for Astrophysical Surveys, National Center for Supercomputing Applications, Urbana, IL, 61801, USA\\
$^5$Center for Astrophysics and Space Sciences, University of California, San Diego, CA, USA\\
$^6$Center for Astrophysics | Harvard \& Smithsonian, Cambridge, MA 02138, USA\\
$^7$Hydrique Ing\'{e}nieurs, Ch. du Rionzi 54, CH-1052, Le Mont-sur-Lausanne, CH}

\date{Accepted \dots Received \dots; in original form \dots}

\pubyear{2020}

\begin{document}
\label{firstpage}
\pagerange{\pageref{firstpage}--\pageref{lastpage}}
\maketitle


\begin{abstract}
While conventional Type Ia supernova (SN Ia) cosmology analyses rely primarily on rest-frame optical light curves to determine distances, SNe Ia are excellent standard candles in near-infrared (NIR) light, which is significantly less sensitive to dust extinction. A SN Ia spectral energy distribution (SED) model capable of fitting rest-frame NIR observations is necessary to fully leverage current and future SN Ia datasets from ground- and space-based telescopes including HST, LSST, JWST, and RST. We construct a hierarchical Bayesian model for SN Ia SEDs, continuous over time and wavelength, from the optical to NIR ($B$ through $H$, or $0.35 -1.8\, \mu$m).  We model the SED as a combination of physically-distinct host galaxy dust and intrinsic spectral components.  The distribution of intrinsic SEDs over time and wavelength is modelled with probabilistic functional principal components and the covariance of residual functions.  We train the model on a nearby sample of 79 SNe Ia with joint optical and NIR light curves by sampling the global posterior distribution over dust and intrinsic latent variables, SED components, and population hyperparameters.   The photometric distances of SNe Ia with NIR data near maximum light obtain a total RMS error of $0.10$ mag with our \textsc{BayeSN} model, compared to $0.14$ mag with SNooPy and SALT2 for the same sample.   Jointly fitting the optical and NIR data of the full sample for a global host dust law, we find $R_V = 2.9 \pm 0.2$, consistent with the Milky Way average. 
\end{abstract}

\begin{keywords}
 supernovae: general -- distance scale -- methods: statistical
\end{keywords}


\section{Introduction}

Type Ia supernovae (SNe Ia) are effective cosmological probes as ``standardiseable candles'': their peak luminosities can be inferred from their optical light curve shapes and colours, so their distances can be estimated from their apparent brightnesses.  Precise and accurate SN Ia distances with small systematic errors are essential to accurate constraints on the cosmic expansion history, including local measurements of the Hubble constant \citep{burns18,riess19}, the late-time cosmic acceleration \citep{riess98,perlmutter99}, and the properties of the dark energy driving it, in particular, its equation-of-state parameter $w$ \citep[e.g.][]{scolnic18,des3yr_19}.  Currently, there is a significant 4.4$\sigma$ tension between the value of $H_0$ locally inferred from SNe Ia and the local distance ladder ($74.03 \pm 1.42 \text{ km s}^{-1} \text{ Mpc}^{-1}$; \citealt{riess19}) and the value derived from Planck CMB analysis and the $\Lambda$CDM cosmological model ($67.4 \pm 0.5 \text{ km s}^{-1}  \text{ Mpc}^{-1}$; \citealt{planck18}). Since this tension could potentially be a sign of new physics, it is imperative to test for systematic errors with empirical cross-checks \citep[e.g.][]{dhawan18}.  With increasing sample sizes, nearby SNe Ia will be able to constrain the growth of structure as probes of the peculiar velocity field (e.g. \citealt{huterer17, howlett17, graziani20}).  Further cosmological constraints can also be derived from strong or weak lensing of SNe Ia \citep[e.g.][]{goldstein18,dhawan20,macaulay20}.  In this paper, we present a new data-driven statistical model, \textsc{BayeSN}, for SNe Ia light curves to extract more precise and accurate SN Ia distances from current and future surveys by exploiting the advantageous properties of SNe Ia in the near-infrared (NIR).

The current global sample used for cosmology, derived from the SDSS-II, SNLS, Pan-STARRS (PS1), low-$z$ and HST surveys, has grown to over a thousand SNe Ia (Pantheon; \citealt{scolnic18}).  Future surveys, such as the Legacy Survey of Space and Time \citep[LSST,][]{Ivezi__2019} with the Vera Rubin Observatory, will boost that number by orders of magnitude.   The constraints on dark energy with the current sample are already limited, not by statistical uncertainties from the numbers of SNe, but by systematic errors. In recent analyses, photometric calibration and SN model uncertainties dominate the systematic error budget (PS1: \citealt{scolnic14b,scolnic18}; DES: \citealt{brout19}). The calibration systematics are now being tamed by better cross-calibration between surveys \citep{scolnic15}, better networks of photometric standards \citep{Narayan16,Narayan19}, and by ongoing efforts to replace the heterogenous low-redshift sample with a large, unbiased, homogenous sample obtained on a precisely-calibrated photometric system (PS1, Foundation Survey; \citealt{foley18,jones19}).  LSST  will increase the cosmologically useful SN Ia sample to $\sim 10^5$ over its 10 year duration.  It will further diminish cross-survey calibration systematics by replacing previous high-redshift SN Ia surveys with a single, homogeneous, and large SN Ia sample taken on a single system. However, systematic errors due to the statistical models and methods used to analyse SN Ia light curve data will remain. 

Observations probing the rest-frame near-infrared (NIR, particularly $\lambda \gtrsim 1 \,\mu$m, e.g. $YJH$-bands) are a route to more precise and accurate distances. NIR observations of SNe Ia significantly improve their cosmological utility. Unlike in the optical, where they must be standardised via correlations of optical luminosity with light-curve shape and colour, SNe Ia are excellent, nearly-standard candles in the NIR, showing little intrinsic luminosity variation ($\sim 0.1$ mag) at peak (e.g. \citealt{krisciunas04a,wood-vasey08,mandel09,contreras10,kattner12,phillips12,barone-nugent12,stanishev18,burns18,avelino19}). The NIR also has significantly reduced sensitivity to dust extinction relative to the optical (by factors of $4-8$, comparing NIR $YJH$ to optical $B$).  \citet{dhawan18} showed how a small set of SNe Ia, used as NIR standard candles to measure $H_0$, can replace a much larger optical sample, while still providing a 4.3\% measurement (consistent with \citealt{riess19}), without any light-curve shape or colour corrections as are required in the optical. We recently compiled a sample of 89 nearby SNe Ia with optical and NIR light curves passing standard quality cuts \citep{avelino19}.  Using 56 SNe Ia with NIR data near peak brightness, where the luminosity dispersion is minimal, we found a 35\% reduction in Hubble Diagram scatter (i.e. more precise distances) when using SNe Ia as NIR standard candles, relative to conventional optical-only fits to the same SNe.  

The combination of optical and NIR data better constrains the host galaxy dust extinction and the shape of the dust law as a function of $\lambda$ (parametrized by $R_V$) \citep{krisciunas07, burns14}, and significantly improves the accuracy and precision of SN Ia distances \citep{mandel11}.  The nature of the dust in SN Ia host galaxies is fundamental to the largest ``correction'' in standardising SNe Ia, that due to colour.  Incorrect modelling interpretation of the SN Ia colour-magnitude relation is therefore a major source of systematic error in SN distances.  However, the correct values(s) of the $R_V$ parametrizing the dust extinction law has long been a matter of confusion, and its proper estimation is fraught with statistical subtleties. 

Very early analyses that found unphysically low values $R_V \lesssim 1$ \citep{branch92} did not account for correlations between the luminosity, colour and light curve shape \citep[later modelled by e.g.][]{phillips93,riess96,phillips99}.  \citet{riess96b} noted that confusing intrinsic colour-luminosity variation with dust effects would lead to mistakenly lower estimated $R_V$ values.  Simple linear regression analyses of SN (extinguished) absolute magnitudes against (apparent) colours and light curve shapes have led to apparent colour-magnitude slopes (e.g. $\beta$ in the Tripp formula) that have sometimes been interpreted as low dust $R_V$ values \citep{tripp98, trippbranch99,guy05,astier06,conley07,freedman09}. \citet{scolnic14a} highlighted the relevance of colour dispersion to estimating a Milky Way dust-like colour-magnitude slope.  \citet{mandel17} showed that statistical confounding of the intrinsic color-luminosity correlation and dispersion with the extrinsic effects of dust leads to estimates of $\beta$ that are biased low relative to the true dust $R_V$, and a probabilistic generative model with explicit parameters for these physically-distinct effects led to a Bayesian estimate of $R_V = 2.8 \pm 0.3$, consistent with the Milky Way average.

Anomalously low $R_V \approx 1.5-1.8$ values have been estimated for a few very highly reddened SNe Ia ($E(B-V) > 1$) \citep[e.g.][]{elias-rosa06,elias-rosa08,wangx08,amanullah14}.  While the origin of these low $R_V$ estimates is still under investigation \citep{lwang05,goobar08,amanullah11,phillips13,amanullah15,johansson17,bulla18a,bulla18b}, these very red SNe are not present in the cosmological sample, due to the standard cut on peak apparent SN colour ($B-V \lesssim 0.3$).  When only low- to moderately-reddened normal SNe Ia with apparent colours consistent with the cosmological sample are considered, values of $R_V \approx 2.5 - 3$ have generally been estimated in nearby samples, often by utilising spectroscopic  or NIR data to break the degeneracy between intrinsic colours and dust in the optical \citep{folatelli10, mandel11,  foleykasen11,chotard11,phillips12,burns14, mandel17,leget20}.

The excellent properties of the NIR have not been fully integrated into and leveraged by the statistical models routinely used for SN Ia cosmology.  We have constructed a new, hierarchical Bayesian model, \textsc{BayeSN}, for time-dependent SN Ia spectral energy distributions (SEDs) from the optical to NIR wavelengths.  With NIR coverage, our model leverages the low luminosity dispersion in the NIR, while its wide optical-to-NIR wavelength range enables it to more stringently constrain the host galaxy dust, and the dust law, affecting the SNe Ia. These two advantages enable us to more accurately improve our model of the intrinsic SED coherently across all wavelengths. While it produces the best distance estimates when fitting complete light curves across the full wavelength range, as a Bayesian model, it also makes the most effective use of the observations available in any partial dataset, e.g. optical-only, NIR-only, while marginalising over the unobserved parts of the SED. 

\textsc{BayeSN} is an essential tool, not only for properly analysing current datasets, but also extracting optimal distances and robust cosmological constraints from future optical and NIR SNe Ia observations. Beyond the datasets analysed in the present work, the ability to effectively leverage joint optical and NIR observations is crucial for fully exploiting a number of recent and current surveys and forthcoming datasets, including the Carnegie Supernova Project-II \citep[CSP-II;][]{phillips19}, the Foundation Supernova Survey \citep{foley18} and Young Supernova Experiment with Pan-STARRS, RAISIN (GO-13046, GO-14216) and SIRAH (GO-15889) with the Hubble Space Telescope (HST), the ESO VISTA Extragalactic Infrared Legacy Survey (VEILS), and the DEHVILS Survey using UKIRT. This is also important for LSST, which will observe SNe Ia in $ugrizy$, and will therefore probe rest-frame $z$ or $y$ to redshifts $z \lesssim 0.3$.  The Nancy Grace Roman Space Telescope (RST, formerly WFIRST) will have a dedicated SN survey and its wide NIR filters will probe rest-frame $YJH$ out to redshifts $z \lesssim 1, 0.7, 0.4$ respectively.

\subsection{Comparison to existing models}\label{sec:introcomparison}

The models used to analyse SN Ia light curves and estimate distances are entirely empirical and are learned from the data. The conventional approach has a number of shortcomings that need to be addressed to exploit fully the data and to control astrophysical and modelling systematics.  The model most commonly used for fitting optical SN Ia light curves is SALT2 \citep{guy07,guy10,betoule14}. It models the SN Ia SED in phase (rest-frame time since peak luminosity) and wavelength, as a function of optical light curve shape ($x_1$) and apparent colour ($c$) at peak. SN Ia light curve fits estimate these parameters and the optical peak apparent magnitude $m_B$. Photometric distances are obtained from a fitted linear dependence of SN Ia absolute magnitude on light curve shape and colour \citep{tripp98}:
\begin{equation}\label{eqn:tripp}
\mu_s = m_{B,s} -M_B + \alpha\, x_{1,s} - \beta\, c_s
\end{equation}
where $\mu_s$ is the distance modulus of an individual SN $s$, $(m_{B,s}, x_{1,s}, c_s)$ are parameters obtained from the SALT2 fit of the individual SN $s$, and $(\alpha, \beta, M_B)$ are global (or population) parameters describing the luminosity trends with light curve shape and colour, and the absolute magnitude intercept at $x_1 = c = 0$, respectively.

Major shortcomings of the conventional approach are: 
\begin{enumerate}
\item Residual (``Intrinsic'')\footnote{\label{fn1}The terminology of ``intrinsic scatter'' here is a confusing misnomer. In the conventional SALT2 framework that is agnostic about the distinction between intrinsic and dust effects, there is no reason to attribute all of its residual scatter to variation intrinsic to the supernovae, even if the model were true.} scatter systematic error: Spectral variations of SN Ia light curve data around the best- fit SED model in excess of measurement error are accounted for by an ``intrinsic scatter model.'' This characterises the covariance of SN Ia spectral residuals unaccounted for by the SALT2 light curve shape and apparent colour parameters.  It is not well constrained, and currently there are two options: one with 30\% chromatic variation and 70\% achromatic variation \citep{guy10}, and the other, based on \citet{chotard11}, with a 75\% : 25\% split.  \citet{scolnic14a} showed that both models are consistent with the cosmological SN Ia data, therefore the current optical data alone cannot discriminate between the two. However, the impact of changing the assumed model for the residual scatter in a cosmological analysis results in a shift $\Delta w \sim 0.04$, and thus is a dominant systematic error. 

Employing the correct residual covariances across phase and wavelength is crucial to the proper quantification of uncertainties and weighting of the SN data.  Our \textsc{BayeSN} SED model coherently estimates the intrinsic residual covariance across phase and wavelength simultaneously with the training of the entire hierarchical model, and this covariance is employed when fitting SN light curves to estimate dust and distance, while marginalising over the SED residuals.

\item Degeneracy between intrinsic vs. dust colour-luminosity variations: The largest ``correction'' in Eq. \ref{eqn:tripp} is due to colour, but the conventional analysis treats it in a simplistic way. Fundamentally, intrinsic variation and dust have physically-distinct effects on the SN Ia SED. However, the SALT2 model assumes that all colour variation can be described by the peak apparent $B-V$ colour parameter $c$ and a single, effective colour law, $CL(\lambda)$.  The conventional approach of fitting a single linear function for (extinguished) absolute magnitude versus apparent colour confounds the two effects.  
 
In contrast, our \textsc{BayeSN} SED model allows for a probabilistic, physically-motivated combination of different spectral effects from intrinsic SN variation and dust across time and wavelength.

\item Lack of NIR coverage: The most recent SALT2.4 model is only specified over rest-frame wavelengths of $0.2-0.9\, \mu$m, though the colour law for $\lambda > 0.7\, \mu$m is an extrapolation.  Although optical surveys, such as Foundation \citep{foley18}, routinely obtain $z$-band data, they cannot be fit by SALT2 for nearby SNe Ia.  SALT2.4 is incapable of leveraging the useful properties of SNe Ia in the rest-frame NIR at $\lambda \gtrsim 1 \mu$m.  

In contrast, our \textsc{BayeSN} SED model is trained on data covering optical to NIR wavelengths extending from $B$ through $H$-band ($0.35-1.8\, \mu$m) and uses Bayesian inference to combine information over the full phase and wavelength range for optimal estimates of dust and distance.

\end{enumerate}

Although SALT2 is the most common SN model used in cosmology, there are alternatives. SNooPy is an optical-NIR model for SN Ia light curves defined in discrete rest-frame $uBVgriYJH$ passbands  \citep{burns11}. It is not a model for the continuous SED; rather, for each discrete rest-frame passband it has a template light curve that varies as a function of a shape parameter (either $\Delta m_{15}(B)$ or $s_{BV}$). It requires the calculation of $K$-corrections of the photometry between each observer-frame passband into a corresponding rest-frame model passband as a preprocessing step. The template light curve model is then fit to the $K$-corrected data in the rest-frame bands.  This 1-to-1 mapping is not ideal, as there are redshifts at which, for example, wide HST WFC3 NIR filters significantly cover two rest-frame model passbands, so the observed light curves are actually sensitive to the statistical properties of the underlying SED in both rest-frame bands.  Furthermore, the $K$-correction calculation employs an ad-hoc ``mangling'' procedure to match a spectral template to observed colours independently at each epoch.  This is prone to overfitting, its uncertainties are difficult to propagate, and is not viable for the noisy, sparse data typical of high-$z$ light curves, in which the light curves in different passbands may be irregularly and asynchronously sampled.  We compare the results from \textsc{BayeSN} to those from applying SALT2 and SNooPy to the same SNe Ia in \S \ref{sec:results}.

SNEMO \citep{saunders18,rose20} and SUGAR \citep{leget20} are recent empirical models built from optical spectrophotometric time series. Whereas SNEMO is a principal components model for the optical SED, SUGAR models the spectral dependence on factors composed of spectral line characteristics.  However, they only cover rest-frame $0.33 < \lambda < 0.86\, \mu$m, and so they cannot leverage the valuable NIR at $\lambda \gtrsim 1\, \mu$m.

\subsection{Outline of paper}

The outline of this paper is as follows. In \S \ref{sec:statmodel}, we describe our new hierarchical Bayesian model for SN Ia SEDs in the optical to NIR.  In \S \ref{sec:data} we describe the compilation of optical and NIR SN Ia light curve data that we analyse.  In \S \ref{sec:implementation}, we describe our computational implementation for training the \textsc{BayeSN} model and fitting SN Ia light curves.  In \S \ref{sec:results}, we present our results, including a Hubble diagram showing the improvement in distances (to $0.10$ mag total RMS error) obtained from \textsc{BayeSN} fits to optical and NIR data compared to current methods applied to the same sample.  We also describe our inferences about host galaxy dust, for which we constrain a global value of $R_V = 2.9 \pm 0.2$ for our sample with $E(B-V)_\text{host} \lesssim 0.4$.  In \S \ref{sec:conclusion}, we conclude.  

In Appendix \S \ref{sec:2dspline}, we describe technical aspects of modelling our SED surfaces, and in Appendix \S \ref{sec:w2}, we describe the extension of our model to a second functional component.

\section{The Statistical Model}\label{sec:statmodel}

To construct and train our SN Ia SED model, we employ a hierarchical Bayesian approach. Hierarchical Bayes provides a principled, coherent framework for modelling multiple uncertain and random effects underlying the data described via a probabilistic generative model.  It is a natural strategy for probabilistic modelling and inference of populations as well as their constituent individuals \citep{gelman_bda,loredohendry10,loredohendry19}. The first applications of hierarchical Bayes to supernova analyses were demonstrated by \citet{mandel09,mandel11}, who developed probabilistic models for SN Ia optical and NIR light curves in discrete passbands.  \citet{mandel14} constructed a hierarchical Bayesian model to disentangle dust reddening from intrinsic colours in the optical by leveraging the velocity-colour relation \citep[VCR;][]{foleykasen11}.

Other hierarchical Bayesian models for SN Ia analysis have focused exclusively on analysing the 3-parameter output from SALT2 fits to SN Ia light curves \citep{march11, rubin15, shariff16, mandel17,hinton19}, rather than the observed data itself.  Since they do not attempt to directly model the irregularly and asynchronously sampled multivariate, multi-band light curve (time series) data, they are dependent on the internal shortcomings of SALT2 described in \S \ref{sec:introcomparison}.  

In constrast, our \textsc{BayeSN} SED model combines the hierarchical Bayesian strategy with techniques from functional data analysis \citep[e.g.][]{ramsay05}
to deal with the full complexity of observed photometric time series, and to perform probabilistic inference on the multiple time- and wavelength-dependent latent functions underlying the observed data.  In particular, we model the modes of variation of the intrinsic SED in terms of a Bayesian formulation of functional principal components.  While principal components analysis (PCA) is a standard tool for dimensionality reduction of multivariate data, in its conventional use, however, it lacks a probabilistic framework. Probabilistic and Bayesian formulations of PCA for multivariate vectorial data were described by \citet{roweis98,tipping99,bishop99,bishop}.  In particular, \citet{tipping99} constructed a probabilistic PCA as a special case of a Gaussian latent variable model for factor analysis, with an associated likelihood function mapping between a low-dimensional latent space and the high-dimensional data space, and a prior distribution over the latent variables. \cite{bishop99} further developed Bayesian PCA by introducing priors on the principal components and residual variance.  These useful probabilistic formulations enable us to embed a principal components model within our hierarchical Bayesian framework while simultaneously accounting for multiple random effects and sources of uncertainty, such as dust, distance, and measurement error.  Thus, we can determine the intrinsic principal components while marginalising over the other uncertainties in the global inference problem.  

A primary goal of \textsc{BayeSN} is to model populations of latent SED functions over time and wavelength, so we extend these concepts to functional data, by incorporating continuity and smoothness constraints on the functional principal components, and by modelling the time- and wavelength-dependent covariance structure of the residual functions.  In this paper, we deal mainly with photometric flux data, which are essentially \emph{integral constraints} (under the passband throughput and with measurement errors) on the latent SED component functions. Embedding the functional inference within a hierarchical Bayesian structure naturally effects regularisation to solve the inverse problem.

\begin{figure*}
	\includegraphics[scale=0.45]{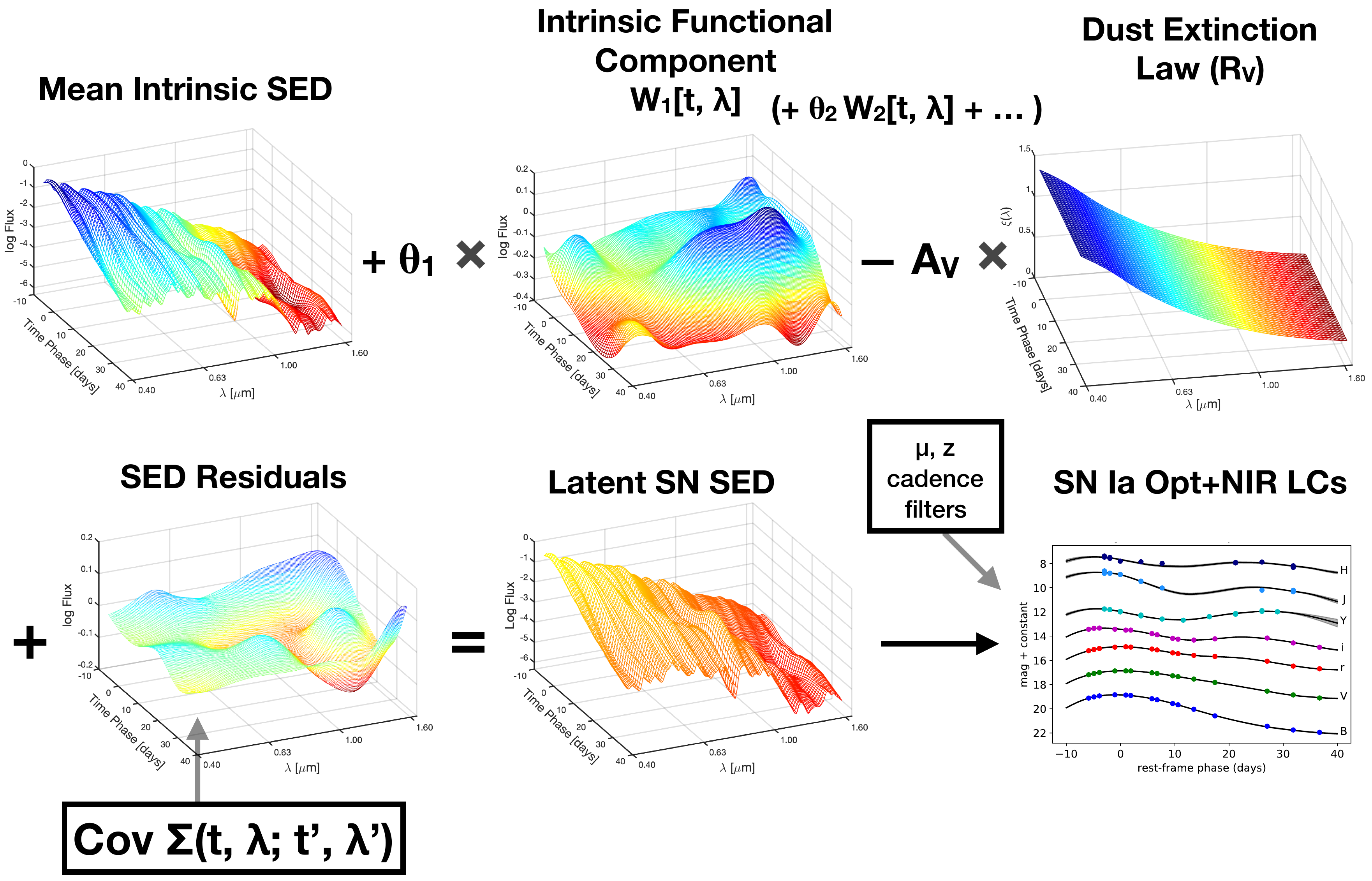}
	\caption{Schematic of the \textsc{BayeSN} forward generative model for the optical and NIR light curve (time series) data of a single SN. The log SN SED across time and wavelength comprises a mean intrinsic SED function modified by intrinsic functional principal components scaled by latent SED shape parameters, and extinguished and reddened by the host galaxy dust law, parametrized by the optical slope $R_V$ and scaled by the visual extinction $A_V$. Variations not captured by these major modes are modelled by residual SED functions whose statistical properties across time and wavelength are captured by a covariance function.  The resulting latent host-dust-reddened rest-frame SED undergoes the effects of distance, redshifting and time dilation, integration of the flux under the observer's filter functions, the survey cadence, and measurement error to yield the observed multi-band optical and NIR time series (light curves) of a SN Ia.}
	\label{fig:forwardmodel}
\end{figure*}

A schematic depiction of the probabilistic forward model of the SED for a single supernova's light curve data is shown in Fig. \ref{fig:forwardmodel}.  We construct a log intrinsic SN SED across time and and optical to NIR wavelengths by modifying a mean intrinsic SED function with functional principal components scaled by latent SED shape parameters. This is further modified by the dust extinction law as a function of wavelength, scaled by the dust extinction parameter.  A random function described by a covariance matrix models the SED residuals, as a function of time and wavelength, that are not captured by the previous main modes of variation.  The combination of these effects yields the latent host-reddened SED in the SN rest-frame. Finally, the effects of distance, redshifting and time dilation, integration of the flux under the observer's filter functions, the observational cadence of the survey, and photometric measurement error yields the observed multi-band optical and NIR time series (light curves) of a SN Ia.

\subsection{Flux Data Model}
Suppose supernova $s$ with spectroscopic redshift $z_s$ has a distance modulus $\mu_s$.  The $i$th photometric observation of SN $s$ is taken at observer-frame Modified Julian Date (MJD) $T^i_s$ through a filter with a transmission function $\mathbb{T}_{s,i}(\lambda_o)$ as a function of observed wavelength $\lambda_o$.  The calibration standard has an SED $F_{\text{std}}(\lambda_o)$, which defines the reference magnitude in the passband. The calibrated flux (``FluxCal'' in SNANA; \citealt{snana09}) is the ratio of the SN flux at the observer $F_\text{obs}^{s,i}$ integrated over the passband to the flux of the standard star through the same passband:
\begin{equation}\label{eqn:synflux}
\begin{split}
f_{s,i} &= 10^{0.4\times Z_{s,i}} \times \frac{\int F^{s,i}_\text{obs}(\lambda_o) \, \mathbb{T}_{s,i}(\lambda_o) \lambda_o \, d\lambda_o}{\int F_{\text{std}}(\lambda_o) \, \mathbb{T}_{s,i}(\lambda_o)  \lambda_o \, d\lambda_o} \\
&= 10^{0.4\times Z_{s,i}} \int F^{s,i}_\text{obs}(\lambda_o) \, \mathbb{B}_{s,i}(\lambda_o) \lambda_o \, d\lambda_o.
\end{split}
\end{equation}
where $Z_{s,i}$ is the zeropoint for this observation\footnote{$Z_{s,i} = 27.5 + m_{s,i}^\text{std}$, where $m_{s,i}^\text{std}$ is the reference magnitude in the passband, and 27.5 is a conventional scaling applied in SNANA files, i.e the flux ratios are multiplied by $10^{0.4\times27.5} = 10^{11}$ for convenience.}.  The passbands used in this analysis are described in \S \ref{sec:filters}.  We define the \emph{normalised} transmission function as:
\begin{equation}
\begin{split}
\mathbb{B}_{s,i}(\lambda_o) &\equiv \frac{\mathbb{T}_{s,i}(\lambda_o)}{\int F_{\text{std}}(\lambda_o) \, \mathbb{T}_{s,i}(\lambda_o)  \lambda_o \, d\lambda_o} \\
&= \frac{\mathbb{T}_{s,i}(\lambda_o)}{\int \mathbb{T}_{s,i}(\lambda_o) \lambda_o \, d\lambda_o}\times \frac{\int \mathbb{T}_{s,i}(\lambda_o) \lambda_o \, d\lambda_o}{\int F_{\text{std}}(\lambda_o) \, \mathbb{T}_{s,i}(\lambda_o)  \lambda_o \, d\lambda_o}.
\end{split}
\end{equation}
The model flux value can be converted to an apparent magnitude, on the system of the standard, like so:
\begin{equation}
\begin{split}
m_{s,i} &= -2.5 \log_{10}\left[f_{s,i}\right] + Z_{s,i} \\
&= -2.5\log_{10} \int F^{s,i}_\text{obs}(\lambda_o) \, \mathbb{B}_{s,i}(\lambda_o) \lambda_o \, d\lambda_o.
\end{split}
\end{equation}

Now we model the observable flux density $F_\text{obs}^{s,i}(\lambda_0)$ (per unit wavelength) for observation $i$ of SN $s$.  If the MJD date of $B$-band maximum is $T^{\text{max}}_s$, then we define the rest-frame phase of this observation as $t_s^i \equiv (T_s^i - T_s^\text{max}) / (1+z_s)$.  We denote the effective SED in the SN rest-frame, extinguished by host galaxy dust, as $S_s(t,\lambda_r)$.  The flux density of the light from SN $s$ at observed wavelength $\lambda_o$ and at time $T_s^i$ at the Earth is:
\begin{equation}\label{eqn:fluxdensity}
\begin{split}
F_{\text{obs}}^{s,i}(\lambda_o) &= (1+z_s)^{-1}\, 10^{-0.4 \,\mu_s}  \times S_s\left(t^i_s, \lambda_r = \frac{\lambda_o}{1+z_s}\right) \\&\times 10^{-0.4 \, A_{\text{MW}}^s \, \xi(\lambda_o; R_{\text{MW}})}
\end{split}
\end{equation}
The last term is the attenuation of flux by dust along the line-of-sight within the Milky Way Galaxy. The $V$-band Milky Way extinction is obtained from the reddening map \citep{schlafly11}, $A^s_{\text{MW}}  = E(B-V)^s_\text{MW}  \times R_{\text{MW}}$, and we adopt $R_{\text{MW}} = 3.1$ and the \citet{fitzpatrick99} extinction law for $\xi(\lambda_o; R_{\text{MW}})$.

The range in observed wavelength $\lambda_o$ over which the transmission is effectively non-zero is denoted as $[\lambda_o^\text{min}, \lambda_o^\text{max}]$.  The effective rest-wavelength range is then $[\lambda_r^{\text{min}} = \lambda_o^{\text{min}} / (1+z_s)$, $\lambda_r^{\text{max}} = \lambda_o^{\text{max}} / (1+z_s)]$.  Combining Eq. \ref{eqn:synflux} with Eq. \ref{eqn:fluxdensity}, we can rewrite the model flux for the $i$th observation of SN $s$ as an integral over the rest-frame wavelength $\lambda_r = \lambda_o/(1+z_s)$:
\begin{equation}\label{eqn:fluxint}
\begin{split}
f_{s,i} &= (1+z_s) \,10^{-0.4 \,\mu_s} \times 10^{+0.4 \cdot Z_{s,i}} \\
&\times \int_{\lambda_r^{\text{min}} }^{\lambda_r^{\text{max}}} S_s(t^i_s, \lambda_r ) \times 10^{-0.4 \, A_{\text{MW}}^s \, \xi[\lambda_r(1+z_s); R_{\text{MW}}]}\\ &\times \mathbb{B}_{s,i}[\lambda_r (1+z_s)] \, \lambda_r \, d\lambda_r.
\end{split}
\end{equation}
This calibrated flux is measured with some photometric noise with a given variance $\sigma^2_{s,i}$, and we assume a Gaussian sampling distribution for the measured flux $\hat{f}_{s,i}$:
\begin{equation}\label{eqn:fluxlkhd}
P(\hat{f}_{s,i}  | f_{s,i}) = N(\hat{f}_{s,i}  | \, f_{s,i}, \sigma^2_{s,i}).
\end{equation}
For all the observations $i$ (across all observation times and filters) of SN $s$, the measurement likelihood is
\begin{equation}\label{eqn:meas_lkhd}
P(\bm{\hat{f}}_s | \, \bm{f}_s ) = \prod_i P(\hat{f}_s^i |\, f_s^i ),
\end{equation}
assuming independence of the flux measurement errors.

\subsection{Dust and Intrinsic Supernova SED Model}

The host-dust-extinguished SED is obtained from the intrinsic SED $S^\text{int}_s(t, \lambda_r)$ in the SN rest-frame via
\begin{equation}
S_s(t,\lambda_r) = S^\text{int}_s(t, \lambda_r) \times 10^{-0.4 \, A_{V}^s \, \xi(\lambda_r; R_V)}
\end{equation}
where $A_{V}^s$ is the host galaxy dust extinction and $\xi(\lambda_r; R_V)$ is the extinction law with parameter $R_V$.  We adopt the extinction law of \citet{fitzpatrick99}.  

Our model intrinsic SN spectral energy distribution is a function of rest-frame phase $t$ and $\lambda_r$.  We decompose it into a \emph{global} spectral template modified by \emph{individual} effects that vary per SN $s$.  
\begin{equation}
\begin{split}
S^\text{int}_s(t, \lambda_r) = {\color{blue}S_0(t,\lambda_r)} &\times {\color{blue}10^{-0.4 M_0}} \times {\color{blue}10^{-0.4 \, W_0(t,\lambda_r)}}\\
&\times {\color{red}10^{-0.4 \, \delta M_s}}\times {\color{red}10^{-0.4 \, \delta W_s(t,\lambda_r)}}
\end{split}
\end{equation}
where $M_0 \equiv -19.5$ is fixed normalisation factor and the fixed function $S_0(t,\lambda_r)$ is the updated spectral template of \citet{hsiaothesis}. This template spans 0.1 $\mu$m to 2.5 $\mu$m from $-$20d to $+$85d past {\it B}-band maximum, and was constructed from over 1,000 spectra, including NIR spectra from \citet{marion09}, using the procedure described in \citet{hsiao07}. It is arbitrarily normalised to have a $B$-band magnitude of zero at peak phase $t = 0$.

The {\color{blue} blue} terms on the top line altogether describe the \emph{global} spectral template.  They model the baseline mean intrinsic SED as the \citet{hsiaothesis} spectral template, normalised by $M_0$ and smoothly warped by the function $W_0(t,\lambda_r)$ to match the inferred mean intrinsic colours of the training sample. 

The {\color{red} red} terms on the bottom line describe the \emph{individual} effects, the modifications to the global SED that are specific to each  supernova $s$.  The $\delta M_s$ term corresponds to an overall shift of the log SED that is independent of phase and wavelength.  The function $\delta W_s(t, \lambda_r)$ corresponds to phase- and wavelength-dependent effects.  We further decompose this function as:
\begin{equation}
\delta W_s(t, \lambda_r) = \left[ \sum_{k=1}^K \theta_k^s \, W_k(t,\lambda_r) \right] + \epsilon_s(t,\lambda_r).
\end{equation}
The $W_k(t,\lambda_r)$ functions are the \emph{functional principal components} describing the major modes of $(t,\lambda_r)$-variation in the log SED underlying the light curves of individual SN $s$.  The $\theta_k^s$ coefficients are scores describing the degree of component $W_k(t,\lambda_r)$ present in SN $s$.  In this work, we mainly take $K = 1$. (In Appendix \ref{sec:w2}, we describe the $W_2(t, \lambda_r)$ inferred for the $K=2$ model.)  The functions $\epsilon_s(t, \lambda_r)$ describe the phase- and wavelength-dependent \emph{SED residuals} that are not captured by the other effects. The total residual SED function of a SN Ia is $\eta_s(t,\lambda_r) = \delta M_s + \epsilon_s(t,\lambda_r)$.

The above equations express a linear model for the logarithm of the host-dust-extinguished SN SED:
\begin{equation}\label{eqn:model_logsed}
\begin{split}
-2.5 &\log_{10} [S_s(t, \lambda_r)/S_0(t,\lambda_r)]  \
= M_0 + W_0(t,\lambda_r) \\
& + \delta M_s + \left[ \sum_{k=1}^K \theta_k^s \, W_k(t,\lambda_r) \right] + \epsilon_s(t,\lambda_r) \\
& + A_{V}^s \, \xi(\lambda_r; R_V) 
\end{split}
\end{equation}
Note that $M_0, W_0(t,\lambda_r), \delta M_s$, and $\delta W_s(t,\lambda_r)$ are in units of magnitude, like $\mu_s$ and $A_V^s$.
The advantage of modelling the logarithm of the SED is that we can easily preserve positive flux at all phases and wavelengths while having priors on the functional components $W_k(t,\lambda_r)$ and the latent principal component scores that span positive and negative reals. 

\subsection{Magnitude Approximation}

For the vast majority of the nearby training set used in this work, the flux data have high signal-to-noise.  Therefore, it is a good approximation to convert these data to magnitudes. The magnitude measurement and the variance of its measurement error are
$\hat{m}_{s,i} = -2.5 \log_{10}[ \hat{f}_{s,i} ]  + Z_{s,i}$ and
\begin{equation}
\sigma^2_{m,s,i} = \left( \frac{2.5}{\ln 10} \frac{\sigma_{s,i}}{\hat{f}_{s,i}} \right)^2.
\end{equation}
Transformation of the model flux (Eq. \ref{eqn:fluxint}) to the model magnitude $m_{s,i}$ yields
\begin{equation}
\begin{split}
m_{s,i} &= \mu_s + M_0 + \delta M_s \\ 
&- 2.5\log_{10}\Big[ (1+z_s) \int_{\lambda_r^{\text{min}}}^{\lambda_r^{\text{max}}}  S_0\left(t_s^i, \lambda_r\right) \\  
&\times 10^{-0.4\,[W_0(t_s^i, \lambda_r) + \delta W_s(t_s^i,\lambda_r) + A_V^s \xi(\lambda_r; R_V)] }\\
& \times 10^{-0.4\, A_{\text{MW}}^s\, \xi[\lambda_o; R_{\text{MW}}]} \times \mathbb{B}_{s,i}(\lambda_o) \, \, \lambda_r \, d\lambda_r \Big].
\end{split}
\end{equation}
Using this, we can change the measurement likelihood function, Eq. \ref{eqn:fluxlkhd}, to:
\begin{equation}\label{eqn:maglkhd}
P(\hat{m}_{s,i} | \, m_{s,i} ) = N( \hat{m}_{s,i}| \, m_{s,i}, \sigma_{m,s,i}^2).
\end{equation}
This form is useful since the model magnitude inside the likelihood is linear in some of the parameters. However, the full flux model (Eq. \ref{eqn:fluxlkhd}) allows us to use low signal-to-noise, or even negative, flux measurements which cannot be reliably converted into magnitudes with Gaussian errors.

\subsection{2D SED Surface Models}

We model the unknown functions $\{W_k(t,\lambda_r): k = 0,\ldots, K\}$, and $\{\epsilon_s(t,\lambda_r) : s = 1,\ldots, N_\text{SN} \}$ in a flexible, data-driven manner.  Each function is represented as a surface defined by a 2D grid of knots.  We specify a 2D grid as the Cartesian product of a 1D grid in rest-frame phase, $\bm{\tau}$, and a 1D grid in rest-frame wavelength $\bm{l}$.  Each 1D grid can be irregularly spaced.   Algorithmic details are described briefly in Appendix \ref{sec:2dspline}.  The essential idea is that a generic, smooth surface $g(t,\lambda_r)$ at any point $(t, \lambda_r)$ in the 2D domain of the SED can be modelled as $g(t, \lambda_r) = \bm{s}(\lambda_r; \bm{l})^T \, \bm{G} \, \bm{s}(t; \bm{\tau})$, where $\bm{s}(x; \bm{\xi})$ denotes the 1D natural cubic spline smoother (column) vector for knots $\bm{\xi}$ at evaluated at point $x$.  The knots matrix $\bm{G}$ has elements $G_{ij} = g( t = \tau_{j}, \lambda_r = l_i)$, which define the values the surface must pass through at the knot locations, and are parameters for inference.

Using this, we model the functions of phase and wavelength in terms of knot matrices $\{\bm{W}_k :  k = 0,\ldots, K\}$ and $\{\bm{E}_s : s = 1,\ldots, N_\text{SN}\}$, like so.  For the global correction to the mean template:
\begin{equation}
W_0(t,\lambda_r) =  \bm{s}(\lambda_r; \bm{l})^T \, \bm{W_0} \, \bm{s}(t; \bm{\tau}).
\end{equation}
For the functional components ($k = 1,\dots,K$),
\begin{equation}
W_k(t,\lambda_r) =  \bm{s}(\lambda_r; \bm{l})^T \, \bm{W_k} \, \bm{s}(t; \bm{\tau}).
\end{equation}
For the residual SED functions of each SN s,
\begin{equation}\label{eqn:epsilon}
\epsilon_s(t, \lambda_r) = \bm{s}(\lambda_r; \bm{l})^T \, \bm{E_s} \, \bm{s}(t; \bm{\tau}).
\end{equation}
These latent functions are determined by the unknown matrices $\{ \bm{W_k} : k = 0,\ldots, K\}$, and $\{ \bm{E_s} :  s = 1,\ldots, N_\text{SN}\}$, which are inferred as hyperparameters and latent variables.   

We specify a set of knots on a grid in rest-frame phase and wavelength.  The phase coordinates are, e.g. $\bm{\tau} = (-10, 0, 10, 20, 30, 40)$ days.  The phase spacing is justified since we know that SN Ia light curves vary smoothly on $\sim 10$ day timescales.  The wavelength coordinates are $\bm{l}$.  We place a knot at the central wavelengths of the filters $BVriYJH$ plus two outer knots bracketting these: $\bm{l} = (0.3, 0.43, 0.54, 0.62, 0.77, 1.04, 1.24, 1.65, 1.85) \, \mu$m.  The purpose of the first and last knots in wavelength is to ensure that our spline surfaces are defined throughout the entire first ($B$) and last ($H$) broadband filters.  To avoid degeneracies, we ``tie down'' the residual knot matrices at the first and last wavelength knots for every phase knot: $E_{s,ij} = 0$ if $i = 1$ or $i = \dim(\bm{l}), \forall\,j$.

\subsection{Population Distributions and Hyperpriors}

We specify the population distributions on the latent parameters of individual supernovae.  

For the latent functional SED effects, following the probabilistic PCA formulation \citep{tipping99}, we adopt the standard Gaussian prior $\theta_s^k \sim N(0,1)$ for the individual score of each SN $s$ in each component $k = 1,\ldots,K$.  Thus, the resulting functions $W_k(t,\lambda_r) (k \ge 1)$ are not scaled to have unit norm, as they would be in standard PCA.  Rather, because the latent scores $\theta_s^k$ are normalised to have a population variance of one, $W_k(t,\lambda_r)$ absorbs a factor of the population standard deviation in that component. A ``$1\sigma$'' effect of the $k$-th component on the SED is thus computed from $\theta_k \, W_k(t, \lambda_r)$ by varying $\Delta \theta_k \pm 1$ around the mean.

For the $\bm{W}_k$ matrices that parametrize our functional components, we use an independent standard normal hyperprior on the value of each knot: $W_{k,ij} \sim N(0,1)$.  This is a weak constraint, since we have scaled the problem to expect these variations to be of order tenths of a magnitude. 

For the residual SED perturbations, we assume a multivariate Gaussian distribution on the column-wise vectorisation of each residual matrix $\bm{E_s}$:
\begin{equation}\label{eqn:resid_pop}
\bm{e}_s = \text{vec}[\bm{E_s}] \sim N(\bm{0}, \bm{\Sigma}_\epsilon).
\end{equation}
A matrix $\bm{\Gamma}(t, \lambda_r; \bm{\tau}, \bm{l})$ can be constructed so that Eq. \ref{eqn:epsilon} can be written equivalently as,
\begin{equation}\label{eqn:espilon_alt}
\epsilon_s(t, \lambda_r) = \bm{\Gamma}(t, \lambda_r; \bm{\tau}, \bm{l}) \, \bm{e}_s.
\end{equation}
While Eq. \ref{eqn:epsilon} and Eq. \ref{eqn:espilon_alt} are equivalent, Eq. \ref{eqn:epsilon} is the more compact representation, since $\bm{\Gamma}(t, \lambda_r; \bm{\tau}, \bm{l})$ tends to a very large (but sparse) matrix.  However, Eq. \ref{eqn:espilon_alt} is useful, because, together with the residual distribution Eq. \ref{eqn:resid_pop}, it implies that the residual functions $\epsilon_s(t, \lambda_r)$ are realisations of a Gaussian process \citep[GP;][]{rasmussen05}:
\begin{equation}
\epsilon_s(t, \lambda_r) \sim \mathcal{GP}[ \bm{0},  k(t,\lambda_r; t', \lambda_r') ]
\end{equation}
with a zero prior mean and a non-stationary kernel for the covariance of the residuals at any two coordinates:
\begin{equation}\label{eqn:gp_kernel}
\begin{split}
k_\epsilon(t,\lambda_r; t', \lambda_r') &\equiv \text{Cov}[ \epsilon_s(t, \lambda_r), \epsilon_s(t', \lambda_r')]  \\
&= \bm{\Gamma}(t, \lambda_r; \bm{\tau}, \bm{l}) \, \bm{\Sigma}_\epsilon \,  \bm{\Gamma}(t', \lambda_r'; \bm{\tau}, \bm{l})^T.
\end{split}
\end{equation}
We adopt this non-stationary covariance structure rather than the more popular stationary kernels, such as squared exponential, since we do not expect the complex physical mechanisms of SN Ia explosions to generate statistical properties that are invariant to phase or wavelength shifts. GPs have been previously used to model spectra, e.g. by \citet{czekala15,czekala17}.

The covariance matrix $\bm{\Sigma}_\epsilon$ encodes the variances and correlation structures of the residual functions: $\bm{\Sigma}_\epsilon = \text{diag}(\bm{\sigma}_\epsilon) \, \bm{R}_\epsilon \, \text{diag}(\bm{\sigma}_\epsilon)$.  Following the separation strategy proposed by \citet*{barnard00}, we specify separate priors on the standard deviation parameters $\bm{\sigma}_\epsilon$ and the correlation matrix $\bm{R}_\epsilon$.  For each $q$-th element $\sigma_{\epsilon,q} \ge 0$, we adopt a weakly-informative half-Cauchy hyperprior with unit scale \citep{gelman06,polson12}, i.e. $P(\sigma_{\epsilon,q}) = HC(\sigma_{\epsilon,q} |\, a=1)$, with probability density
\begin{equation}\label{eqn:halfcauchy}
HC(x | \, a) \propto  (a^2+x^2)^{-1}
\end{equation}
for $x \ge 0$, and zero otherwise. This hyperprior is proper, and relatively flat for small $x$.  It is sensible because we have scaled the problem to expect $\sigma_{\epsilon,q}$ to be less than a magnitude.  For the correlation matrix, we adopt the LKJ hyperprior as implemented in \textsc{Stan} and derived from \citet*{lkj09}.
\begin{equation}
P(\bm{R}_\epsilon) \propto | \bm{R}_\epsilon |^{\eta-1}
\end{equation}
with $\eta = 1$.  This places a uniform prior on positive semi-definite correlation matrices.

The $\delta M_s$ terms model a phase- and wavelength-independent shift of the SED in overall log luminosity.  Since these shifts are indistinguishable from the effect of distance on the apparent light curves, this propagates into an uncertainty floor on photometric distance estimates.  We model the population of these shifts as $\delta M_s \sim N(0, \sigma_0^2)$ and estimate their variance $\sigma_0^2$ as a hyperparameter.  We use a weak half-Cauchy prior  (Eq \ref{eqn:halfcauchy}) on $\sigma_0$ with scale $a = 0.1$, since we expect this to be of order a tenth of magnitude.

We assume that host galaxy extinction $A_V^s$ is drawn from an exponential distribution with mean extinction hyperparameter $\tau_A$:
\begin{equation}
P(A_V^s | \tau_A) = \tau_A^{-1} \, \exp(-A_V^s / \tau_A),
\end{equation}
for $A_V \ge 0$ and zero otherwise. 
This is a sensible choice, since the true $A_V^s$ must be non-negative, and we expect the most lines of sight through the host galaxies to pass through little dust, with the probability density decreasing with increasing column density.  This model distribution has been used before by, e.g. \citet{jha07} and \citet{mandel09}.
The hyperprior we adopt for $\tau_A$  is also a unit half-Cauchy, $P(\tau_A) = HC(\tau_A, 1)$, reflecting our expectations that the typical $\tau_A$ is on the order of tenths of a magnitude.  For the unknown $R_V$, we assume a single global value with a uniform hyperprior $R_V \sim U(1,5)$ reflecting a wide range of possible values.  In the future, we can expand our framework to allow per-SN variation in $R_V^s$ by modelling and inferring their population distribution, as was done previously by \cite{mandel11}.

\subsection{External Distance Constraints}

In the training phase, we use estimates $\hat{\mu}_{\text{ext},s}$ of the SN distance moduli that are external to the photometric SN data, as described in \citet{avelino19}.  We assume they have Gaussian errors around the true distance modulus.

For the vast majority of the training set, we utilise the redshift as an indicator of distance conditional on the fiducial cosmological model $\hat{\mu}_{\text{ext},s} = \mu_{\Lambda CDM}(z_s)$ with $\Omega_M = 0.28$ and $\Omega_\Lambda = 0.72$. However, at these redshifts $z < 0.04$, these distances are relatively insensitive to the cosmological parameters, other than $H_0$ which only sets an overall scale for all absolute magnitudes, and for which we adopt $73.24 \text{ km s}^{-1} \text{ Mpc}^{-1}$ \citep{riess16}.  These redshifts are corrected to the CMB frame and corrected for bulk flows.  The distance modulus uncertainty, due to errors in observed redshift $z_s$ as estimates for the cosmological redshift $z_s^c$, from redshift and peculiar velocity uncertainties is
\begin{equation}\label{eqn:sigma_ext}
\hat{\sigma}^2_{\text{ext},s} \approx \left(\frac{5}{z_s \ln 10}\right)^2 \left[ \sigma_\text{pec}^2 / c^2 + \sigma^2_{z,s} \right],
\end{equation}
where we have adopted $\sigma_\text{pec} = 150 \text{ km s}^{-1}$ \citep{carrick15}.  The external distance constraint can be expressed as $P(\mu_s | \, z_s) \propto N(\hat{\mu}_{\text{ext},s} |\, \mu_s, \hat{\sigma}^2_{\text{ext},s})$ after marginalising out the unknown $z_s^c$.

For eight SNe Ia in our training set at $z < 0.01$, we use external distance estimates $\hat{\mu}_{\text{ext},s}$ from available redshift-independent measures (e.g. Cepheids), and their uncertainties $\hat{\sigma}^2_{\text{ext},s}$, as listed in Table 4 of \citet{avelino19}.   These external distance constraints can be expressed as $P(\mu_s | \, \hat{\mu}_{\text{ext},s}) \propto N(\hat{\mu}_{\text{ext},s} | \mu_s, \hat{\sigma}^2_{\text{ext},s})$.

\subsection{The Global Joint Posterior Distribution}

For an individual SN $s$, the joint probability density of its flux light curve data $\bm{\hat{f}}_s$ and its latent parameters $(\bm{\theta}_s,  A_V^s, \mu_s, \bm{e}_s)$ conditional on the population hyperparameters and redshift is
\begin{equation}
\begin{split}
P(\bm{\hat{f}}_s&, \bm{\theta}_s,  \bm{e}_s, \delta M_s, A_V^s, \mu_s | \,\bm{W}_{0:K}, \bm{\Sigma}_\epsilon, \sigma_0, \tau_A, R_V; z_s) \\
&= P(\bm{\hat{f}}_s | \, \bm{\theta}_s, \bm{e}_s, \delta M_s, A_V^s, \mu_s; \bm{W}_{0:K}, R_V)  \\
&\times  P(\bm{\theta}_s) P(\bm{e}_s | \,\bm{\Sigma}_\epsilon) P(\delta M_s |\, \sigma_0) P(A_V^s | \,\tau_A)  P(\mu_s | \, z_s)\\
\end{split}
\end{equation}
where $\bm{W}_{0:K} \equiv  \{\bm{W}_0,  \bm{W}_1, \ldots, \bm{W}_K \}$ is the collection of matrices describing the intrinsic mean and $K$ functional components of the SED, and $\bm{\theta}_s \equiv (\theta_1^s,\ldots, \theta_K^s)^T$ are the intrinsic coefficients of SN $s$.  The first factor on the right-hand side is the data likelihood defined by Eqs. \ref{eqn:meas_lkhd}, \ref{eqn:fluxint}, and \ref{eqn:model_logsed}.  For the eight SN with redshift-independent distance measurements, we replace $P(\mu_s | \, z_s)$ with $P(\mu_s | \, \hat{\mu}_{\text{ext},s})$.

The global posterior distribution of all the latent variables of individual supernovae and the population hyperparameters given the data, external distance constraints, and redshifts is 
\begin{equation}\label{eqn:globalpost}
\begin{split}
&P(\{\bm{\theta}_s, \bm{e}_s, \delta M_s, A_V^s, \mu_s \}; \bm{W}_{0:K}, \bm{\Sigma}_\epsilon, \sigma_0, \tau_A, R_V\, | \, \{ \bm{\hat{f}_s}; z_s \}) \propto \\
&\Big[ \prod_{s=1}^{N_{\text{SN}}} P(\bm{\hat{f}}_s, \bm{\theta}_s,\bm{e}_s, \delta M_s,  A_V^s, \mu_s,  | \,\bm{W}_{0:K}, \bm{\Sigma}_\epsilon, \sigma_0, \tau_A, R_V; z_s)   \Big] \\
&\times P(\bm{W}_{0:K}) \,P(\bm{\sigma}_{\epsilon,q}) \, P(\bm{R}_\epsilon) \, P(\sigma_0) \,P(\tau_A) \,P(R_V).
\end{split}
\end{equation}
This global posterior distribution is the objective function for training our model to learn the population hyperparameters, covariance structure, and SED components while marginalising over the latent variables of individual SNe Ia.  It provides a coherent, probabilistic quantification of uncertainty of over all parameters and hyperparameters.

\subsection{Photometric Distance Estimation}

The training process gives us posterior estimates of the hyperparameters $\bm{\hat{H}}  \equiv (\bm{\hat{W}}_{0:K}, \bm{\hat{\Sigma}}_\epsilon, \hat{\sigma}_0^2, \hat{\tau}_A, \hat{R}_V)$ marginalised over all latent variables in the sample.  For simplicity, we take the posterior means of these hyperparameters as point estimates.  Under distance-fitting mode, we condition on the hyperparameters, and the posterior density of the latent parameters of any given SN $s$ is
\begin{equation}\label{eqn:distfitpost}
\begin{split}
P&(\bm{\theta}_s, \bm{e}_s, \delta M_s, A_V^s, \mu_s | \,\bm{\hat{f}}_s; \bm{\hat{H}} ) \\
&\propto P(\bm{\hat{f}}_s | \, \bm{\theta}_s, \bm{e}_s, \delta M_s, A_V^s, \mu_s; \bm{\hat{W}}_{0:K}, \hat{R}_V) \\
& \times P(\bm{\theta}_s) \times P(\bm{e}_s | \,\bm{\hat{\Sigma}}_\epsilon) \times P(\delta M_s | \, \hat{\sigma}_0) \times P(A_V^s | \,\hat{\tau}_A) , \\
\end{split}
\end{equation}
where we omit any external distance constraint.  By sampling this joint posterior, we can approximate the marginal posterior density of the photometric distance modulus,
\begin{equation}\label{eqn:photdist}
\begin{split}
P&(\mu_s | \,\bm{\hat{f}}_s; \bm{\hat{H}} ) = \\
& \int P(\bm{\theta}_s, \bm{e}_s, \delta M_s, A_V^s, \mu_s | \,\bm{\hat{f}}_s; \bm{\hat{H}} ) \, d\bm{\theta}_s \, d\bm{e}_s \, d\delta M_s \, dA_V^s  ,
\end{split}
\end{equation}
as well as its posterior summaries such as the mean and variance via marginalisation.  However, this distribution is not necessarily Gaussian, nor is it required to be, since dust effects introduce some asymmetry.

\begin{figure*}
	\includegraphics[scale=0.4]{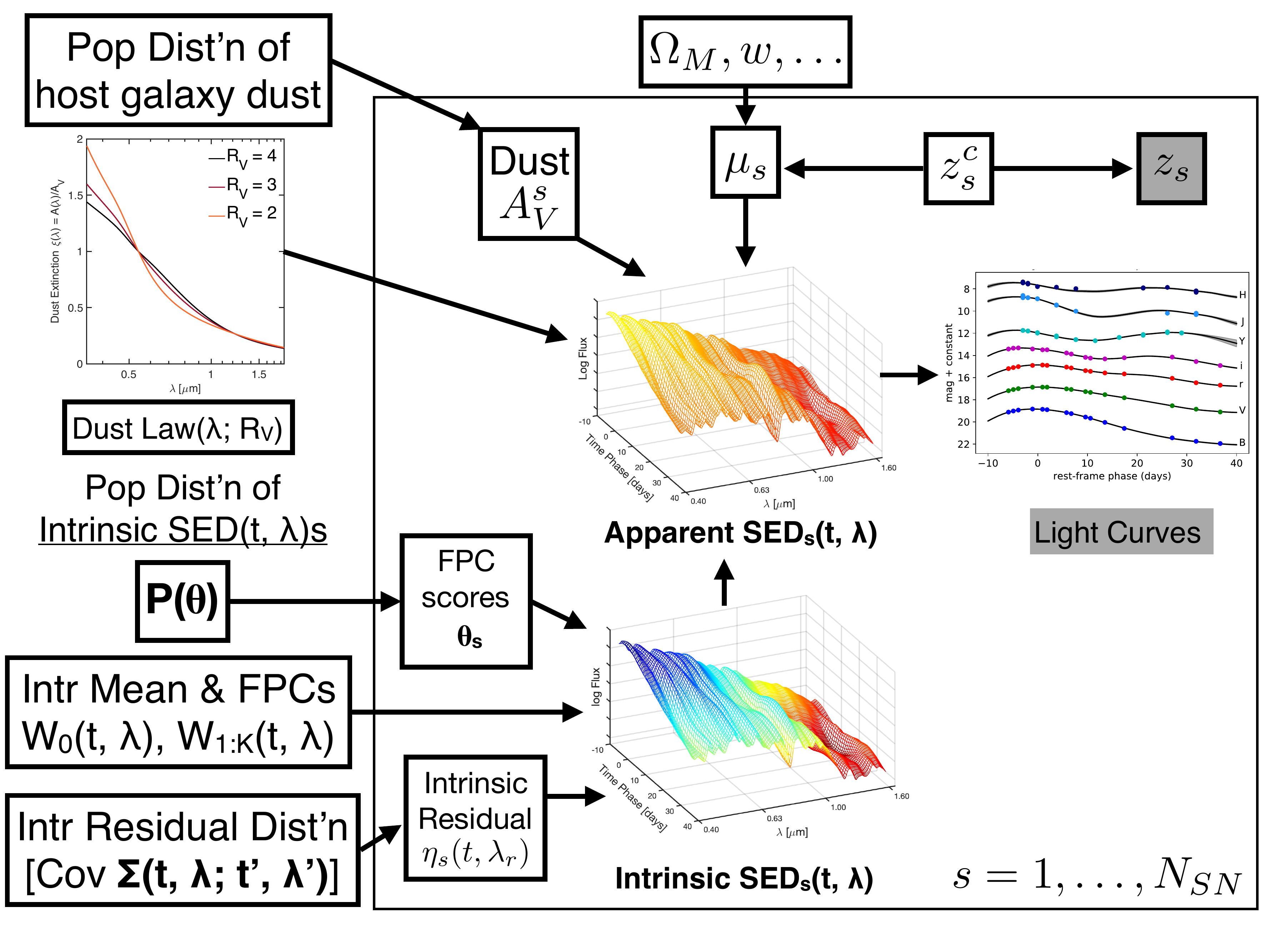}
	\caption{ Probabilistic graphical model depicting the hierarchical \textsc{BayeSN} model for optical-NIR SN Ia light curve data. Each open box presents a set of unknown parameters or hyperparameters, each grey-shaded box represents observed data, and the arrows indicate relations of conditional probability.  Parameters within the plate, labelled $s = 1,\ldots,N_\text{SN}$, are repeated for every SN $s$, whereas parameters outside the plate represent global or population parameters. The intrinsic SED of a single SN Ia $s$ is constructed from the mean SED and functional principal components $W_{0:K}(t,\lambda_r)$, a draw of the FPC scores $\bm{\theta}_s$ from its population distribution, and a draw of an intrinsic residual SED function from its population distribution described by a covariance function over time and wavelength. The host galaxy dust extinction $A_V^s$ of a SN $s$ is drawn from a population distribution of extinction values, and the dust law is parametrized by an unknown $R_V$. The effects of dust and distance modulus $\mu_s$ on the intrinsic SED combine (with appropriate redshifting and time dilation) to yield the apparent SN SED.  This is observed with some observational cadence and noise through the observer's filter functions to yield the optical and NIR light curve data.  During training, the distance is constrained externally to the light curve by the cosmological redshift and the fiducial cosmological parameters (fixed in this low-$z$ analysis). The redshift is observed with some error, including uncertainties in local peculiar velocities. The hierarchical global posterior density (Eq. \ref{eqn:globalpost}) estimates the unknown latent variables and hyperparameters conditional on the observed data  of the entire SN Ia sample.}
	\label{fig:pgm}
\end{figure*}

In principle, instead of fixing the hyperparameters to their posterior means from training, we could use the samples of the joint posterior over the hyperparameters (Eq. \ref{eqn:globalpost}) to incorporate their uncertainties into the photometric distance estimates.  However, this is a more computationally burdensome process, and we have found the posterior means to be sufficient for our purposes.

\subsection{Probabilistic Graphical Model}

Our hierarchical Bayesian model can be depicted with a type of probabilistic graphical model called a directed acyclic graph, shown in Fig. \ref{fig:pgm}.  Each open box presents a set of unknown parameters or hyperparameters, each grey-shaded box represents observed data, and the arrows indicate relations of conditional probability.  Parameters within the plate, labelled $s = 1,\ldots,N_\text{SN}$, are repeated for every SN $s$, whereas parameters outside the plate represent global or population parameters. The intrinsic SED of a single SN Ia $s$ is constructed from the mean SED and functional principal components $W_{0:K}(t,\lambda_r)$, a draw of the FPC scores $\bm{\theta}_s$ from its population distribution, and a draw of an intrinsic residual SED function from its population distribution described by a covariance function over time and wavelength. The host galaxy dust extinction $A_V^s$ of a SN $s$ is drawn from a population distribution of extinction values, whereas the unknown $R_V$, parametrizing the dust law, is given a wide prior.    The effects of dust and distance modulus $\mu_s$ on the intrinsic SED combine (with appropriate redshifting and time dilation) to yield the apparent SN SED.  This is observed with some cadence and noise through the observer's filter functions to yield the optical and NIR light curve data.  During training, the distance is constrained externally to the light curve by the cosmological redshift and the fiducial cosmological parameters (fixed in this low-$z$ analysis). The redshift is observed with some uncertainty due to local peculiar velocities. Bayesian inference with the hierarchical model solves the inverse problem through the computation of the posterior probability density (Eq. \ref{eqn:globalpost}) of the unknown latent variables and hyperparameters conditional on the observed data of the entire SN Ia sample.

\section{Data}\label{sec:data}

\subsection{Optical and NIR Light Curve Data}

We use the compilation of low-$z$ SNe Ia with joint optical and NIR light curves described in \citet{avelino19}.   For our purposes, we define the optical as the $BVRI$ filters and the NIR as the $YJH$.  In our various analyses below, we fit either the available optical ($BVRI$) or optical+NIR ($BVRIYJH$) for a given SNe Ia.  The selection criteria and cuts were described in \citet{avelino19} and detailed information on the specific SNe is listed in their Tables 2 \& 3.  In particular, a colour excess cut $E(B-V)_\text{host} \le 0.4$ was applied for consistency with the cosmological sample.  For each SN Ia, we use published optical and NIR data only from the same survey; we do not mix data sources within a single SN.  Consequently, SN 2005bo, SNF20080514-002, SN 2010iw, SN 2010kg, SN 2011ao, SN 2011B, SN 2011by, SN 2011df were removed because they had NIR data, but no published optical data, from the CfA.  SN 2006bt was removed because it is a known peculiar supernova \citep{foley10_06bt}.   We failed to fit the light curves of SN 2000E \citep{valentini03} with our current model (although it was not an outlier in the Hubble Diagram), so we have omitted it to avoid biasing the training.

The resulting sample comprises 79 SNe Ia with joint optical and NIR light curves.  \citet{avelino19} further defined a subset with NIR data near maximum light (see their Table 13). To enter into this subset, a SN Ia was required to have at least one NIR observation at least 2.5 days before maximum light.   We have a total of 48 SNe Ia in this cut, which we refer to as ``NIR@max''.  All SNe Ia in the full sample have some NIR data available regardless of the phase of the first NIR observation.  These SNe Ia and cuts are listed in Table \ref{table1}.  Additional information can be found in \citet{avelino19}.  The full dataset consists of 22 SN from the CfA Supernova Program \citep[CfA;][]{jha99,wood-vasey08, hicken09a, hicken12, friedman15}, 44 from the Carnegie Supernova Project \citep[CSP;][]{krisciunas17}, 8 from the Las Campanas Observatory \citep[K04a,b:][]{krisciunas04b,krisciunas04c}, as well as 5 others from individual papers in the literature (K03: \citealt{krisciunas03}; P08: \citealt{pignata08}; St07: \citealt{stanishev07}; L09: \citealt{leloudas09}; K07: \citealt{krisciunas07}).

The size of our training set reflects the recent progress of ground-based surveys in accumulating quality joint optical and NIR SN Ia light curve data \citep{friedman15, krisciunas17}.  The number of SNe Ia in our current compilation more than doubles those used to train previous NIR-capable light curve models.  The training set for the first \textsc{BayeSN} models included $37$ SNe Ia with both optical and NIR coverage \citep{mandel09, mandel11}, and the training set for SNooPy comprised $\lesssim 30$ SNe Ia \citep{burns11}.  Further increases in the training set will soon be possible with forthcoming data from CSP-II \citep{phillips19} and the \textit{Supernovae in the Infrared avec Hubble} (SIRAH) program (HST GO-15889, P.I. S. Jha).

\subsection{Passband Throughput}\label{sec:filters}

For each observation in the data compilation, we specify a model for the passband throughput.  A model passband throughput is needed to forward model the observed flux, regardless of whether that flux is reported in the natural system of the telescope or transformed on to a ``standard'' system such as SDSS \textit{ugriz} \citep{Fukugita96}. For a measurement reported on the natural system of a telescope, the total passband throughput must include all terrestrial elements of the measurement chain -- site atmospheric transmission, mirror reflectivity, filter transmission, transmission of camera optics, and detector quantum efficiency. For measurements reported on a standard system, the passband throughput must reflect the original measurement chain used to observe the standard stars that were employed in calibrating the SN flux, rather than the measurement chain of the facility used to observe the SN itself.

For the Carnegie Supernova Project and related objects observed at Las Campanas Observatory \citep{krisciunas17, krisciunas04b,krisciunas04c}, we use the total natural system passband throughputs\footnote{\url{https://csp.obs.carnegiescience.edu/data/filters}} as defined in the implementation of the SNooPy \citep{burns11}. We take care to include any changes in the CSP passband throughputs when a filter was replaced. For the NIR SNe observed by the CfA using the 1.3m PAIRITEL telescope at Mt. Hopkins \citep{wood-vasey08, friedman15}, we use the natural system passband throughputs measured by the 2MASS project\footnote{\url{https://old.ipac.caltech.edu/2mass/releases/second/doc/sec3_1b1.html\#s18}}, which used the same facility. Finally, for objects observed by the CfA Supernova Program \citep{jha99, hicken09a, hicken12} and remaining literature objects (K03, K07, St07, P08, and L09) we use published the \emph{standard} system photometry and model the passband throughput using the shifted Bessell filters described in \citet{Stritzinger05}. While the CfA SN program published both natural and standard system photometry, and the former is generally preferred as it avoids some potential systematic errors in transforming the flux, using the natural system photometry relies on having a good description of the passband throughput of the natural system. Unfortunately, there are no determinations of all the elements in the measurement chain for objects observed by the CfA SN survey, and the current model for passband throughput included in the SNDATA repository\footnote{\url{http://snana.uchicago.edu/downloads/SNDATA_ROOT.tar.gz}} does not include any model for the site atmosphere at all. The CfA Supernova measurements were the result of an extensive effort over almost two decades with four separate cameras,  through a variety of filters, using a telescope that underwent numerous mirror coatings, and the provenance of each measurement cannot easily be determined retrospectively. By contrast, the standard system photometry for CfA objects is known to be consistent with standard system photometry measured by the CSP and LOSS \citep{ganeshalingam10}. Thus, we prefer to use the standard system photometry over the natural system photometry in this work. 
\begin{table*}
	\centering
	\caption{Table of supernovae}
	\begin{threeparttable}\label{table1}
		\begin{tabular}{l c c c c c c c c} \toprule
			SN & source & cut & filters & $z_\text{CMB}$\tnote{a} & $\hat{\mu}_\text{ext}\tnote{b}$ & $\hat{\mu}_\text{phot} \text{ (resub)}$\tnote{c} &  $\hat{\mu}_\text{phot} \text{ (CV)}$\tnote{d} \\ 
			\midrule
SN1998bu & CfA & NIR@max & $BVRIJH$ & 0.003 & $30.07 \pm 0.20$ & $29.99 \pm 0.10$ & $29.96 \pm 0.09$ \\
SN1999ee & K04a & NIR@max & $BVRIJH$ & 0.011 & $33.33 \pm 0.14$ & $33.25 \pm 0.10$ & $33.21 \pm 0.09$ \\
SN1999ek & K04b & NIR@max & $BVRIJH$ & 0.018 & $34.34 \pm 0.09$ & $34.18 \pm 0.10$ & $34.18 \pm 0.10$ \\
SN2000bh & K04a & - & $BVRIYJH$ & 0.024 & $35.00 \pm 0.06$ & $34.94 \pm 0.10$ & $34.93 \pm 0.08$ \\
SN2000ca & K04a & NIR@max & $BVRIJH$ & 0.024 & $34.99 \pm 0.06$ & $35.00 \pm 0.10$ & $34.99 \pm 0.10$ \\
SN2001ba & K04a & NIR@max & $BVIJH$ & 0.030 & $35.51 \pm 0.05$ & $35.66 \pm 0.09$ & $35.66 \pm 0.09$ \\
SN2001bt & K04b & NIR@max & $BVRIJH$ & 0.014 & $33.85 \pm 0.11$ & $33.79 \pm 0.10$ & $33.80 \pm 0.09$ \\
SN2001cn & K04b & - & $BVRIJH$ & 0.015 & $34.03 \pm 0.10$ & $33.94 \pm 0.10$ & $33.96 \pm 0.10$ \\
SN2001cz & K04b & NIR@max & $BVRIJH$ & 0.017 & $34.25 \pm 0.09$ & $33.96 \pm 0.10$ & $33.95 \pm 0.10$ \\
SN2001el & K03 & NIR@max & $BVRIJH$ & 0.004 & $31.31 \pm 0.04$ & $31.28 \pm 0.10$ & $31.17 \pm 0.09$ \\
SN2002dj & P08 & NIR@max & $BVRIJH$ & 0.008 & $32.65 \pm 0.40$ & $32.95 \pm 0.10$ & $32.95 \pm 0.10$ \\
SN2003du & St07 & - & $BVRIJH$ & 0.009 & $32.92 \pm 0.06$ & $32.87 \pm 0.10$ & $32.86 \pm 0.09$ \\
SN2003hv & L09 & - & $BVRIYJH$ & 0.005 & $31.15 \pm 0.25$ & $31.30 \pm 0.09$ & $31.34 \pm 0.10$ \\
SN2004S & K07 & - & $BVRIJH$ & 0.011 & $33.23 \pm 0.14$ & $33.27 \pm 0.10$ & $33.24 \pm 0.10$ \\
SN2004ef & CSP & - & $BVriYJH$ & 0.030 & $35.50 \pm 0.05$ & $35.52 \pm 0.09$ & $35.50 \pm 0.08$ \\
SN2004eo & CSP & NIR@max & $BVriYJH$ & 0.015 & $34.00 \pm 0.10$ & $33.82 \pm 0.10$ & $33.88 \pm 0.09$ \\
SN2004ey & CSP & NIR@max & $BVriYJH$ & 0.015 & $34.02 \pm 0.10$ & $34.12 \pm 0.10$ & $34.11 \pm 0.08$ \\
SN2004gs & CSP & - & $BVriYJH$ & 0.029 & $35.39 \pm 0.05$ & $35.38 \pm 0.10$ & $35.39 \pm 0.09$ \\
SN2005cf & CfA & NIR@max & $BVr'i'JH$ & 0.007 & $32.26 \pm 0.10$ & $32.30 \pm 0.09$ & $32.31 \pm 0.10$ \\
SN2005el & CSP & NIR@max & $BVriYJH$ & 0.015 & $34.00 \pm 0.10$ & $33.98 \pm 0.09$ & $34.02 \pm 0.09$ \\
SN2005iq & CSP & NIR@max & $BVriYJH$ & 0.034 & $35.74 \pm 0.05$ & $35.88 \pm 0.09$ & $35.90 \pm 0.09$ \\
SN2005kc & CSP & NIR@max & $BVriYJH$ & 0.015 & $33.89 \pm 0.11$ & $33.75 \pm 0.10$ & $33.74 \pm 0.09$ \\
SN2005ki & CSP & NIR@max & $BVriYJH$ & 0.020 & $34.63 \pm 0.08$ & $34.62 \pm 0.10$ & $34.62 \pm 0.09$ \\
SN2005lu & CSP & - & $BVriY$ & 0.032 & $35.62 \pm 0.05$ & $35.71 \pm 0.12$ & $35.72 \pm 0.11$ \\
SN2005na & CfA & - & $BVr'i'JH$ & 0.027 & $35.28 \pm 0.06$ & $35.23 \pm 0.11$ & $35.24 \pm 0.11$ \\
SN2006D & CfA & NIR@max & $BVr'i'JH$ & 0.009 & $32.84 \pm 0.17$ & $32.91 \pm 0.09$ & $32.89 \pm 0.09$ \\
SN2006N & CfA & - & $BVr'i'JH$ & 0.015 & $33.89 \pm 0.11$ & $33.82 \pm 0.10$ & $33.78 \pm 0.10$ \\
SN2006ac & CfA & - & $BVr'i'JH$ & 0.024 & $34.98 \pm 0.06$ & $35.08 \pm 0.10$ & $35.06 \pm 0.10$ \\
SN2006ax & CSP & NIR@max & $BVriYJH$ & 0.018 & $34.36 \pm 0.09$ & $34.31 \pm 0.09$ & $34.30 \pm 0.10$ \\
SN2006bh & CSP & NIR@max & $BVriYJH$ & 0.011 & $33.24 \pm 0.14$ & $33.34 \pm 0.09$ & $33.33 \pm 0.10$ \\
SN2006cp & CfA & - & $BVr'i'JH$ & 0.022 & $34.84 \pm 0.07$ & $34.97 \pm 0.10$ & $34.98 \pm 0.12$ \\
SN2006ej & CSP & - & $BVriYJH$ & 0.021 & $34.66 \pm 0.07$ & $34.67 \pm 0.10$ & $34.67 \pm 0.10$ \\
SN2006kf & CSP & NIR@max & $BVriYJH$ & 0.019 & $34.53 \pm 0.08$ & $34.71 \pm 0.10$ & $34.72 \pm 0.08$ \\
SN2006lf & CfA & NIR@max & $BVr'i'JH$ & 0.012 & $33.49 \pm 0.13$ & $33.52 \pm 0.10$ & $33.53 \pm 0.09$ \\
SN2007A & CSP & NIR@max & $BVriYJH$ & 0.017 & $34.27 \pm 0.09$ & $34.25 \pm 0.10$ & $34.24 \pm 0.10$ \\
SN2007af & CSP & NIR@max & $BVriYJH$ & 0.006 & $31.79 \pm 0.05$ & $31.94 \pm 0.09$ & $31.98 \pm 0.09$ \\
SN2007ai & CSP & NIR@max & $BVriYJH$ & 0.033 & $35.69 \pm 0.05$ & $35.52 \pm 0.09$ & $35.52 \pm 0.09$ \\
SN2007as & CSP & NIR@max & $BVriYJH$ & 0.018 & $34.41 \pm 0.08$ & $34.42 \pm 0.09$ & $34.43 \pm 0.09$ \\
SN2007bc & CSP & NIR@max & $BVriYJH$ & 0.021 & $34.72 \pm 0.07$ & $34.74 \pm 0.10$ & $34.73 \pm 0.09$ \\
SN2007bd & CSP & NIR@max & $BVriYJH$ & 0.031 & $35.57 \pm 0.05$ & $35.60 \pm 0.10$ & $35.57 \pm 0.10$ \\
SN2007ca & CSP & NIR@max & $BVriYJH$ & 0.015 & $33.89 \pm 0.11$ & $34.04 \pm 0.10$ & $34.04 \pm 0.08$ \\
SN2007co & CfA & - & $BVr'i'JH$ & 0.027 & $35.30 \pm 0.06$ & $35.43 \pm 0.10$ & $35.42 \pm 0.10$ \\
SN2007cq & CfA & - & $BVr'i'JH$ & 0.025 & $35.11 \pm 0.06$ & $34.87 \pm 0.10$ & $34.86 \pm 0.11$ \\
SN2007jg & CSP & NIR@max & $BVriYJH$ & 0.038 & $36.02 \pm 0.04$ & $36.14 \pm 0.10$ & $36.16 \pm 0.09$ \\
SN2007le & CSP & NIR@max & $BVriYJH$ & 0.006 & $32.13 \pm 0.24$ & $32.20 \pm 0.09$ & $32.20 \pm 0.10$ \\
SN2007qe & CfA & - & $BVr'i'JH$ & 0.024 & $34.96 \pm 0.07$ & $35.18 \pm 0.10$ & $35.19 \pm 0.09$ \\
SN2007sr & CSP & - & $BVriYJH$ & 0.004 & $31.29 \pm 0.11$ & $31.62 \pm 0.09$ & $31.63 \pm 0.09$ \\
SN2007st & CSP & - & $BVriYJH$ & 0.021 & $34.72 \pm 0.07$ & $34.42 \pm 0.10$ & $34.40 \pm 0.09$ \\
SN2008C & CSP & - & $BVriYJH$ & 0.018 & $34.31 \pm 0.09$ & $34.37 \pm 0.10$ & $34.39 \pm 0.09$ \\
SN2008af & CfA & - & $BVr'i'JH$ & 0.034 & $35.78 \pm 0.05$ & $35.66 \pm 0.12$ & $35.63 \pm 0.11$ \\
SN2008ar & CSP & NIR@max & $BVriYJH$ & 0.029 & $35.42 \pm 0.05$ & $35.30 \pm 0.10$ & $35.29 \pm 0.10$ \\
SN2008bc & CSP & NIR@max & $BVriYJH$ & 0.016 & $34.05 \pm 0.10$ & $34.14 \pm 0.09$ & $34.12 \pm 0.08$ \\
SN2008bf & CSP & NIR@max & $BVriYJH$ & 0.025 & $35.13 \pm 0.06$ & $35.13 \pm 0.09$ & $35.10 \pm 0.09$ \\
SN2008fl & CSP & - & $BVriYJH$ & 0.020 & $34.59 \pm 0.08$ & $34.49 \pm 0.09$ & $34.50 \pm 0.09$ \\
SN2008fr & CSP & - & $BVriYJH$ & 0.038 & $36.04 \pm 0.04$ & $36.10 \pm 0.09$ & $36.08 \pm 0.10$ \\
SN2008fw & CSP & - & $BVriYJH$ & 0.009 & $32.76 \pm 0.18$ & $33.05 \pm 0.10$ & $33.05 \pm 0.09$ \\
SN2008gb & CfA & NIR@max & $BVr'i'JH$ & 0.038 & $36.03 \pm 0.04$ & $35.94 \pm 0.10$ & $35.87 \pm 0.11$ \\
SN2008gg & CSP & - & $BVriYJH$ & 0.031 & $35.58 \pm 0.05$ & $35.66 \pm 0.10$ & $35.65 \pm 0.09$ \\
SN2008gl & CSP & - & $BVriYJH$ & 0.033 & $35.72 \pm 0.05$ & $35.79 \pm 0.10$ & $35.84 \pm 0.09$ \\
SN2008gp & CSP & NIR@max & $BVriYJH$ & 0.034 & $35.74 \pm 0.05$ & $35.71 \pm 0.09$ & $35.70 \pm 0.09$ \\
SN2008hj & CSP & NIR@max & $BVriYJH$ & 0.037 & $35.97 \pm 0.04$ & $36.01 \pm 0.10$ & $36.00 \pm 0.08$ \\
SN2008hm & CfA & - & $BVr'i'JH$ & 0.021 & $34.70 \pm 0.07$ & $34.76 \pm 0.10$ & $34.75 \pm 0.10$ \\
SN2008hs & CfA & NIR@max & $BVr'i'JH$ & 0.019 & $34.47 \pm 0.06$ & $34.70 \pm 0.10$ & $34.80 \pm 0.10$ \\
SN2008hv & CSP & NIR@max & $BVriYJH$ & 0.014 & $33.81 \pm 0.11$ & $33.85 \pm 0.10$ & $33.85 \pm 0.09$ \\
\bottomrule
		\end{tabular}
			\end{threeparttable}
\end{table*}

\begin{table*}
	\centering
	\contcaption{Table of supernovae}
	\begin{threeparttable}
		\begin{tabular}{l c c c c c c c c} \toprule
			SN & source & cut & filters & $z_\text{CMB}$\tnote{a} & $\hat{\mu}_\text{ext}\tnote{b}$ & $\hat{\mu}_\text{phot} \text{ (resub)}$\tnote{c} &  $\hat{\mu}_\text{phot} \text{ (CV)}$\tnote{d} \\ 
			\midrule
SN2008ia & CSP & - & $BVriYJH$ & 0.022 & $34.86 \pm 0.07$ & $34.84 \pm 0.10$ & $34.82 \pm 0.09$ \\
SN2009D & CSP & NIR@max & $BVriYJH$ & 0.024 & $35.03 \pm 0.06$ & $35.03 \pm 0.09$ & $35.00 \pm 0.09$ \\
SN2009Y & CSP & NIR@max & $BVriYJH$ & 0.009 & $32.95 \pm 0.16$ & $33.01 \pm 0.09$ & $32.95 \pm 0.09$ \\
SN2009aa & CSP & NIR@max & $BVriYJH$ & 0.029 & $35.40 \pm 0.05$ & $35.27 \pm 0.10$ & $35.27 \pm 0.09$ \\
SN2009ab & CSP & - & $BVriYJH$ & 0.010 & $33.14 \pm 0.15$ & $33.47 \pm 0.10$ & $33.49 \pm 0.08$ \\
SN2009ad & CSP & NIR@max & $BVriYJH$ & 0.029 & $35.40 \pm 0.05$ & $35.33 \pm 0.10$ & $35.31 \pm 0.10$ \\
SN2009ag & CSP & NIR@max & $BVriYJH$ & 0.010 & $33.12 \pm 0.15$ & $33.09 \pm 0.09$ & $33.07 \pm 0.10$ \\
SN2009al & CfA & NIR@max & $BVr'i'JH$ & 0.023 & $34.94 \pm 0.07$ & $34.84 \pm 0.09$ & $34.83 \pm 0.09$ \\
SN2009an & CfA & NIR@max & $BVr'i'JH$ & 0.011 & $33.23 \pm 0.14$ & $33.32 \pm 0.09$ & $33.31 \pm 0.09$ \\
SN2009bv & CfA & NIR@max & $BVr'i'JH$ & 0.038 & $36.05 \pm 0.04$ & $36.13 \pm 0.10$ & $36.13 \pm 0.10$ \\
SN2009cz & CSP & NIR@max & $BVriYJH$ & 0.022 & $34.79 \pm 0.07$ & $34.79 \pm 0.10$ & $34.78 \pm 0.09$ \\
SN2009kk & CfA & - & $BVr'i'JH$ & 0.012 & $33.51 \pm 0.13$ & $33.96 \pm 0.10$ & $33.97 \pm 0.09$ \\
SN2009kq & CfA & - & $BVr'i'JH$ & 0.013 & $33.58 \pm 0.12$ & $33.72 \pm 0.10$ & $33.75 \pm 0.10$ \\
SN2010ai & CfA & NIR@max & $BVr'i'JH$ & 0.024 & $34.99 \pm 0.05$ & $34.96 \pm 0.10$ & $34.95 \pm 0.10$ \\
SN2010dw & CfA & - & $BVr'i'JH$ & 0.039 & $36.09 \pm 0.04$ & $35.99 \pm 0.10$ & $35.95 \pm 0.09$ \\
\bottomrule
		\end{tabular}
		\begin{tablenotes}
			\item[a] Redshift with corrections for local flows and CMB as described in \citet{avelino19}.  For 8 SN with available redshift-independent distance estimates from Cepheids, Tully-Fisher, or surface brightness fluctuations, this is an effective redshift as described in \citet{avelino19}.
			\item[b] External distance estimate and standard deviation, either from redshift-independent distance estimate or from redshift and assumed $H_0 = 73.24 \text{ km s}^{-1} \text{ Mpc}^{-1}$. See \citet{avelino19} Tables 2 and 4.
			\item[c] Optical+NIR \textsc{BayeSN} photometric distance estimate obtained by resubstitution.
			\item[d] Optical+NIR \textsc{BayeSN} photometric distance estimate obtained by cross-validation.
		\end{tablenotes}
	\end{threeparttable}
\end{table*}
 
Ultimately, we plan on training a version of \textsc{BayeSN} exclusively on SNe Ia observed by the Foundation Survey and the Young Supernova Experiment (YSE; D. Jones et al. 2020, in prep.), which have well-determined measurements of the natural system passband throughput.

\section{Implementation}\label{sec:implementation}

\subsection{BayeSN}

We have implemented our Bayesian model in the \textsc{Stan} probabilistic programming language \citep{stan17} to specify and sample the global posterior density over all latent variables and hyperparameters conditional on the training set data.  \textsc{Stan} implements a variant of dynamic Hamiltonian Monte Carlo \citep[HMC; ][]{neal11,betancourt17}, originally based on the No-U-Turn Sampler (NUTS) \citep{hoffman14}.  \textsc{Stan} utilises automatic differentiation to compute gradients of the log posterior (Eqs. \ref{eqn:globalpost}, \ref{eqn:distfitpost}) and guide efficient exploration and convergence to the target density in high-dimensional parameter spaces.  We typically run 4 chains in parallel, each initialised with random jitter to start at a different point in parameter space.  We follow standard procedures to assess convergence and mixing of the chains \citep{gelman92, gelman_bda}. The first half of the iterations, which are used for adaptation of the HMC algorithm and burn-in, are discarded. The algorithm adapts the integration time to yield samples that are nearly serially uncorrelated, and we run it long enough so that the effective sample size is approximately 1,000.

We discretise the integrals over wavelength (Eq. \ref{eqn:fluxint}) as numerical Riemann sums with resolution $\Delta \lambda_r = 20$ \AA.  This provides sufficient precision for evaluating the model fluxes (with discretisation error $< 0.2\%$ and therefore much smaller than typical photometric error).

We can employ the model and Bayesian inference code in two modes.  In \emph{training} mode, we condition on the external distance estimates and their uncertainties, along with the SN Ia light curves and redshifts, to sample the joint posterior of all hyperparameters and latent variables.  Trying to find the single optimal point of the global posterior in the high-dimensional parameter space is vulnerable to overfitting.  Instead, we use the Bayesian approach to sample the global joint posterior Eq. \ref{eqn:globalpost}, which allows us to marginalise over the posterior uncertainties in the latent variables when estimating the hyperparameters, including the SED components.  In \emph{distance-fitting} mode, we use posterior estimates of the hyperparameters of the already-trained model, and we remove the external distance constraint.  Redshifts are only used to shift the SED between the rest-frame and observer-frame and to account for time-dilation. We then compute posterior inference on the latent parameters of individual SNe, and marginalise to obtain the posterior the photometric distance from the SN Ia light curve (Eq. \ref{eqn:photdist}).

For the purposes of this analysis, we define optical as the $BVRI$ bands and the NIR as $YJH$.  For \textsc{BayeSN} and \textsc{SNooPy} we fit the $BVRIYJH$ bands, where $RI$ includes $ri$ and $r'i'$ filters, where applicable.  The version of \textsc{BayeSN} described here has not been trained on $U$-band data; preliminary analysis with a \textsc{BayeSN} prototype including the $U$-band does not show a significant improvement in results on this sample.  We apply our current model either to the available $BVRI$ (optical) or $BVRIYJH$ (optical+NIR) data.

\subsection{SALT2 and SNooPy fitting}

We used the \texttt{SALT2.4} model of \citet{betoule14} and use the calibration therein of the Tripp estimator (Eq. \ref{eqn:tripp}) for distances. The specific implementation of SALT2 used is available in the \texttt{sncosmo} package \citep{sncosmo}. For each object, we use literature estimates of the time of $B$-band maximum to select observations between -10d and +40d in phase with $S/N > 3$. This ensures that the same observations are used by both SALT2 and \textsc{BayeSN}. \textsc{SALT2} has a range of 2000--9000\AA\, and therefore can fit the $UBVRI$ bands, but as with \textsc{BayeSN}, we do not fit the $U$-band and restrict the comparison to $BVRI$. We compared the \textsc{SALT2} results with or without $U$-band, and found that the $U$-band data did not improve the results for our sample. The limited template range of \textsc{SALT2} also prevents us from any comparison with $BVRIYJH$ fits. \citet{pierel18} created a NIR extension to the SALT2 model that is suitable for simulations, but that did not use the same training procedure as that used to create the SALT2.4 model templates. Therefore, it is not suitable for fitting real light curves and does not yield calibrated distances. 

For each object, we begin with an initial guess for the parameters, which we refine with \texttt{Minuit} \citep{minuit}. We use the result from \texttt{Minuit} to set the initial positions of 32 walkers used to sample the posterior distribution with the \texttt{emcee} Markov Chain Monte Carlo package \citep{Foreman_Mackey_2013}. We generate 2000 samples per walker after discarding the first 500 steps as burn-in. We visually inspect the parameter chains and 2D marginalized posterior distributions. We report the median value of the samples as the ``best-fit'' estimate and use the \nth{16} and \nth{84} percentiles of the samples as a credible interval. The official procedure for SALT2 light curve fitting weights the fit by the photometric errors only; although a residual scatter model exists \citep{guy10}, when we have tried to use it for fitting, we obtained a much larger Hubble diagram scatter and results with substantially larger uncertainties than plausible for these low-redshift, high $S/N$ objects.  For consistent comparisons, we have adjusted the the SALT2 distance estimates to a scale of $H_0 = 73.24 \text{ km s}^{-1} \text{ Mpc}^{-1}$.

We use the \textsc{SNooPy} \texttt{EBV\_model2}\footnote{ \textsc{SNooPy} also has a ``max\_model'' mode which allows one to fit ($K$-corrected) light curve data to a template light curve model in a single rest-frame filter to find a single magnitude at maximum.  We do not compare against this mode, since the purpose of \textsc{BayeSN} is to fit the SED over the entire phase and wavelength range covered by the available data in multiple passbands simultaneously, without using $K$-corrections to compute a 1-to-1 map between photometry in observer-frame and rest-frame filters.} 
to fit the observations using templates parametrized by the light curve stretch, $s_{BV}$. The \texttt{EBV\_model2} uses the same algorithm as \citet{prieto06} to build the templates together with the updated calibration of 24 CSP supernovae presented in \citet{burns11}. The resulting \texttt{EBV\_model2}   rest-frame light curves templates cover $uBgVriYJH$.  To be consistent with our comparison to \textsc{SALT2}, which is restricted to modeling only the optical observations, we fit $BVRI$ (optical-only) as well as $BVRIYJH$ (optical+NIR) data with \textsc{SNooPy}. We use the same initial guesses for the \textsc{SNooPy} fit parameters as used for the \textsc{SALT2} fits. \textsc{SNooPy} uses a non-linear least-squares Levenberg-Marquadt algorithm to minimize the variance weighted residuals to the model. As with \textsc{SALT2}, we report the statistical uncertainties on the fit parameters derived from inverting the Hessian matrix at the best-fit parameters, and we have adjusted the SNooPy distance estimates to a scale of $H_0 = 73.24 \text{ km s}^{-1} \text{ Mpc}^{-1}$. Our low-redshift SNe have well-sampled light curves with high S/N and we do not find any significant differences between the Levenberg-Marquadt results and those using MCMC sampling. The SNooPy light curve fitting procedure weights the light curve fit only by the photometric errors; there is no residual covariance model.

\section{Results \& Discussion}\label{sec:results}

\subsection{Light Curve Inference for Individual SNe Ia}

\begin{figure}
	\includegraphics[scale=0.6]{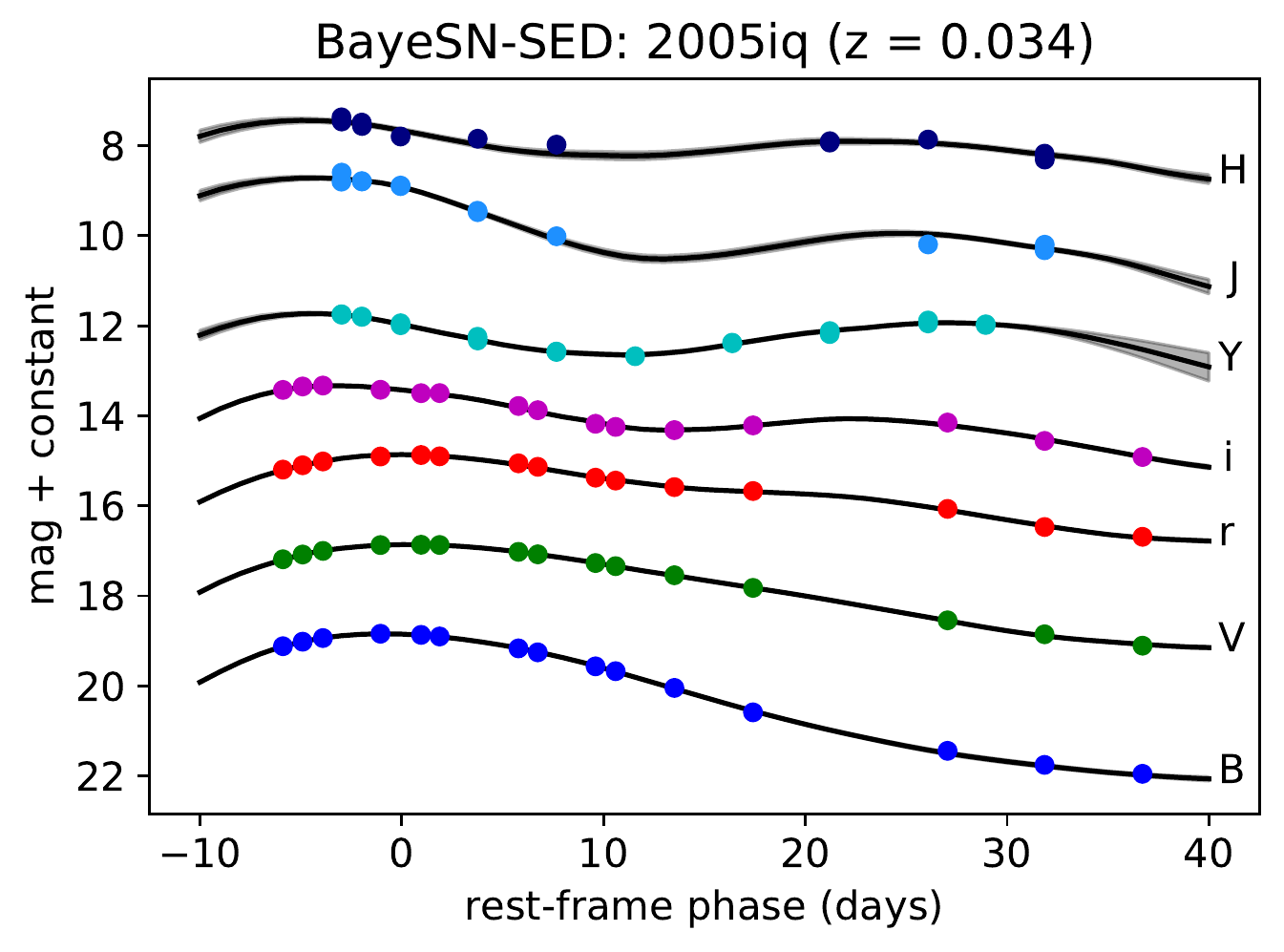}
	\includegraphics[scale=0.35]{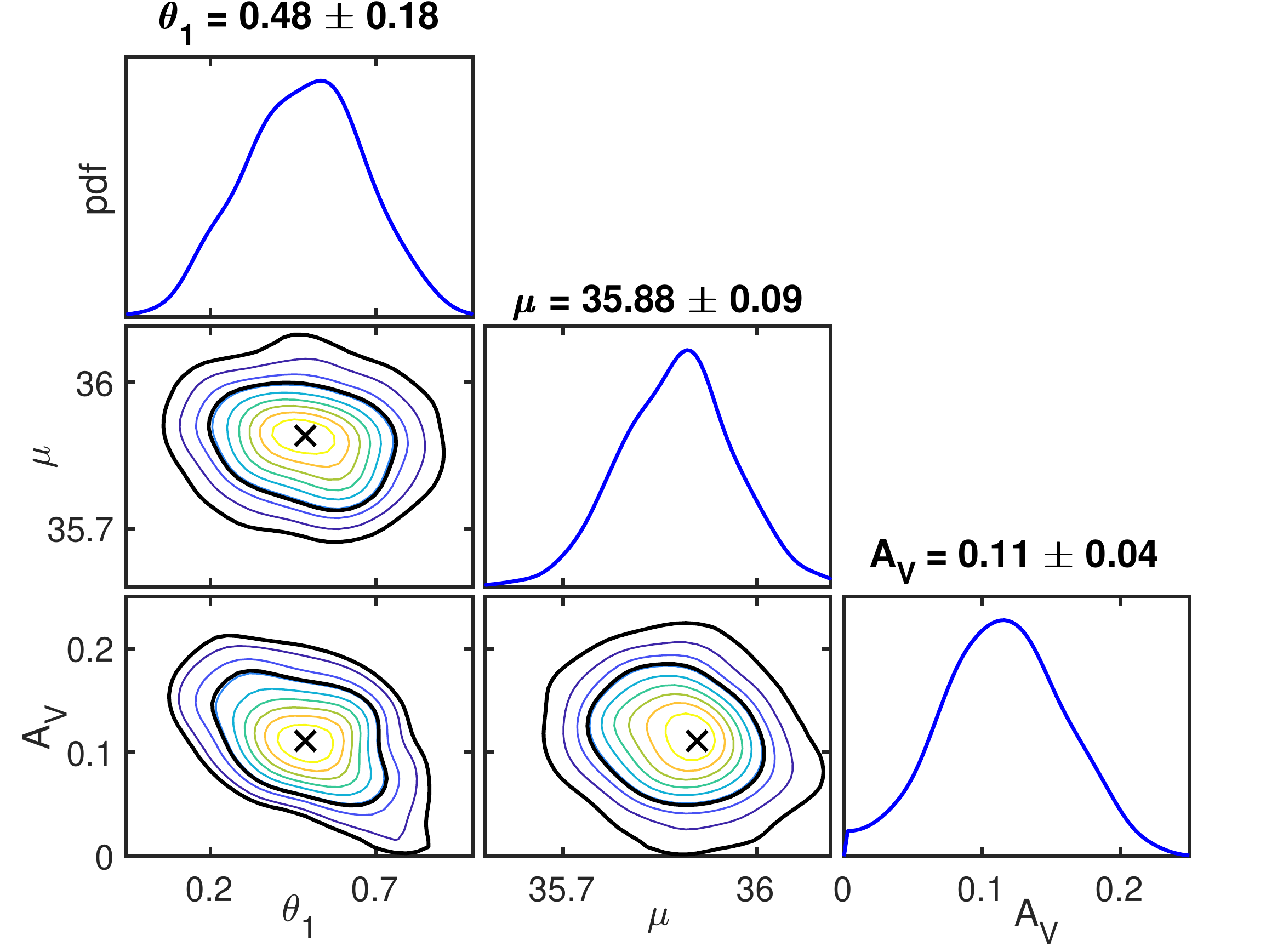}
	\caption{(top) Example \textsc{BayeSN} light curve fit of optical and NIR $BVriYJH$ CSP observations of the Type Ia SN 2005iq. (bottom) Posterior distribution of latent parameters of light curve fit to CSP observations of SN 2005iq.  In the 2D contour plots, the black contours contan 68\% and 95\% of the marginal posterior probability, and the mode is indicated. The 1D marginal plots depict a kernel density estimate applied to the MCMC samples for each parameter.  The SED shape parameter $\theta_1$ and host galaxy dust extinction $A_V$ are marginalised over to obtained the posterior distribution of the photometric distance modulus $\mu_s$.}
	\label{fig:sn2005iq_lc}
\end{figure}

As an example, Fig. \ref{fig:sn2005iq_lc} demonstrates a \textsc{BayeSN} light curve fit to optical and NIR observations of SN 2005iq (CSP, $z = 0.034$).  It also shows the posterior distribution of the latent parameters $(\theta_1, A_V, \mu)$ obtained under distance-fitting mode.  To obtain the marginal distribution of the photometric distance modulus $\mu_s^\text{phot}$, the other latent parameters of the SN (including the residuals $\bm{e}_s$) are integrated over.  The photometric distance modulus is well constrained to $\pm 0.09$ mag using the joint optical and NIR data at all phases.

\begin{figure}
	\includegraphics[scale=0.37]{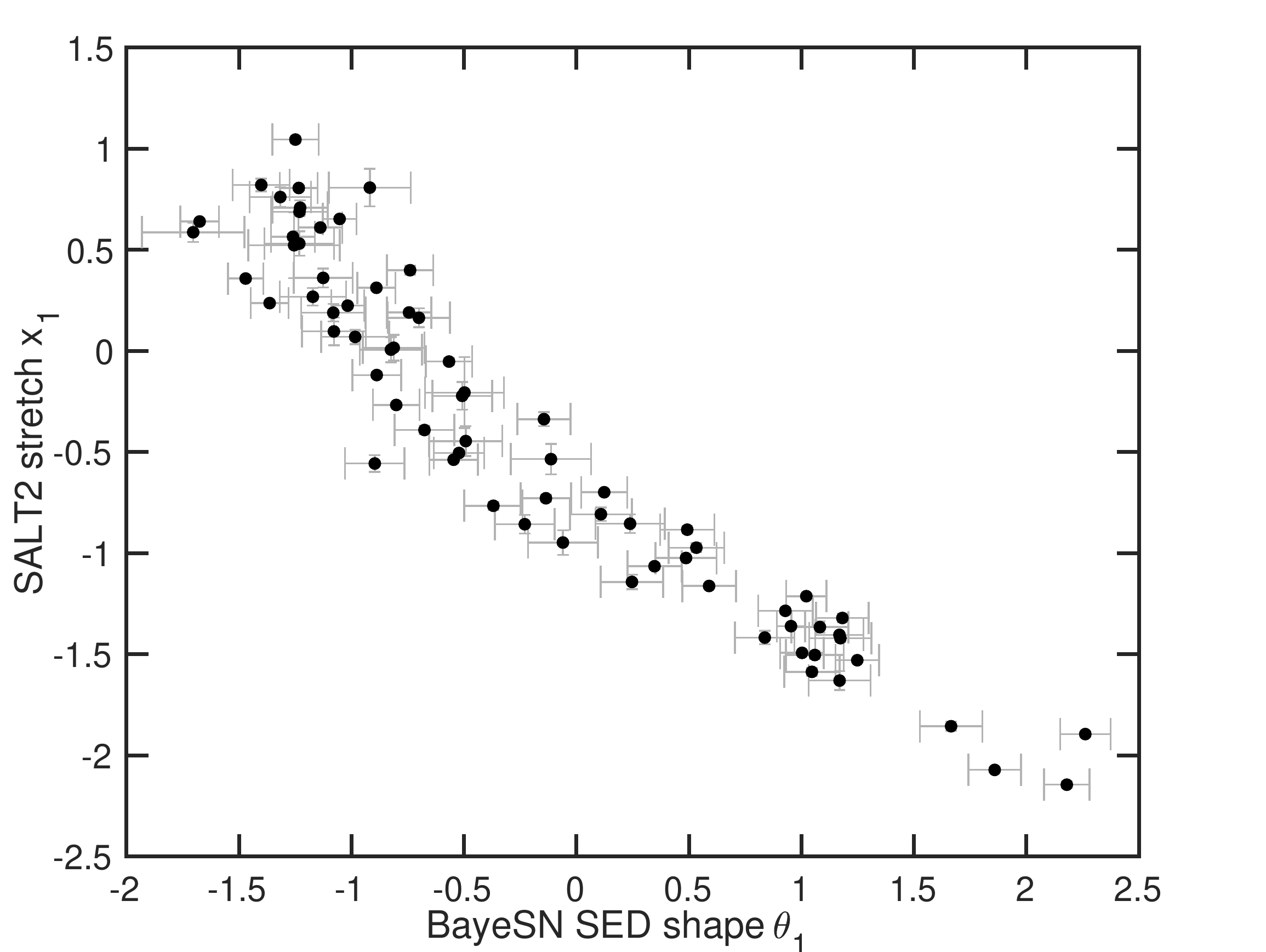}
	\includegraphics[scale=0.37]{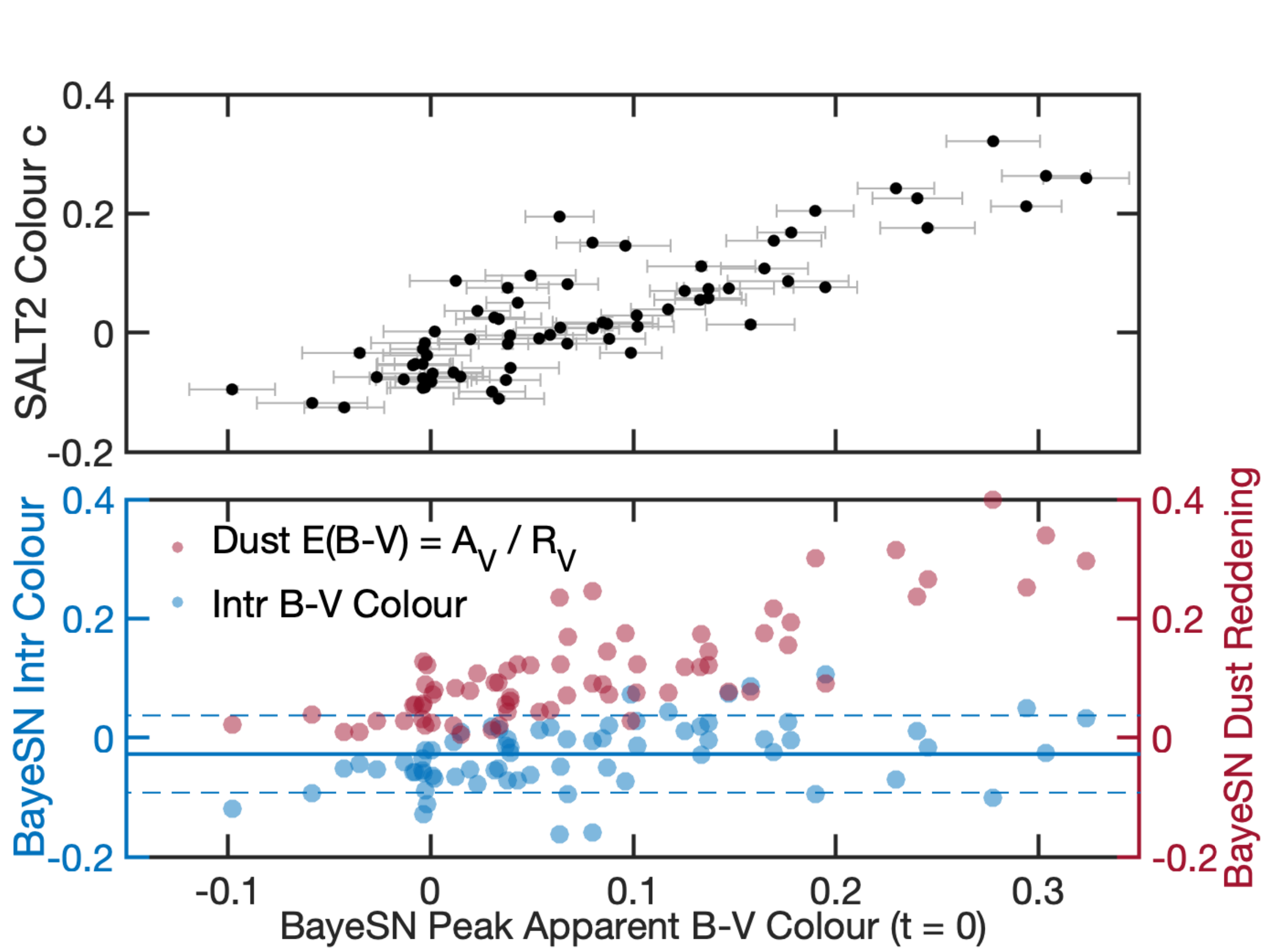}
	\caption{Comparison of BayeSN and SALT2 parameters. (upper panel) Strong correlation between the $\theta_1$ coefficient of the first principal SED component and the SALT2 light curve shape ''stretch'' parameter $x_1$.  (lower panel, top) correlation between the peak ($t = 0$) $B-V$ apparent colour from \textsc{BayeSN} light curve fit and the SALT2 colour parameter $c$. (lower panel, bottom) \textsc{BayeSN} models the apparent colour as the sum of two latent components: the intrinsic colour (blue) and the positive reddening due to dust, $E(B-V) = A_V / R_V$ (red).  We plot the SNe with $B$ and $V$ measurements within $\pm 5$ days of $B$ maximum light.}
	\label{fig:bayesn_vs_salt2}
\end{figure}

In Fig. \ref{fig:bayesn_vs_salt2}, we show a visual comparison between the \textsc{BayeSN} and \textsc{SALT2} parameter estimates. In the top panel, we plot the SED shape parameter $\theta_1$, which is the score of the first functional component, against the SALT2 $x_1$ ``stretch'' parameter for the same SNe Ia.  The sign of $\theta_1$ has been chosen to be in the same sense as the decline rate $\Delta m_{15}(B)$ of \citet{phillips93}, which is the magnitude change in $B$-band between $B$-band peak and 15 days afterwards. Larger values of $\theta_1$ correspond to faster (larger) post-maximum optical decline rates. Larger $x_1$ values correspond to broader optical light curves, which have slower (smaller) optical decline rates. There is a fairly tight, slightly non-linear correlation between $\theta_1$ and $x_1$, suggesting that they are capturing the same underlying major mode of variation.

In the bottom panel of Fig. \ref{fig:bayesn_vs_salt2}, we compare the SALT2 colour parameter $c$ and the \textsc{BayeSN} fitted value of the apparent $B-V$ colour at peak $t=0$.  The latter is determined by evaluating the rest-frame SED model (at redshift $z_s = 0$) with the fit parameters $(\theta_1^s, A_V^s, \bm{e}_s)$ for each SN, and integrating it under reference $B$ and $V$ bandpasses, which we take to be those of the CSP.  There is a strong but not exactly 1-to-1 correlation between the two. The \textsc{BayeSN} model is able to leverage the optical and NIR data of the full light curve to probabilistically decompose the apparent colour into an intrinsic $B-V$ colour and dust reddening $E(B-V)$. The former is estimated from the light curve fit by evaluating the rest-frame SED with the light curve fit parameters $(\theta_1^s, \bm{e}_s)$ and setting $A_V^s = 0$, and integrating it under the reference passbands, and the latter is determined by $E(B-V)_s =  A_V^s / R_V$.  Our model finds that the apparent colours are the sum of two different effects and captures these two different sources of variation, which are each correlated with the rest of the SED (and thus luminosity as a function of wavelength) in different ways.  In contrast, the conventional Tripp formula (Eq. \ref{eqn:tripp}) assumes that the apparent color-magnitude relation is described by a single factor depending on the SALT2 apparent colour parameter $c$.

\subsection{Population Inference}

The statistical properties of the latent SED, captured by the intrinsic FPC, residual covariance, and dust distribution, are learned during the \textsc{BayeSN} model training phase by sampling the global posterior density, Eq. \ref{eqn:globalpost}.

\subsubsection{Intrinsic SED components}

\begin{figure*}
\includegraphics[scale=0.42]{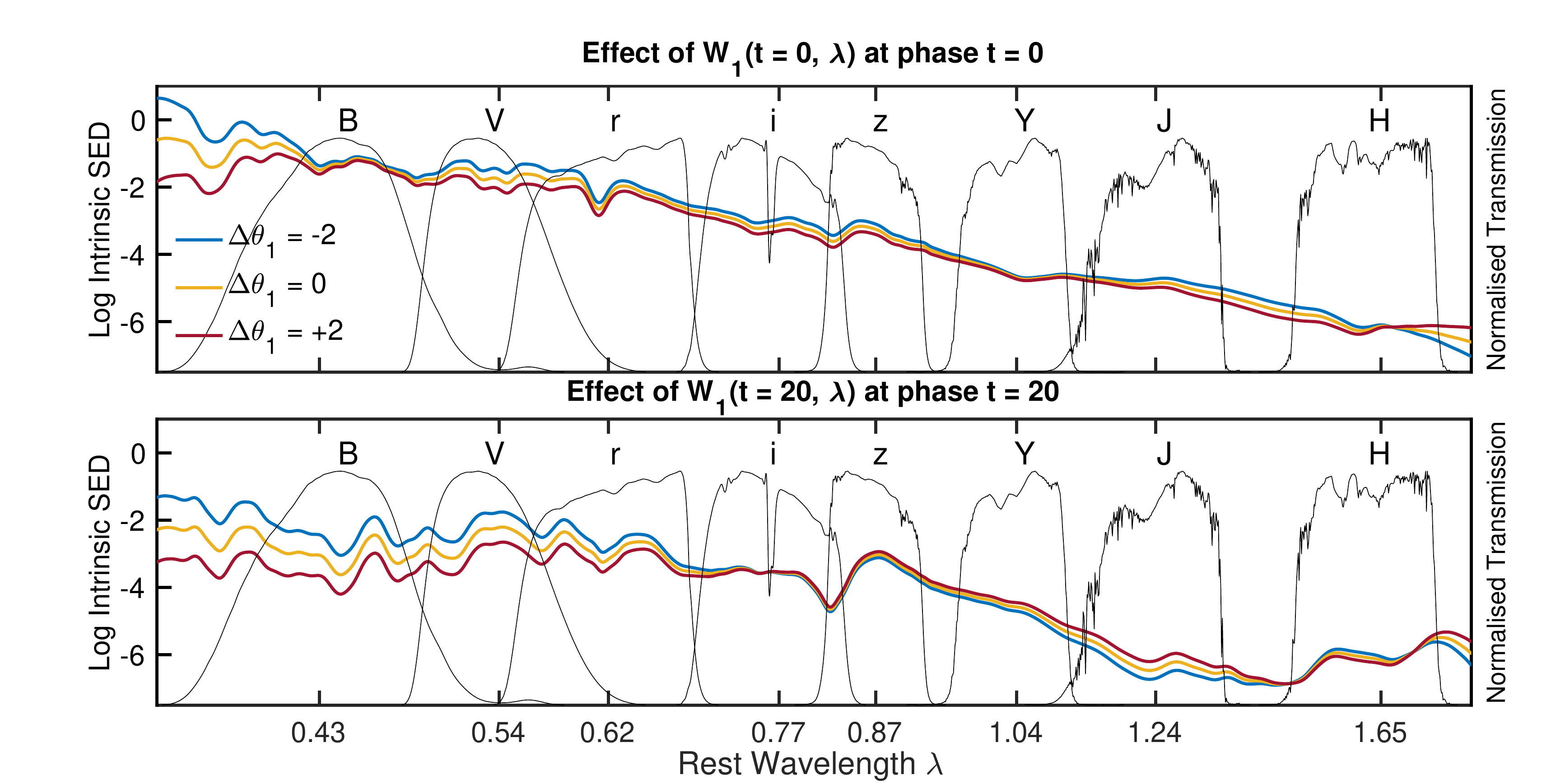}
\includegraphics[scale=0.42]{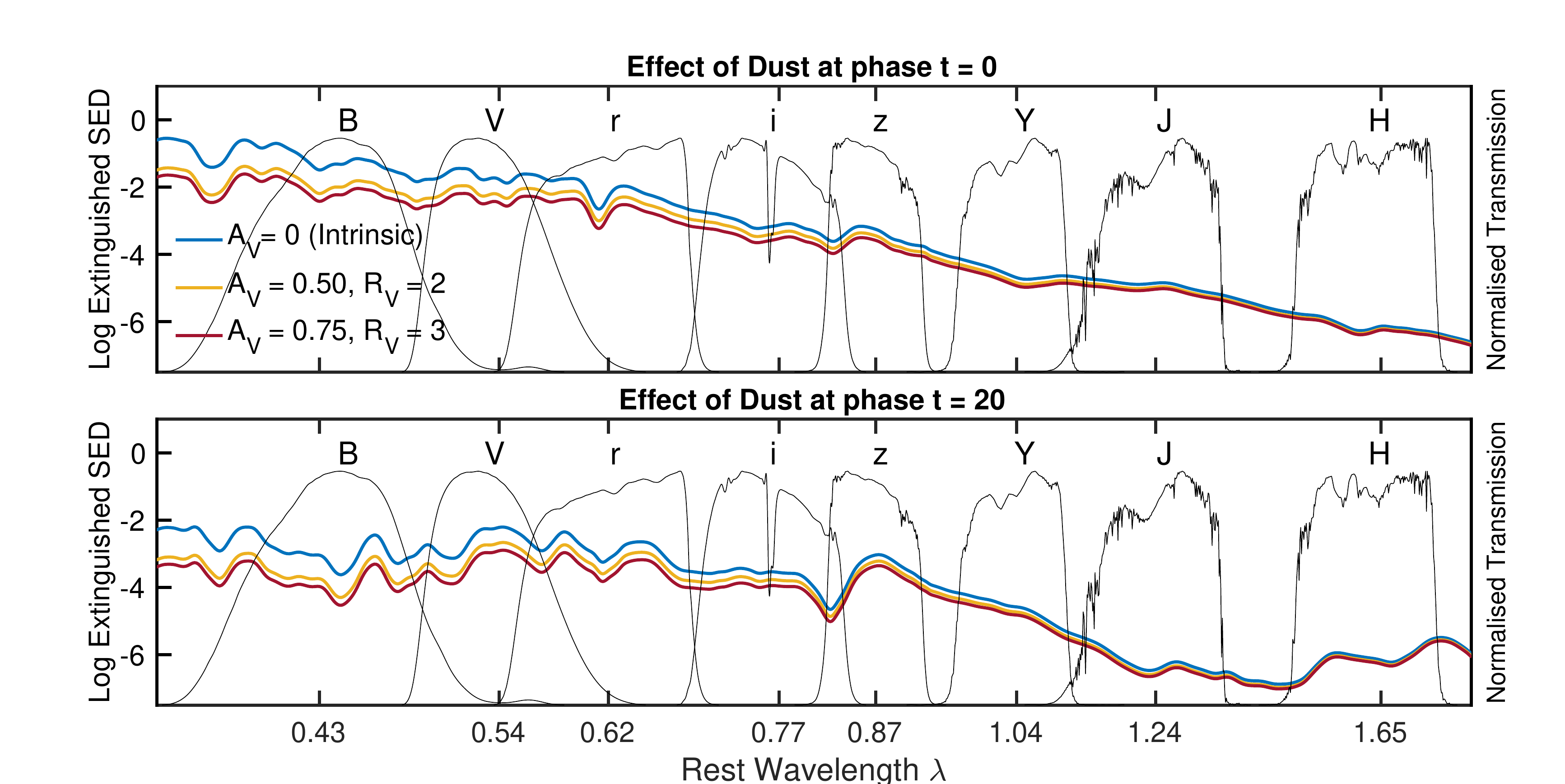}
	\caption{(top) Variation in the optical and NIR intrinsic SED captured by the first functional component $W_1(t,\lambda)$ at $t = 0$ and $20$ days. We vary the value of $\theta_1$ by $\bar{\theta}_1 \pm 2\sigma$, holding all other SN parameters to zero. (bottom) The effect of dust extinction on the optical and NIR SED.  We apply dust extinction to the baseline mean intrinsic SED with different combinations of $A_V, R_V$ that produce the same optical colour excess $E(B-V) = A_V / R_V$.}
	\label{fig:W1_SED}
\end{figure*}

The baseline intrinsic SED depicted in Fig. \ref{fig:forwardmodel} is obtained with $\theta_1^s = A_V^s = \bm{e}_s = \delta M_s = 0$, and is equal to $S_0(t,\lambda_r)10^{-0.4[M_0 + W_0(t,\lambda_r)]}$.  The first functional principal component (FPC) $W_1(t,\lambda)$ is also shown in Fig. \ref{fig:forwardmodel}. The top panel of Fig. \ref{fig:W1_SED} shows the effect of our first functional component $W_1(t,\lambda)$ on the baseline intrinsic SED at phases $t = 0$ and $t = 20$ as one changes the coefficient $\theta_1$.  In the bottom panel, for comparison, we show the effect of the dust extinction on the SED.  An interesting difference between the two is the sign flip of the effect of $\theta_1$ in the NIR at phase $t = 20$.  Under this effect, SNe Ia that are dimmer in the optical are actually brighter in the NIR $YJ$ bands at this later phase.  This is an indication of the correlation of dimmer SNe Ia having earlier rises to the secondary NIR maximum.  In contrast, the effect of dust is to make SNe Ia dimmer at all phases.  This sign-flip distinction may help break the degeneracy between intrinsic SN and extrinsic dust effects.

The figure also shows a reference set of filter passband functions (with arbitrary scaling for visualisation purposes).  We visualise the effect of our functional components on rest-frame photometric light curves by integrating our SED model with various parameter values under this reference set.  The reference set we choose for illustration are the CSP $BVriYJH$ passbands and the $z$-band filter from Pan-STARRS1 (PS1).  The rest-frame $z$-band (at $\approx 0.9 \,\mu$m, between $i$ and $Y$) region of SN Ia SEDs is regularly probed by low-$z$ surveys such as Foundation and YSE, but is not modelled by either SALT2 or SNooPy.  In \S \ref{sec:fdn}, we demonstrate an example of \textsc{BayeSN} fitting of a rest-frame $z$-band SN Ia light curve from Foundation DR1 \citep{foley18}.

\begin{figure*}
\includegraphics[scale=0.58]{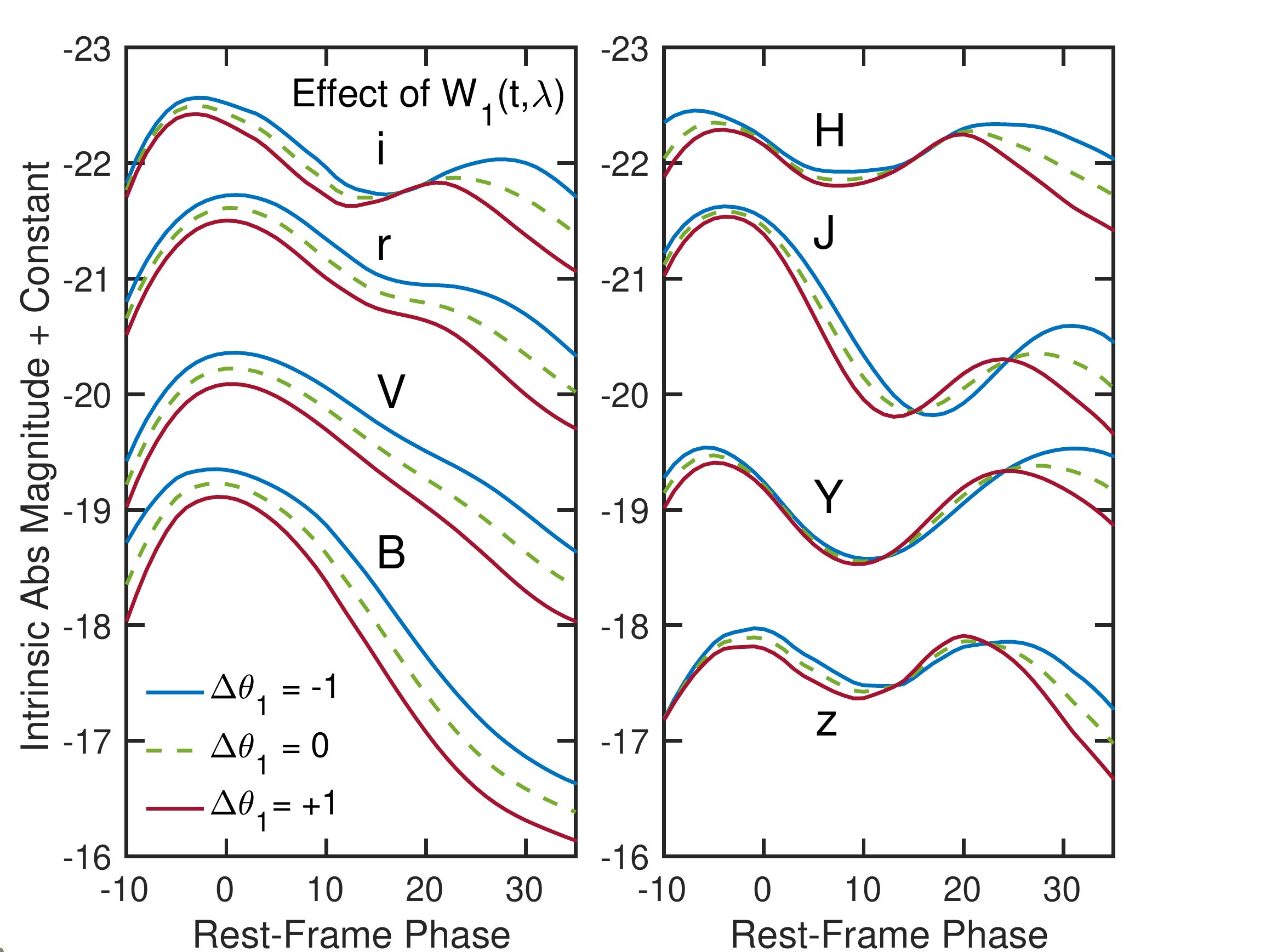}
	\caption{Intrinsic variation in optical and NIR intrinsic absolute light curves captured by the first functional component $W_1(t,\lambda)$. Variation in $\theta_1 W_1(t,\lambda)$ captures the  width-luminosity relation in the optical \citep{phillips93}. Variation in this component simultaneously modulates the amplitude and timing of the second peak in the near-infrared. For visual clarity, the absolute light curves have been shifted vertically by arbitrary constants ($B: 0, V: -1, r:-2.5,i: -4, z: 0.5, Y: -1, J: -3, H:-4$).}
	\label{fig:W12_lcs}
\end{figure*}

By integrating the SED model under these reference optical and near-infrared passbands, we show in Fig. \ref{fig:W12_lcs} the effect of the 1st FPC $W_1(t,\lambda_r)$ on the intrinsic optical and and NIR light curves. We see that this intrinsic component captures the optical width-luminosity relation \citep{phillips93}: intrinsically brighter supernovae have more slowly-declining (or broader) light curves, whereas dimmer ones decline faster.  This effect is seen most clearly in the $B$ and $V$ bands.  In the redder optical bands ($r$ and $i$) and into the NIR $zYJH$ bands, we see that this same effect is also correlated with the timing of the second peak at $t = 20-30$ days: brighter supernovae tend to have later secondary NIR peaks, while dimmer SNe Ia have earlier ones, which is a further reflection of the trend seen in Fig. \ref{fig:W1_SED}.  In $iYJH$ bands, the effect also correlates to more pronounced second peaks.  The empirical relation we capture correlates strongly with the theoretical models of \citet{kasen06}, who found that brighter SNe Ia should have more pronounced NIR secondary maxima at later phases due to role of the ionisation evolution of iron group elements in the SN ejecta in redistributing energy from the optical to the NIR.  Similar trends have been seen by \citet{dhawan15}, and \citet{shariff16b} explored the use of the phase of the secondary NIR maximum for standardising SN Ia optical magitudes.

The first NIR peak typically occurs a few days before the optical ($B$) peak ($t = 0$). Estimation of the 1st FPC at early pre-maximum phases in the NIR is somewhat limited by the relative scarcity of quality NIR observations there in the current dataset (particularly in the $H$-band).  Future data releases with greater NIR coverage at early phases will help us improve the model.

\begin{figure}
\includegraphics[scale=0.37]{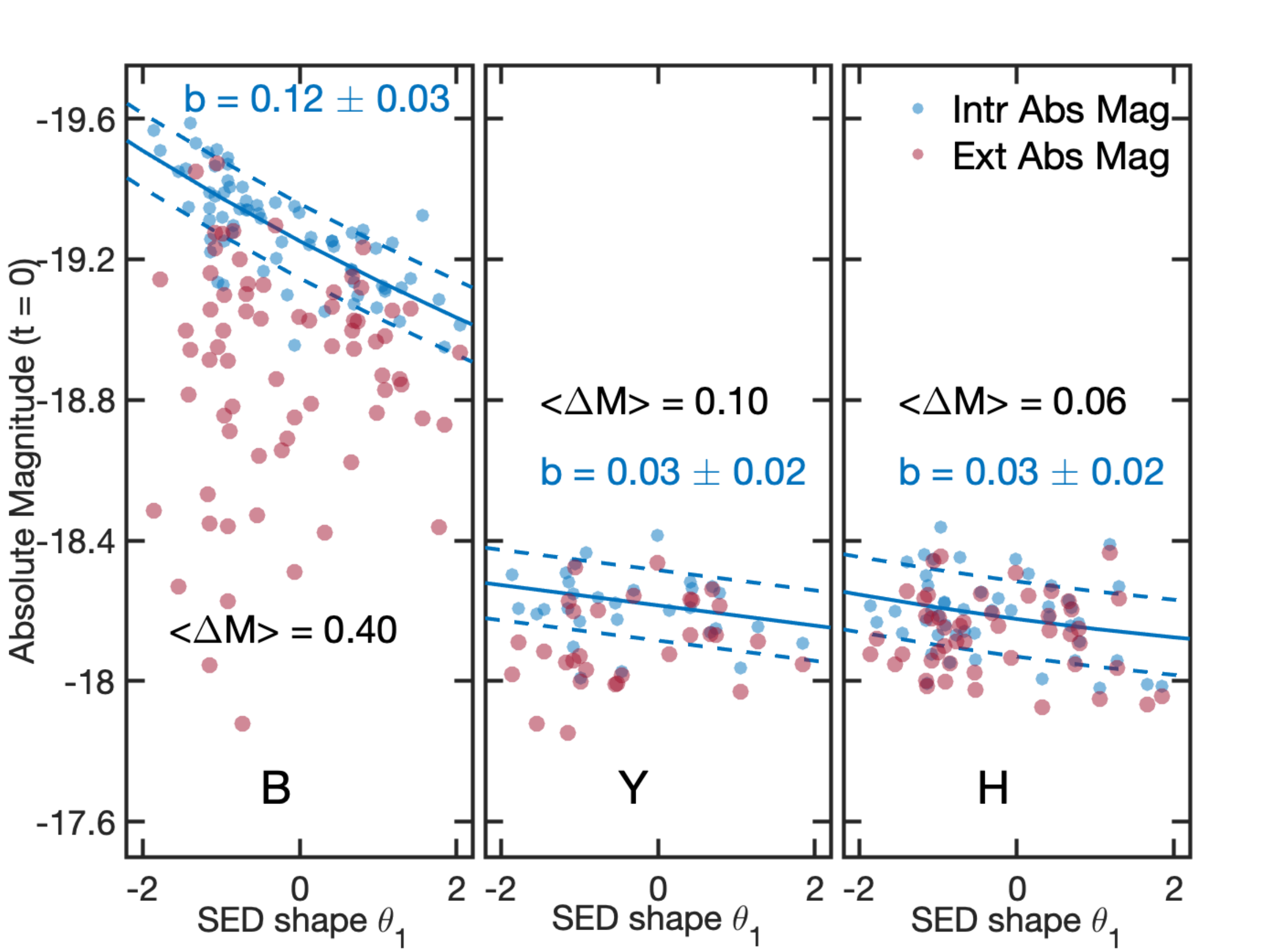}
	\caption{Intrinsic variation and host galaxy dust effects on peak absolute magnitudes at $T_{\text{B,max}}$ (phase $t=0$) in the rest-frame optical $B$ and NIR $Y, H$ bands. Each point is a posterior realisation of the intrinsic absolute magnitude $M_s^\text{int}$ (blue) or host dust-extinguished absolute magnitude $M_s^\text{ext}$ (red) of each SN.  In each panel, we plot the SNe with data in a given filter.  The solid line indicates the mean effect of the intrinsic $W_1(t, \lambda)$ model component on the intrinsic absolute magnitude through the coefficient $\theta_1$.  The slope of this line is indicated as $b$.  The dashed lines indicate $\pm 1$ standard deviation captured by the intrinsic residual covariance.  The mean effect of host galaxy dust extinction in each band, quantified by the sample average difference between each SN's extinguished and intrinsic absolute magnitude, is shown.}
	\label{fig:absmags_vs_t1}
\end{figure}

In Fig. \ref{fig:absmags_vs_t1}, we illustrate the dependence of optical and NIR absolute magnitudes on the SED shape parameter $\theta_1$ of the FPC.  The extinguished absolute magnitudes of a SN $s$ are obtained by evaluating the model SED with its fitted parameters $(\theta_1^s, \bm{e}_s, \delta M_s, A_V^s)$, setting $\mu_s = 0$, and integrating it under the reference passbands in the SN rest-frame.  The intrinsic absolute magnitudes are obtained in the same way but by setting $A_V^s = 0$.  In the optical $B$-band the average dust extinction correction is a 0.40 mag shift in absolute magnitude for the sample.  In the NIR $Y,H$-bands, the mean shift is 0.10 and 0.06 mag, respectively, which reflects the much diminished effect of dust extinction in the NIR compared to the optical (Fig. \ref{fig:forwardmodel}).  The relatively steep mean dependence of the $B$ intrinsic absolute magnitude on $\theta_1$ captures the optical width-luminosity relation \citep{phillips93}. In the NIR, the slopes of the dependence of $Y$ and $H$ intrinsic absolute magnitudes with $\theta_1$ are consistent with zero, after marginalising over the posterior uncertainties. The scatter about the mean intrinsic relation due to the SED residual functions is approximately 0.10 mag.  We note that the scatter around the mean intrinsic relation is not necessarily identical to the photometric distance uncertainty nor the expected scatter in the Hubble diagram.  This is because the SED shape $\theta_1$ and the dust extinction $A_V$ factors must themselves be estimated from the data, and their uncertainties are themselves influenced by the intrinsic residual covariance.  Instead, proper inference of the photometric distance uncertainty comes from the marginalisation in Eq. \ref{eqn:photdist}.  However, the diminished effect of dust $A_V$ and the insensitivity to $\theta_1$ in the NIR do significantly reduce their contributions to the derived photometric distance uncertainties.

\begin{figure*}
\includegraphics[scale=0.58]{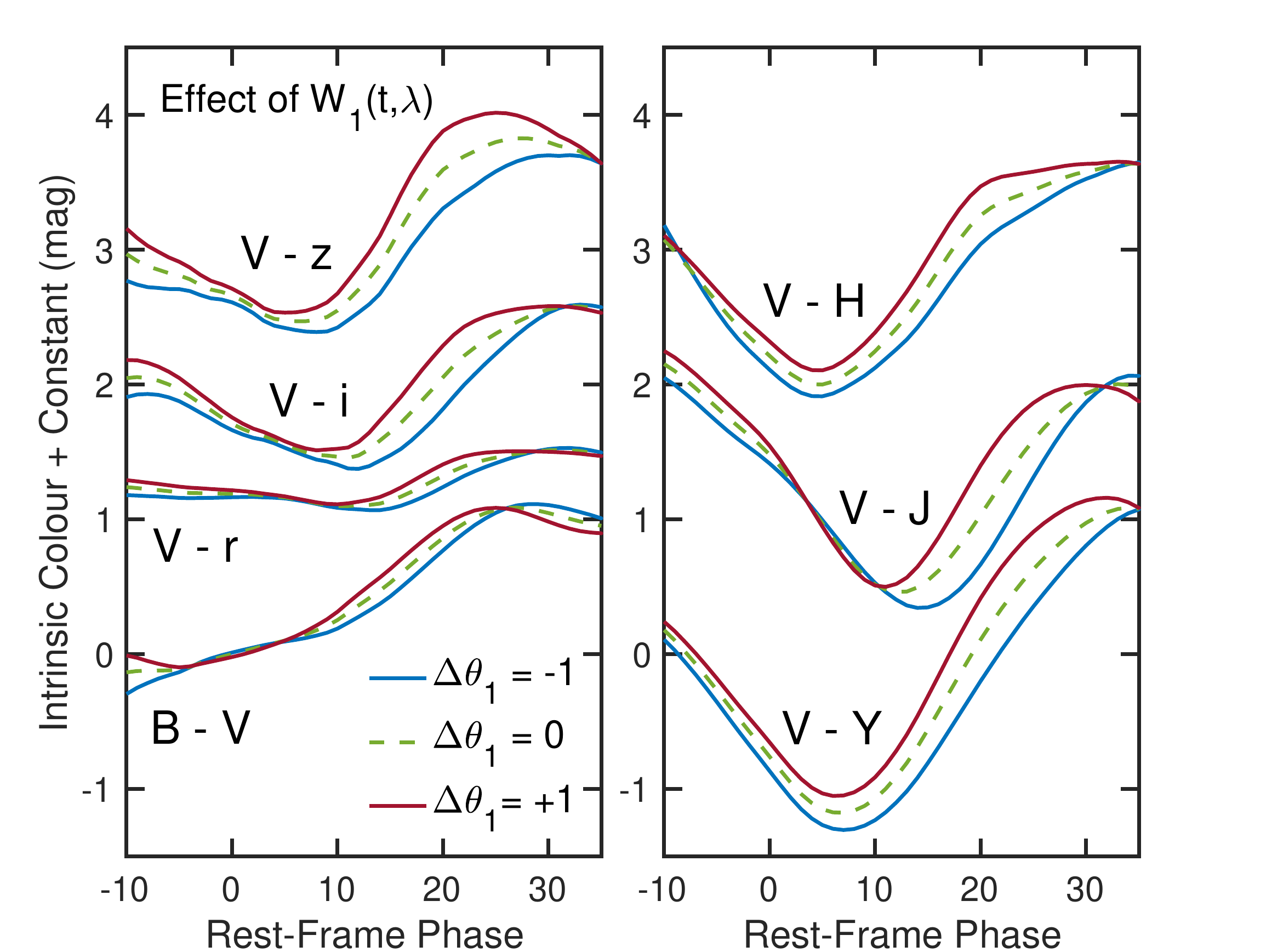}
\includegraphics[scale=0.58]{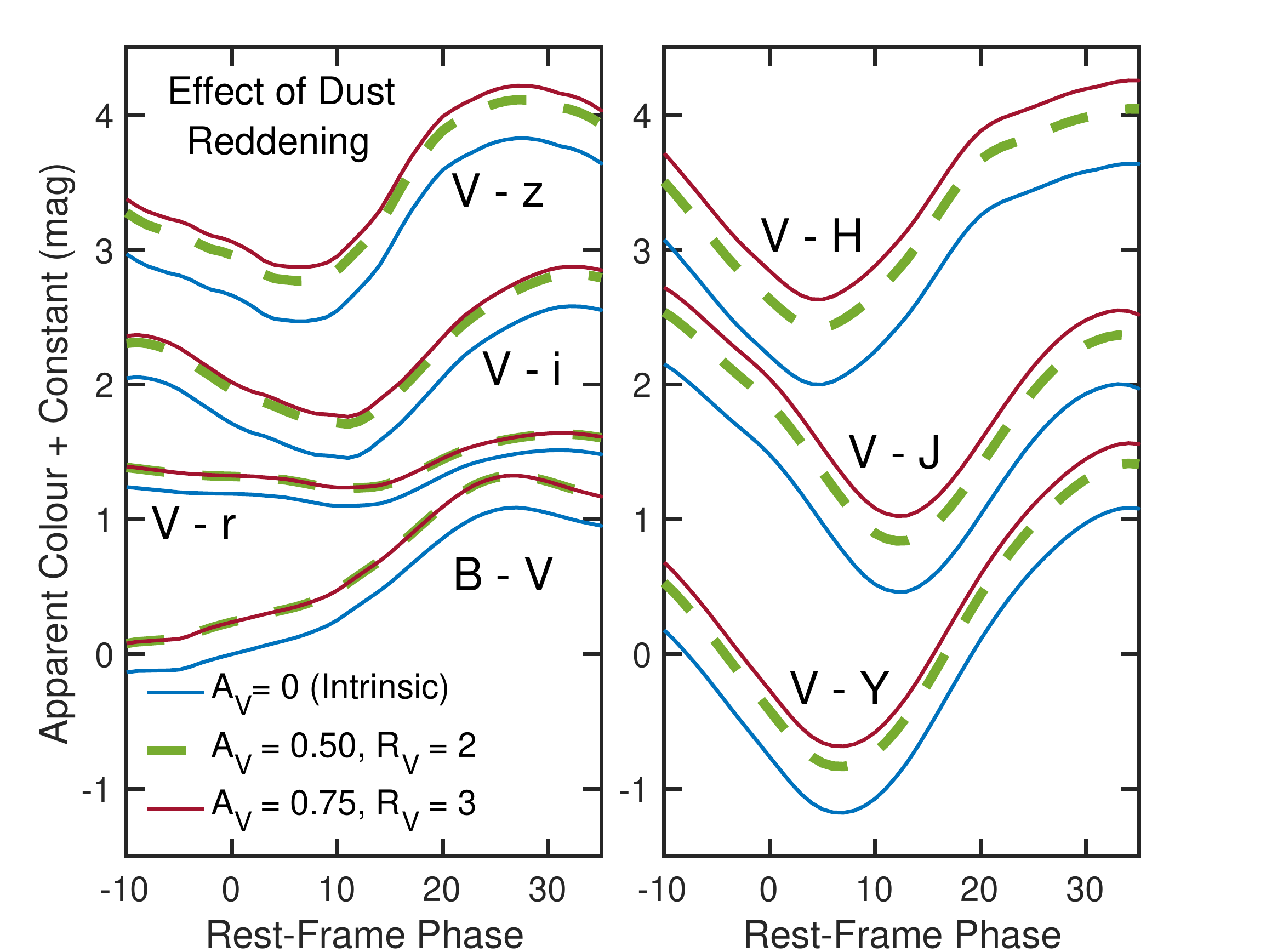}
	\caption{(top) Intrinsic variation in optical and NIR colour curves captured by the first functional component $W_1(t,\lambda)$.  We vary the value of $\theta_1$ by $\bar{\theta}_1 \pm 1\sigma$, while fixing $A_V$ and other SN parameters to zero. (bottom) Effect of host galaxy dust extinction on optical and NIR colour curves. We show unreddened, intrinsic colour curves (blue), and two apparent colour curves with the same amount of optical $E(B-V)$ colour excess due to dust, but two different values of the dust law $R_V = 2$ or $3$.  We fix $\theta_1$ and other SN parameters to zero.  The phase-dependence of the $W_1(,t,\lambda)$ component on intrinsic colour curves makes it distinguishable from dust.  The effect of dust reddening on colour curves is approximately constant with phase.  (bottom left) With optical data only, it is difficult to distinguish between two different combinations of host dust $A_V, R_V$ that produce the same colour excess $E(B-V) = A_V / R_V$.  (bottom right) Since the dust extinction in the NIR is smaller and less dependent on $R_V$, the optical-NIR colour curves help to break this degeneracy.  For visual clarity, the colour curves have been shifted vertically by arbitrary constants ($B-V: 0, V-r: 1.3, V-i: 2.5, V-z: 3.5, V-Y: 0.25, V-J: 2.25, V-H: 3.25$).}
	\label{fig:colcurves}
\end{figure*}

Colour curves, derived from flux ratios or magnitude differences between different filters, provide a useful window for understanding SNe Ia, since they are independent of the distance estimate and its errors.  In the top panel of Fig. \ref{fig:colcurves}, we illustrate the effect of the $W_1(t, \lambda_r)$ on the intrinsic optical-NIR colour curves by varying $\theta_1$.  At each epoch $t$, these are obtained by integrating the resulting rest-frame SED under each passband taking the difference with respect to the $V$-band magnitude.  The general trend is that the intrinsically brighter, and more slowly declining, SNe Ia (more negative $\theta_1$) tend to have bluer (more negative) colour curves in each of the colours shown.  The first FPC $W_1(t,\lambda_r)$ modulates the colour curves in a time-dependent fashion. While there are fixed points in phase when particular intrinsic colours are fairly insensitive to $\theta_1$, at phases $10 < t < 20$ days, there is significant intrinsic colour variation in all optical-NIR colours relative to $V$-band.

In the bottom panel of Fig. \ref{fig:colcurves}, we compare this with the impact of host galaxy dust reddening on the optical-NIR colour curves. In contrast to the intrinsic FPC, the effect of dust on colour curves is relatively constant in phase\footnote{In principle, it is not exactly time-independent: since the intrinsic SN SED is time-evolving, even if the amount of dust extinction $A_V$ is truly constant, the reddening effect on each magnitude has some time-dependence \citep[e.g.][]{phillips99, jha07}). However, this effect is too small to be seen on the plot.}, and the main effect is across different colours. We show the mean intrinsic colour curves with no dust $A_V = 0$ (thin blue), as well as two combinations of the dust parameters $[(A_V, R_V) = (0.75, 3) \text{ or } (0.50, 2)]$ that result in the same colour excess $E(B-V) = A_V / R_V = 0.25$.  The plot demonstrates that, with apparent colour information in optical $BVr$ data alone, it is very difficult to distinguish between the two possibilities.  In contrast, the optical-NIR $V - YJH$ colour information helps to break the degeneracy and distinguish between the two values of the dust law $R_V$.

\subsubsection{Intrinsic SED Residual Distribution}

The model captures the population distribution of residual SED variations that are unexplained by the intrinsic FPC, the host galaxy dust extinction, peculiar velocities or other external distance uncertainties, or measurement error, through the residual covariance.  The total residual SED function of a SN Ia $s$ is $\eta_s(t,\lambda_r) = \delta M_s + \epsilon_s(t,\lambda_r)$.  An example of an SED residual function is shown in Fig. \ref{fig:forwardmodel}.

\begin{figure*}
\includegraphics[scale=0.58]{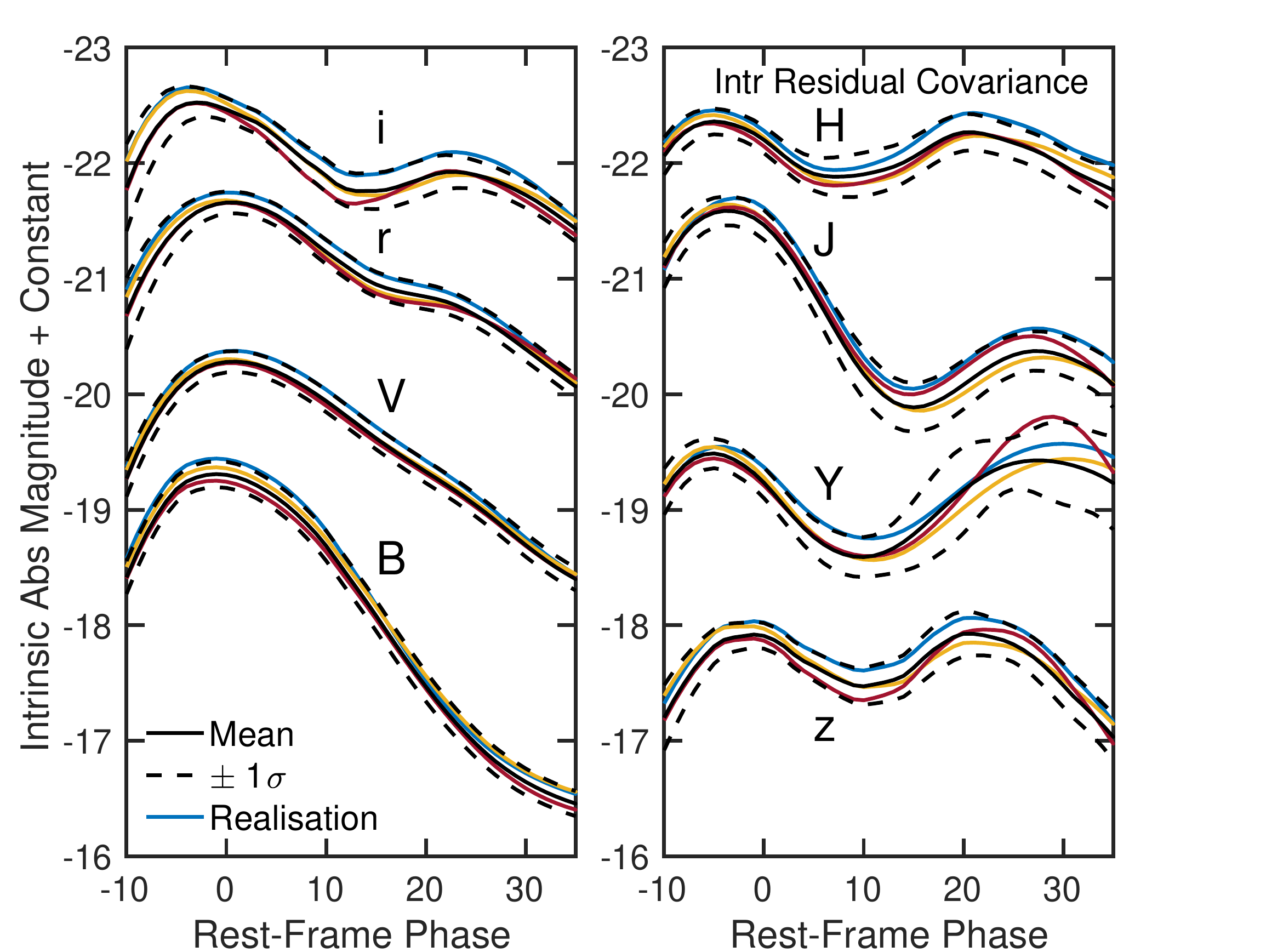}
\includegraphics[scale=0.58]{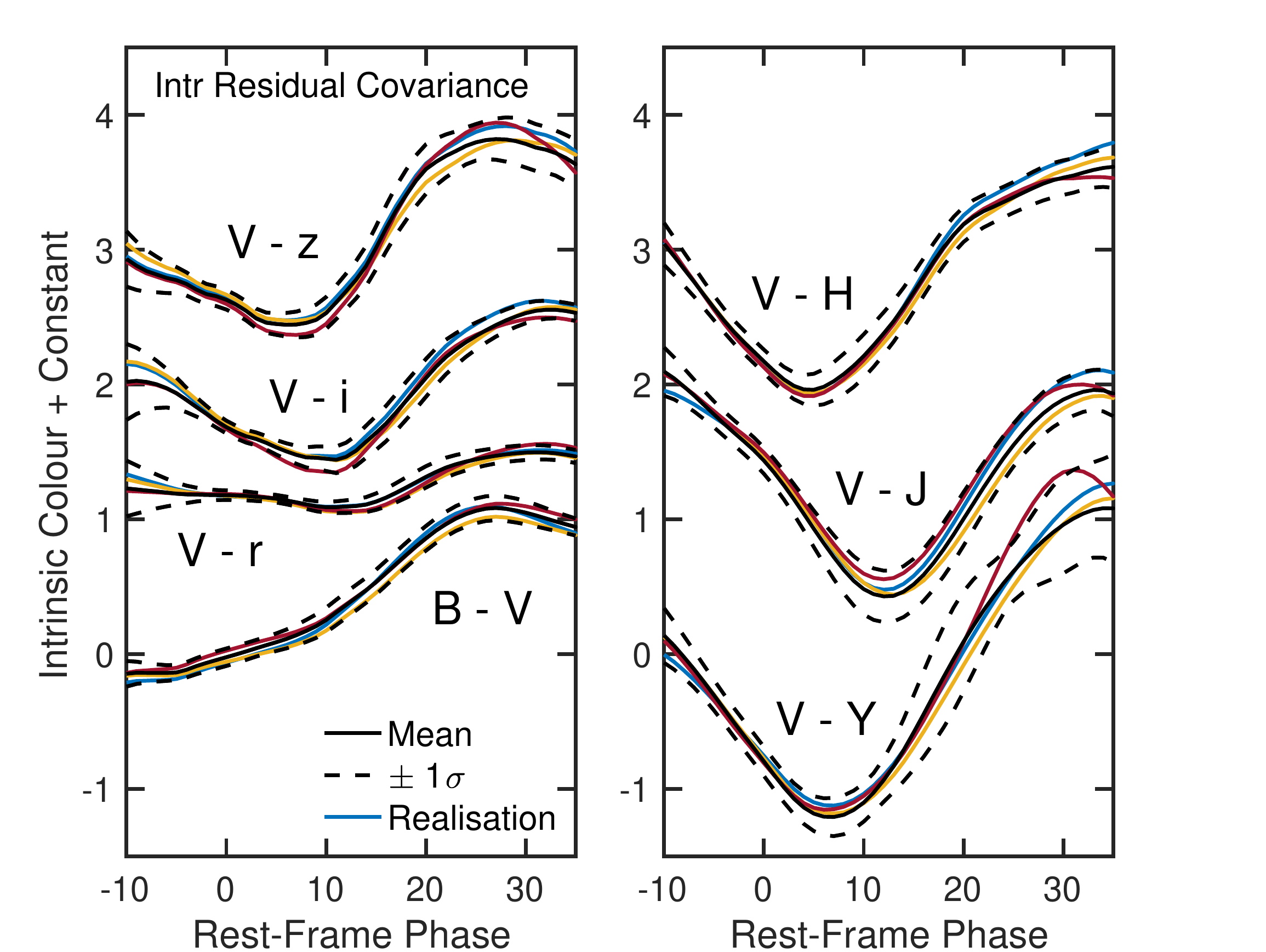}
	\caption{Effects of the covariance of phase- and wavelength-dependent intrinsic SED residuals on optical and NIR light curves (top) and colour curves (bottom).  We fix the main effects $\theta_1 = A_V = 0$.  The black solid lines represent the light curves or colour curves generated from the mean intrinsic SED model.  The dashed lines correspond to $\pm 1$ population standard deviation around the mean curves captured by the intrinsic SED residual covariance. The light curves or colour curves corresponding to the effects of three sample realisations of intrinsic SED residual functions $\eta_s(t,\lambda)$ are shown as blue, yellow, or red curves. For example, the red curves in all the panels correspond to the effect of a single realisation of a intrinsic SED residual function.}
	\label{fig:intrcov}
\end{figure*}

Fig. \ref{fig:intrcov} shows the effect of intrinsic SED residuals on rest-frame optical and NIR light and colour curves.  We hold $\theta_1 = A_V = 0$, and we compute the impact of the distribution of SED residuals on the light curves and colour curves by integrating through the reference passbands.  We compute the $\pm 1\sigma$ range at each epoch $t$. We do not unrealistically assume the residuals are statistically independent at each phase or in each filter; rather the residuals manifest as continuous perturbations around the main effects. The model captures continuous residual SED functions correlated across phase and wavelength. To illustrate this, we show the effect of three realisations of the intrinsic residual functions on the light curves. The residual variance is generally narrow at phases around the first peak.  In later phases, particularly in the NIR, there is more intrinsic residual variation because the 1st FPC does not capture the full range of variation of the second peak.  In \S \ref{sec:w2}, we show that some of the additional structure there may be captured with higher order FPCs. 

\subsubsection{Host Galaxy Dust Population}

\begin{figure}
\includegraphics[scale=0.5]{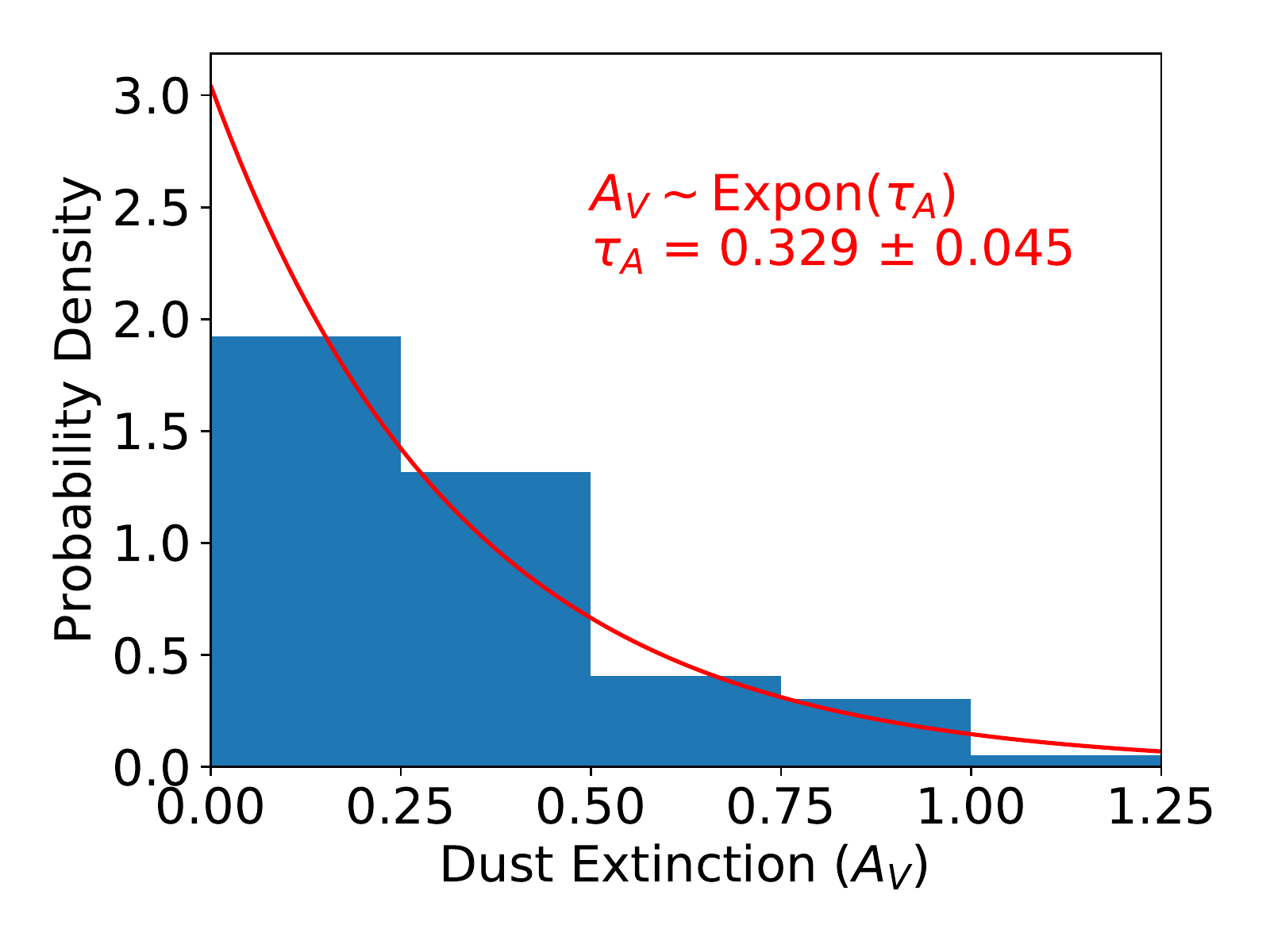}
	\caption{Distribution of posterior mean estimates of $A_V^s$ for the SNe Ia sample.  The model exponential distribution with the inferred scale (average) $\tau_A$ is shown.}
	\label{fig:AVdistr}
\end{figure}

\begin{figure}
\includegraphics[scale=0.35]{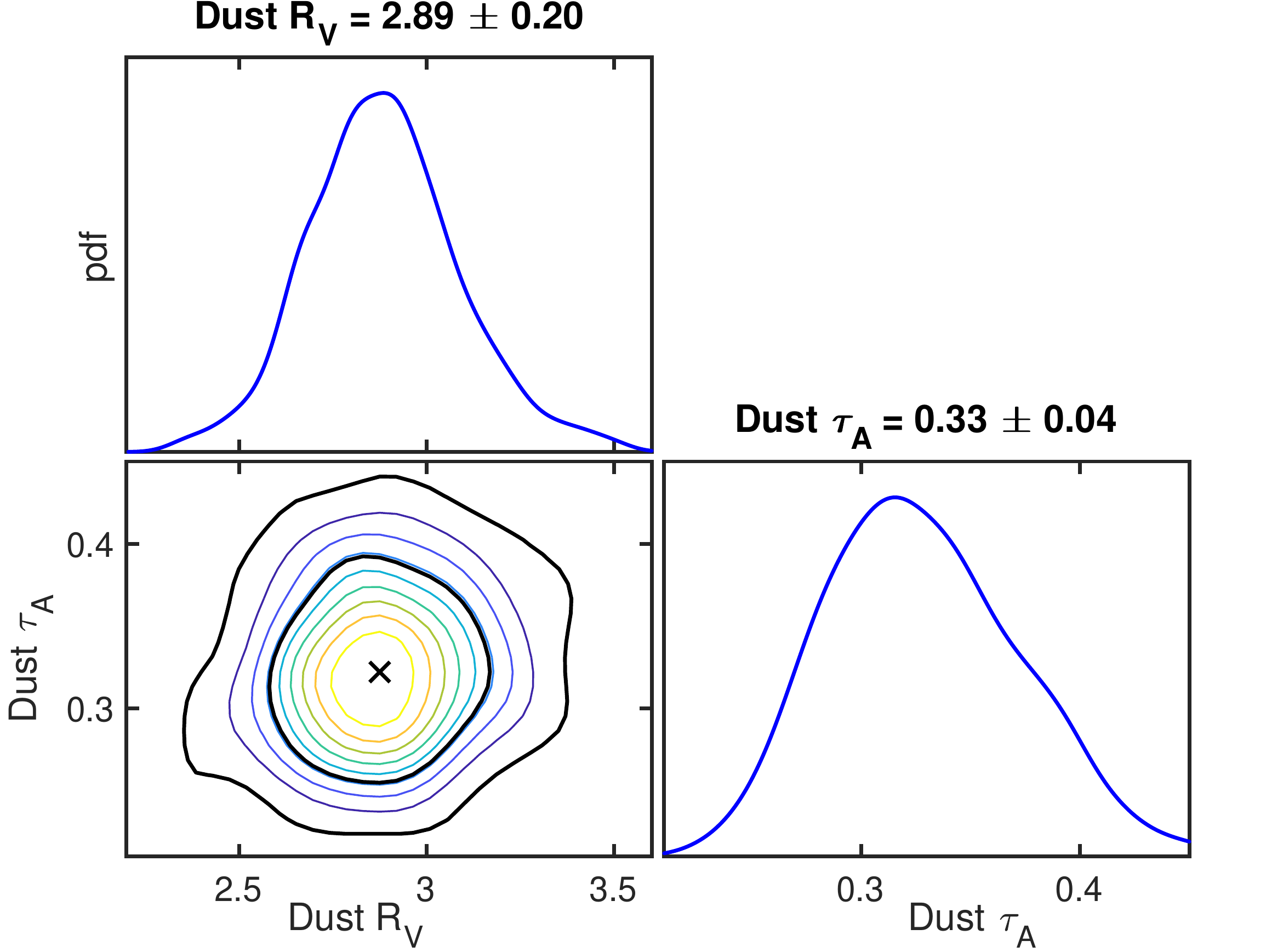}
	\caption{Posterior distribution of the inferred global $R_V$ of the host galaxy dust law and $\tau_A$, the population mean $A_V$.  The black contours of the 2D contour plot contain 68\% and 95\% of the posterior probability, and the mode is marked.  The 1D marginals are depicted by kernel density estimates of the MCMC samples.}
	\label{fig:dustRVtau}
\end{figure}

Fig. \ref{fig:AVdistr} shows the distribution of posterior mean estimates of the individual dust extinction $A_V$ values.  It is well described by an exponential distribution with an average value of $\tau_A  = 0.329 \pm 0.045$ mag.  Fig. \ref{fig:dustRVtau} shows posterior inferences of the average $\tau_A$ and the global value of the dust law slope $R_V$.   The posterior constraints are determined during the training phase, and thus are obtained by marginalising over all other components and hyperparameters of the hierarchical model.   These posterior estimates are well constrained fairly independently.  In particular, for this sample with colour excess $E(B-V)_\text{host} \lesssim 0.4$, the estimated global $R_V = 2.89 \pm 0.20$ is consistent with the average for normal Milky Way dust.  This is in good agreement with previous analyses of nearby samples, which have found, at these relatively low-to-moderate values of reddening (which are similar to those found in the cosmological sample), average values of the host dust $R_V \approx 2.5 - 3$ \citep{mandel11, chotard11, foleykasen11, burns14, mandel17, leget20}.


The hierarchical model constrains $R_V$ by analysing and weighing the entire distribution of SEDs over phase and optical to NIR wavelengths using the entire training set of SNe Ia.  For visualisation purposes, however, it is useful to inspect a ``slice'' of this inference by examining a low-dimensional summary.  Multi-dimensional colour information is useful as it provides constraints on the dust distribution while being insensitive to the distance estimate (and its errors). We exploit the fact that the optical and NIR data allows us to constrain the dust effects over a much larger wavelength range than is possible conventionally with the optical data alone.  The plot of the dust law in Fig. \ref{fig:pgm} shows that the extinction at NIR $H$-band ($\approx 1.6 \, \mu$m, cf. Fig. \ref{fig:W1_SED}) is only $16\%$ of that in optical $V$-band ($\approx 0.54\, \mu$m), and very insensitive to $R_V$.  Thus, the differential extinction (the colour excess) between $V$- and $H$-bands probes a large net dust effect ($\approx 0.83 \,A_V$), while itself being insensitive to $R_V$.   Meanwhile, the colour excess between $B$ ($\approx 0.43\, \mu$m) and $H$-bands similarly covers a large wavelength range and therefore a large dust effect, but because of the high sensitivity of $A_B$ to $R_V$ (for a given $A_V$), this colour excess is very sensitive to $R_V$.  The complementary optical $B-V$ colours cover only a narrow range in the optical, and therefore captures a smaller differential effect of dust, but is also highly sensitive to $R_V$.  The advantage of measurements spanning optical to NIR is that we can leverage the joint colour information in these SNe Ia to constrain and break the degeneracy between $A_V$ and $R_V$ in the optical-only colours (Fig. \ref{fig:colcurves}).

\begin{figure}
\includegraphics[scale=0.37]{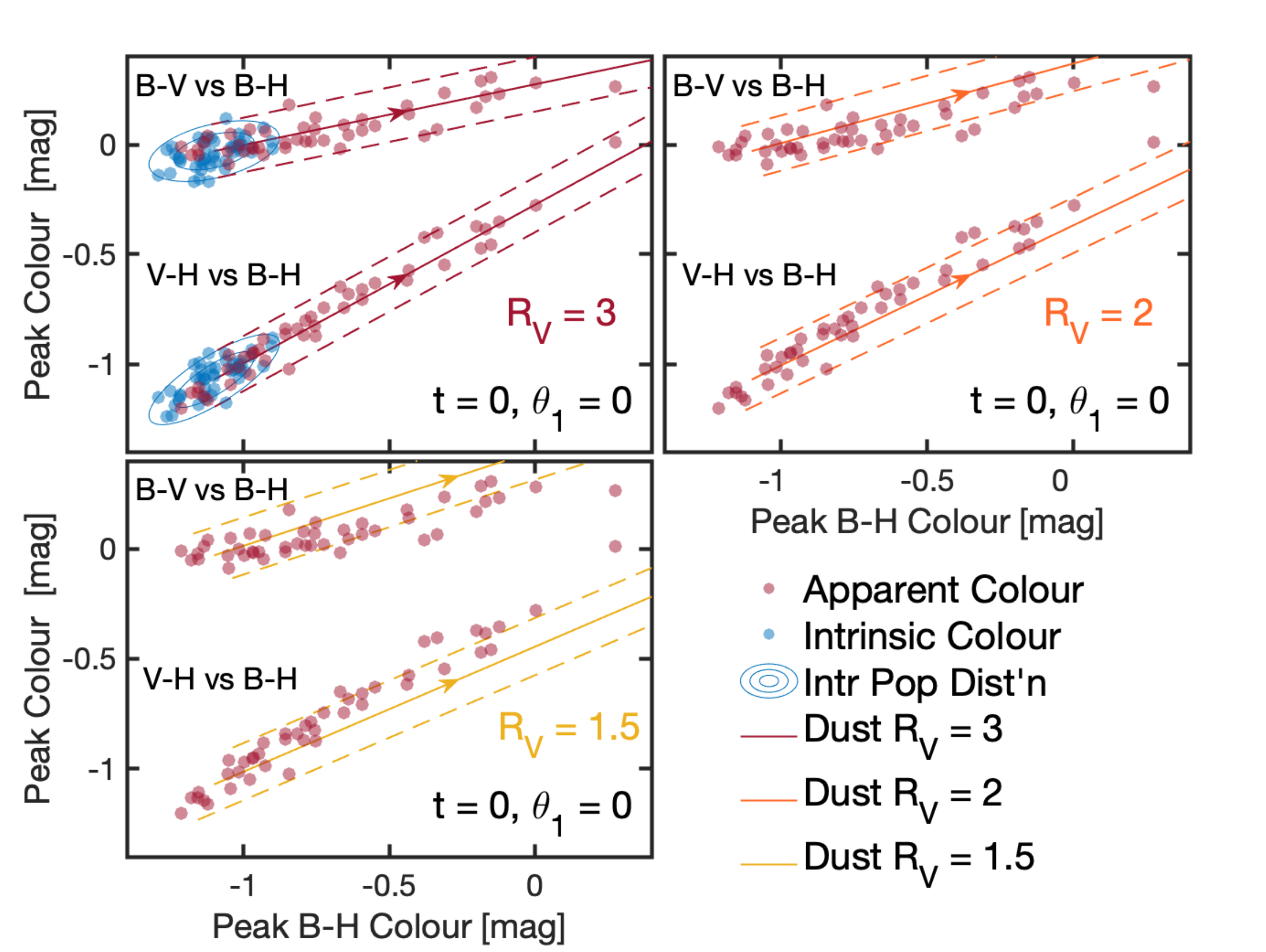}
	\caption{ Constraints on the host galaxy dust $R_V$ from the optical and NIR colour-colour diagram of SNe Ia observed in $B$, $V$, and $H$ near peak ($t = 0$).  (top left) Each point is a posterior realisation of the peak apparent colours (red) or intrinsic colours (blue) of a SN, corrected for the inferred intrinsic colour-shape relation to $\theta_1 = 0$.  Blue ellipses are (68\%, 95\%) contours of the intrinsic colour population distribution inferred during the training phase, which estimated a global dust law parameter $R_V = 2.9 \pm 0.2$. For comparison, the red solid (dashed) lines have the slope of the reddening vector for $R_V = 3$ in these colours, and intercept the mean (are tangent to the 95\% contour) of the intrinsic distribution. Nearly all of the SNe Ia apparent colours should lie within the dashed lines under the correct dust reddening law.  (top right) Comparison of the apparent colour distribution with the inconsistent dust reddening vector for $R_V = 2$.  (bottom left) Comparison of the apparent colour distribution with the inconsistent dust reddening vector for $R_V = 1.5$.}
\label{fig:optnir_cols}
\end{figure}

Fig. \ref{fig:optnir_cols} shows a ``slice'' of the constraints on $R_V$ in these colours from training the \textsc{BayeSN} model (Eq. \ref{eqn:globalpost}).  The top left panel shows posterior realisations of the SNe Ia peak ($t = 0$) apparent colours (red) and intrinsic colours (blue) inferred by the model, and corrected for the inferred intrinsic colour-shape relation (Fig. \ref{fig:colcurves}) to $\theta_1 = 0$.  The blue contours indicate the (68\%, 95\%) contours of the inferred intrinsic colour distribution, which is anchored by the SNe Ia with the least inferred amount of dust.  The red solid (dashed) lines have the slope of the reddening vector for $R_V = 3$, and intercept the mean (are tangent to the 95\% contour) of the intrinsic distribution. Nearly all of the SNe Ia apparent colours should lie within the dashed lines under the correct dust reddening law.  The arrows indicate the dust reddening vector for each colour pair for a given dust law $R_V$ and illustrate a shift corresponding to $A_V = 0.57$ mag from the centre of the intrinsic colour distribution.  In the other panels, we show that the apparent colour distribution in these colour pairs are inconsistent with the dust reddening vectors for $R_V = 2$ or $R_V = 1.5$.  That is, given the apparent colours of the low-reddening set (low $B-H$), assuming a low $R_V$ would predict bluer (more negative) average $V-H$ apparent colours for a given $B-H$ for more reddened SNe Ia  (e.g. $B-H> 0.5$) than is observed. Conversely, the apparent colours of the high-reddening set (high $B-H$) would imply that the apparent $V-H$ colours of the low-reddening set ought to be redder (more positive) than is observed, when assuming a low $R_V$. For $B-V$ colours, the same inconsistencies persist but in the opposite sense.  The high- and low-reddening ends of the apparent colour distribution are most consistent with each other for $R_V \approx 3$.

Because the estimation of $R_V$ hinges on the comparison of the colours of high-reddening SNe to those of low-reddening SNe, the most highly reddened SNe have the most leverage.  In our sample, SN 1998bu has the largest extinction estimate ($A_V = 1.15 \pm 0.08$).  To test that our $R_V$ estimate is not entirely driven by this SN, we retrained the full hierarchical model omitting SN 1998bu.  We found $R_V = 2.83 \pm 0.19$, indicating that our estimate is robust to the reddest SN.

\subsubsection{Covariance Structure of Optical and NIR Peak Absolute Magnitudes}

During the training phase, we estimate the population covariance structure of SN Ia SEDs.  The covariance structure is implied by the model Eq. \ref{eqn:model_logsed} and the population distribution of the latent parameters.  The total population covariance of the log latent SED at two different rest-frame coordinates $(t, \lambda_r)$ and $(t', \lambda_r')$ is captured by the model as
\begin{equation}
\begin{split}
\text{Cov}[\log S(t, \lambda_r)&, \log S(t', \lambda_r')] = \text{Var}[A_V]\, \xi(\lambda_r; R_V)\, \xi(\lambda_r'; R_V) \\
&+ \Big[ \sum_{i=1}^K W_i(t, \lambda_r) W_i(t', \lambda_r') \Big] \\ 
&+ \sigma_0^2 + k_\epsilon(t,\lambda_r; t', \lambda_r') 
\end{split}
\end{equation}
where $k_\epsilon(t,\lambda_r; t', \lambda_r') $ is given by Eq. \ref{eqn:gp_kernel}, and we invoke the statistical independence and normalisation of the FPC scores: $\text{Cov}[\theta_i, \theta_j] = \delta_{ij}$.  On the right-hand side, the top line describes the covariance across rest-frame wavelength induced by the dust extinction and the dust law $\xi(\lambda)$, which depends on $R_V$.  The second line describes covariance across both phase and wavelength induced by the $K$ intrinsic functional principal components of the SED.   The third line describes the covariance of the intrinsic residual terms $\eta_s(t,\lambda_r) = \delta M_s + \epsilon_s(t,\lambda_r)$.  Because the absolute magnitude in any one passband at some phase $t$ is obtained by exponentiating Eq. \ref{eqn:model_logsed} and then performing an integral of the SED under the transmission function, the covariance between any pair of absolute magnitudes in different filters at different phases is not analytic and must be computed numerically.

\begin{figure}
\includegraphics[scale=0.35]{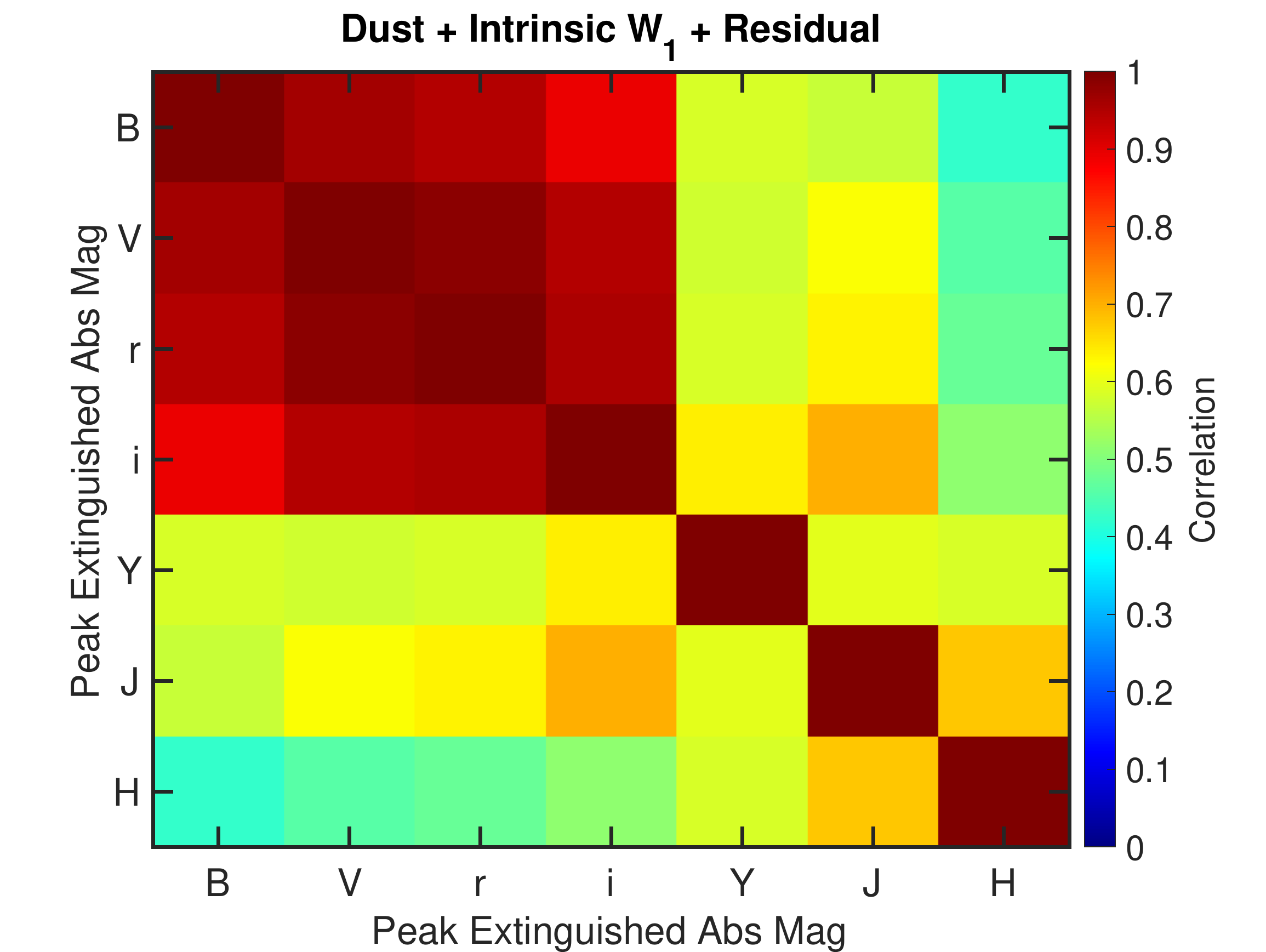}
	\caption{Map of population correlations between peak ($t = 0$) extinguished absolute magnitudes in optical and NIR passbands.  These include all modelled sources of latent SED variation, including dust extinction, the intrinsic FPC $\theta_1 W_1(t,\lambda_r)$, and the residual SED covariance. Dust effects induce significant wavelength-dependent correlations in the optical, but have significantly diminished effect in the NIR.  While the optical magnitudes are significantly correlated with themselves, they are less so with the NIR magnitudes, with optical-NIR cross-correlations as low as $\approx 40\%$.  This indicates there is additional information in the NIR that helps improve distance estimates.}
	\label{fig:corrmaps}
\end{figure}

The full covariance structure over rest-frame phase and wavelength learned by the model is complex, and we defer a detailed discussion to future work. Here, we distill some of its key aspects.  Fig. \ref{fig:corrmaps} depicts the population cross-correlation structure between peak (at $t = 0$) optical and NIR absolute magnitudes. The variation in absolute magnitudes is generated by the combination of the various latent component effects on the SED, and is obtained by integrating the SED through the reference filters. The map shows the correlation of the peak extinguished absolute magnitudes across optical and NIR passbands, inclusive of dust, intrinsic $\theta_1$ SED variation, and residual covariance.  The peak absolute magnitudes in the optical have a very strong total correlation, whereas the cross-correlation between optical and NIR peak absolute magnitudes is as low as $\approx 40\%$.  This is caused in part by the strong, coherent wavelength-dependence of the host galaxy dust extinction. However, the dust extinction is significantly diminished in the NIR. This reduced cross-correlation indicates there is additional information in the NIR magnitudes that helps to improve distance estimates.

\subsection{Hubble Diagram Analysis}

After training the model by sampling Eq. \ref{eqn:globalpost}, we obtain posterior estimates of the FPC and population hyperparameter $\bm{\hat{H}}  \equiv (\bm{\hat{W}}_{0:K}, \bm{\hat{\Sigma}}_\epsilon, \hat{\sigma}_0^2, \hat{\tau}_A, \hat{R}_V)$.  We then use these to evaluate the photometric distances, derived from the light curves alone, using Eq. \ref{eqn:photdist}.  We take the posterior mean and standard deviation of the posterior probability density of the photometric distances.  Table \ref{table1} lists the redshifts, external distance estimates, and \textsc{BayeSN} photometric distance moduli for the sample.

We assess the accuracy and precision of our photometric distance estimate by comparison to the external distance estimates, via the Hubble residuals, $\hat{\mu}_s^\text{phot}- \hat{\mu}_s^\text{ext}$.  We compare them using two summary statistics, listed in Table \ref{table2}.  First, we report the simple total RMS the differences between our posterior mean estimate photometric distance modulus $\hat{\mu}_s^\text{phot}$ and the external distance estimate $\hat{\mu}_s^\text{ext}$.  Second, we report a statistic we denote $\hat{\sigma}_{\text{-pv}}$, obtained by minimising
\begin{equation}
\hat{\sigma}_{\text{-pv}} = \underset{\sigma_{\text{-pv}}}{\arg\max} \log\left[ \prod_s N(\hat{\mu}_s^\text{phot} | \, \hat{\mu}_s^\text{ext}, \sigma_{\text{ext},s}^2 + \sigma_{\text{-pv}}^2) \right].
\end{equation}
This is a maximum likelihood estimate of the amount of dispersion in the Hubble residuals not accounted for by the uncertainties in the external distance estimate, which is dominated by the peculiar velocity uncertainty $\sigma_\text{pec} = 150 \text{ km s}^{-1}$ for the vast majority of this low-$z$ sample.

\begin{figure*}
	\includegraphics[scale=0.55]{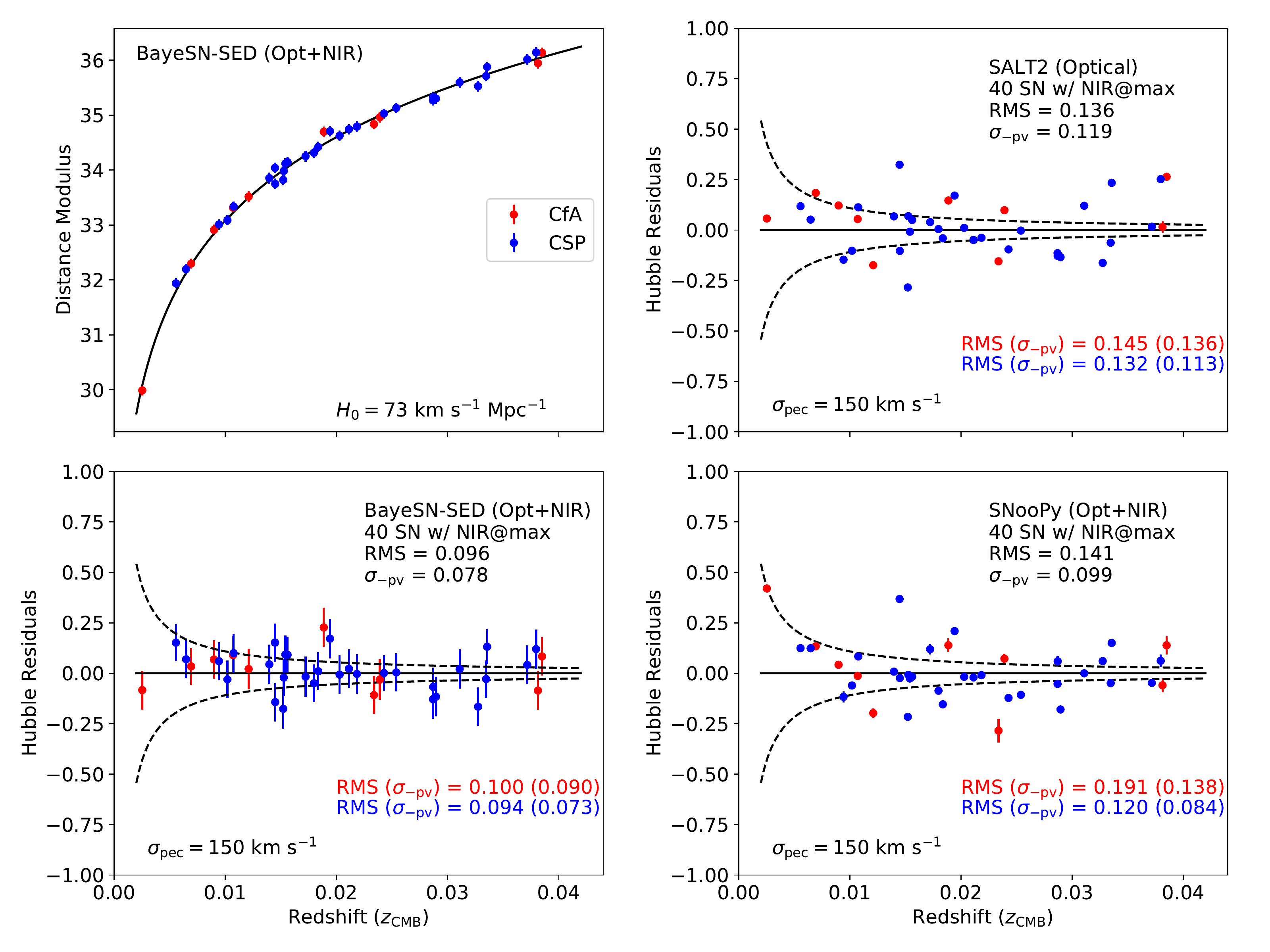}
	\caption{Comparison of Hubble diagrams and Hubble residuals from \textsc{BayeSN}, \textsc{SNooPy}, and \textsc{SALT2}, applied to the same set of CfA and CSP SNe Ia with NIR data near maximum light.  (top left) Hubble Diagram of photometric distances obtained by fitting the optical and NIR light curves, compared to the local distance-redshift relation under standard cosmological parameters.  (bottom left) Hubble residuals for \textsc{BayeSN}.  The simple total RMS is 0.096 mag.  After removing the expected variance due to peculiar velocity uncertainty (dashed, $\sigma_\text{pec} = 150 \text{ km s}^{-1}$), the remaining dispersion is $\hat{\sigma}_{\text{-pv}} = 0.078$ mag.  The distance uncertainties are determined via marginalisation accounting for the residual covariance (Eq. \ref{eqn:photdist}). (top right) Hubble residuals from SALT2 applied to the optical-only data ($BVRI$) of the same sample. (bottom right) Hubble residuals from SNooPy applied to the optical and NIR data of the same sample.}
	\label{fig:hd1}
\end{figure*}

It is conventional in the SALT2 analysis to compute an ``intrinsic dispersion''\footnote{But see footnote \ref{fn1}.} $\sigma_\text{int}$ of the Hubble residuals, by estimating the amount of scatter in the Hubble residuals in excess of the expected contributions of ``measurement error'' (which is really the estimated uncertainty on the fit parameters $m_B, x_1, c$), and the peculiar velocity uncertainties. This is necessary because only the light curve fitting uncertainties on the SALT2 parameters are propagated through the Tripp formula, Eq. \ref{eqn:tripp}, to compute the distance modulus uncertainties, and the results are typically much smaller than the total RMS in the Hubble diagram. Similarly, SNooPy only uses the photometric measurement uncertainties in the light curve fit. In contrast, \textsc{BayeSN} produces distance uncertainties via Bayesian marginalisation of the SED fit to the light curve data, coherently incorporating $\theta_1$ and $A_V$ uncertainties and the residual covariance over phase and wavelength (Eq. \ref{eqn:photdist}).  Since each method has a different way of reporting the distance errors, we do not ``subtract'' the reported distance errors from the total RMS. Instead, to ensure consistent comparisons across methods, we use $\hat{\sigma}_{\text{-pv}}$ to remove from the total RMS only the expected contribution from external distance errors (e.g. peculiar velocities), which are the same for each method applied to the same set of SNe Ia.

Table \ref{table2} lists these Hubble residual dispersion measures for different subsets of the SN Ia sample. The vast majority comes from two large surveys with homogeneously reduced data, the CfA  \citep{hicken09a, hicken12, friedman15} and CSP-I \citep{krisciunas17}.  We label this set ``CfA+CSP''.  Including the minority of other SNe Ia drawn from the more heterogenous data sources in the literature results in the ``All'' sample.  Furthermore, a subset of the full ``AnyNIR'' sample with NIR observations near maximum light is labelled ``NIR@max.''  We run \textsc{BayeSN} and \textsc{SNooPy} on either optical-only ($BVRI$) or optical+NIR ($BVRIYJH$) light curve data, while SALT2 is only run on optical $BVRI$ data.

\begin{table*}
	\centering
	\caption{Summary of Hubble Residuals}
	\begin{threeparttable}\label{table2}
		\begin{tabular}{c c c c c c c c c c c} 
			\toprule
			SN source\tnote{a}& NIR cut\tnote{b} & $N_\text{SN}$ & $\lambda$\tnote{c}  & Model\tnote{d} & total rms\tnote{e} & $\sigma_\text{-pv} (150)$\tnote{f} \\ 
			\midrule
			CfA+CSP & NIR@max & 40 & $BVRIYJH$ & BayeSN-tr & 0.096 & 0.078  \\
			CfA+CSP & NIR@max & 40 & $BVRIYJH$  & BayeSN-cv & 0.108 & 0.097   \\
			CfA+CSP & NIR@max & 40 & $BVRIYJH$  & SNooPy & 0.141 & 0.099 \\
			CfA+CSP & NIR@max & 40 & $BVRI$ & SALT2 & 0.136 & 0.119   \\
			\midrule
			All & NIR@max & 48 & $BVRIYJH$ & BayeSN-tr & 0.113 & 0.083   \\
			All & NIR@max & 48 & $BVRIYJH$ & BayeSN-cv & 0.123 & 0.103 \\
			All & NIR@max & 48 &  $BVRIYJH$ & SNooPy & 0.148 & 0.099 \\
			All & NIR@max & 48 & $BVRI$ & SALT2 & 0.141 & 0.117  \\
			\midrule
			CfA+CSP & AnyNIR & 66 & $BVRIYJH$ & BayeSN-tr & 0.135 & 0.097  \\
			CfA+CSP & AnyNIR & 66 & $BVRIYJH$  & BayeSN-cv & 0.145 & 0.114   \\
			CfA+CSP & AnyNIR & 66 & $BVRIYJH$  & SNooPy & 0.157 & 0.115 \\
			CfA+CSP & AnyNIR & 66 & $BVRI$ & SALT2 & 0.159 & 0.131 \\
			\midrule
			All & AnyNIR & 79 & $BVRIYJH$ & BayeSN-tr & 0.137 & 0.096 \\
			All & AnyNIR & 79 & $BVRIYJH$ & BayeSN-cv & 0.147 & 0.113  \\
			All & AnyNIR & 79 &  $BVRIYJH$ & SNooPy & 0.161 & 0.113 \\
			All & AnyNIR & 79 & $BVRI$  & SALT2 & 0.159 & 0.126   \\
			\midrule
			CfA+CSP & AnyNIR & 66 & $BVRI$  & BayeSN-tr & 0.149 & 0.118   \\
			CfA+CSP & AnyNIR & 66 & $BVRI$ & BayeSN-cv & 0.156 & 0.132 \\
			CfA+CSP & AnyNIR & 66 & $BVRI$  & SNooPy & 0.158 & 0.132 \\
			CfA+CSP & AnyNIR & 66 & $BVRI$ & SALT2 & 0.159 & 0.131   \\
			\midrule
			All & AnyNIR & 79 & $BVRI$  & BayeSN-tr & 0.150 & 0.115   \\
			All & AnyNIR & 79 & $BVRI$  & BayeSN-cv & 0.157 & 0.128  \\
			All & AnyNIR & 79 & $BVRI$  & SNooPy & 0.158 & 0.130 \\
			All & AnyNIR & 79 & $BVRI$  & SALT2 & 0.159 & 0.126   \\
			\bottomrule
		\end{tabular}
		\begin{tablenotes}
			\item[a] Data Source.  ``All'' $=$ CfA+CSP+Others
			\item[b] The ``NIR@max'' cut requires NIR data near maximum light. ``AnyNIR'' does not.
			\item[c] In optical+NIR fitting, all available data in $BVRIYJH$ is used. In optical-only fitting, only available data in $BVRI$ is used, where $R$ and $I$ can also include $r,r'$ and $i,i'$. 
			\item[d] ``Bayesn-tr'' refers to the error of photometric distances from resubstitution of the whole training set. ``BayeSN-cv'' refers to the error of photometric distances from 10-fold cross-validation. We cannot do equivalent cross-validation with SALT2 or SNooPy.
			\item[e] Simple total RMS of the Hubble residuals.
			\item[f]  Dispersion estimate after removing expected variance due to peculiar velocity uncertainties, assuming $\sigma_\text{pec} = 150 \text{ km s}^{-1}$.	
		\end{tablenotes}
	\end{threeparttable}
\end{table*}

\subsubsection{Resubstitution or Training Error}

The resubstitution, or training error, is obtained by training the model on the entire dataset, and then applying it to determine the photometric distance estimates to the SNe Ia that were in the training set.  In Table \ref{table2}, these estimates are labelled ``BayeSN-tr''.  Fig. \ref{fig:hd1} shows the Hubble diagram obtained with \textsc{BayeSN} fits of optical and NIR data of the CfA+CSP NIR@max sample.  With joint optical and NIR data, \textsc{BayeSN} achieves a low total RMS $= 0.096$ mag on this set. Removing the expected contribution from external distance error and peculiar velocities, we obtain $\hat{\sigma}_{\text{-pv}} = 0.077$ mag.  Meanwhile, on the same set of SNe Ia, \textsc{SNooPy} and \textsc{SALT2} have larger RMS $\approx 0.14$ mag, with $\hat{\sigma}_{\text{-pv}} \approx 0.10 - 0.12$ mag.  Notably, the photometric distance modulus uncertainties of individual SNe Ia from \textsc{SNooPy} or \textsc{SALT2} with the standard procedure are small in comparison to the total RMS, because they only propagate the uncertainties due to photometric light curve errors. In contrast, the \textsc{BayeSN} photometric distance uncertainties are obtained in a principled manner by marginalisation of the latent components including the residual covariance (Eq. \ref{eqn:photdist}).  The individual photometric distance uncertainties from \textsc{BayeSN} listed in Table \ref{table1} already reflect the scatter in the Hubble diagram.

We assess the significance of the difference between the RMS Hubble residual of distance from our model compared to those from SALT2 using bootstrap.  From the full training set SNe Ia, we construct a bootstrapped set by sampling with replacement.  For each method, we compute the Hubble residual RMS of the SNe Ia within the bootstrapped set.  We compute the difference in RMS between the two methods within the bootstrapped set.  We repeat this 1,000 times and then compute the variance of the differences in RMS across the bootstraps.  This procedure accounts for the fact that each method is analysing the \emph{same} set of SNe Ia, and therefore the joint sampling distribution of both methods' RMS over bootstraps is correlated. For the CfA+CSP NIR@max subset, we compare SALT2 using optical (which has the lowest RMS of the alternate methods) versus \textsc{BayeSN} using optical+NIR, and we find a $\Delta \text{RMS} = 0.040 \pm 0.012\, (3.3\sigma)$.

Table \ref{table2} summarises of Hubble diagram dispersions of the other subsets of the SN Ia sample.  We find that the addition of the literature sample to the CfA+CSP sample (to constitute All) increases the dispersion slightly in nearly all cases, which is to be expected since these SNe Ia come from more heterogenous data sources. \textsc{BayeSN} optical+NIR distances are still more precise than SNooPY and SALT2 in the AnyNIR sample, when we do not require NIR measurements near maximum light, but the advantage is reduced.  On optical-only data ($BVRI$), all three methods perform similarly, with total RMS $\approx 0.15-0.16$ mag.

\subsubsection{Cross-Validation}

Cross-validation techniques to test the sensitivity of SN Ia models and their distance estimates to the finite training set have been previously employed by \citet{mandel09,mandel11} and \citet*{blondin11}.  These procedures address the double use of the data inherent in resubstitution. We performed 10-fold cross-validation to assess the out-of-training sample distance error.   We equally divided the full training set into 10 folds, each with a roughly similar redshift distribution.  First, we hold out one fold, and train a new \textsc{BayeSN} SED model on the SNe Ia in the other 9 folds.  Then we used the new trained model to estimate the photometric distances of the SNe Ia in the held-out fold.   We repeated this procedure 10 times, each time holding out a different fold, training a new model on the complement, and using it to evaluate the photometric distances of the held-out SNe.  The Hubble residual summaries of the cross-validation out-of-training sample photometric distances thus obtained are listed in Table \ref{table2} as ``BayeSN-cv''.

In the best case, for the CfA+CSP NIR@max subset, the total RMS of the photometric distances relative to the external distances is $0.108$ mag. As expected, this is slightly higher than the RMS training error ($0.096$ mag) because the cross-validated distance of each SN is obtained using a model trained on a set that excludes that SN.  This is an  overestimate of the true error of the fully-trained model, since each model under CV is trained on a 10\% smaller training set than the full sample. We expect the difference between the training and cross-validation error to narrow as more training data becomes available.  Still, the difference between the two numbers is already small (0.012 mag), so it is reasonable to conclude that the typical distance error for similar optical and NIR light curves with peak NIR data is $\approx 0.10$ mag.  

A large fraction of our training set SNe Ia were also used in the training sets for both SNooPy \citep{burns11} and SALT2 \citep{guy10}.  To our knowledge, there has been no equivalent cross-validation analysis, including hold-out and iterative retraining, for these other models.  Since we are unable to retrain these other models on partitions or resampled subsets, it is difficult to make equivalent, direct comparisons of these models to our cross-validation results.

Our cross-validation runs demonstrate the capability of our training code to straightforwardly and repeatedly train new models on different SN Ia datasets automatically without human intervention.  We will be able to use this modularity to train and compare new BayeSN SED models based on datasets partitioned by survey or by astrophysical classes (e.g. SN Ia host galaxy properties or spectroscopic subclasses) to investigate the statistical and physical differences in the learned SED components and latent variables.

\subsection{Application to Foundation SN Ia light curves}\label{sec:fdn}

The optical and NIR light curves in our training set listed in Table \ref{table1} are mainly from the Carnegie Supernova Project and CfA Supernova Program, which typically measured high-quality light curves with relatively frequent time sampling (c.f. Fig. \ref{fig:sn2005iq_lc}).  However, most SN Ia light curves used for cosmology are not sampled as well in phase or wavelength.  To test our model on SN Ia light curves outside of our training set with more typical sampling, we have fit $griz$ light curves obtained by the Foundation Supernova Survey using the Pan-STARRS1 (PS1) telescope \citep{foley18} .

\begin{figure}
	\includegraphics[scale=0.6]{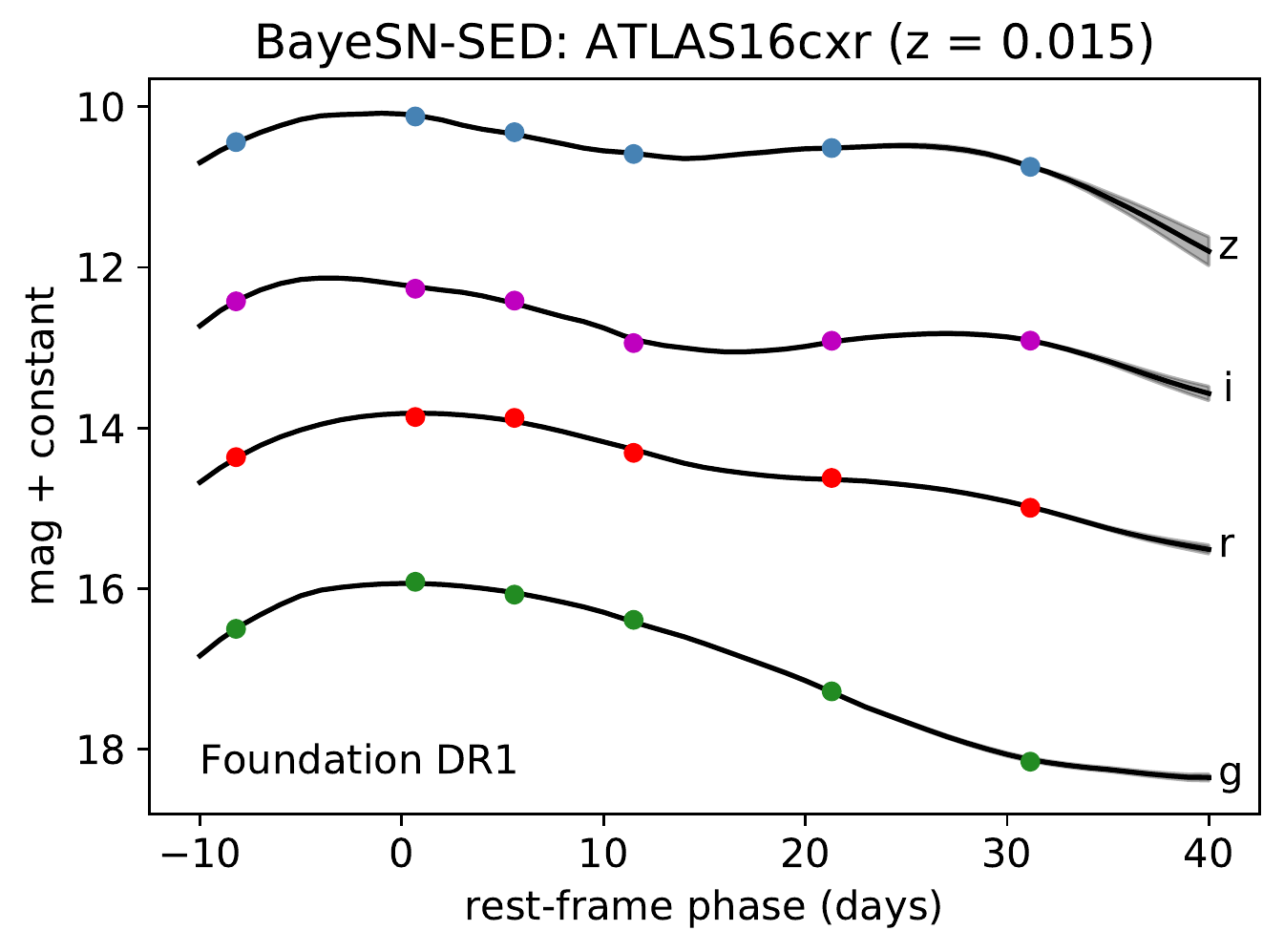}
	\includegraphics[scale=0.35]{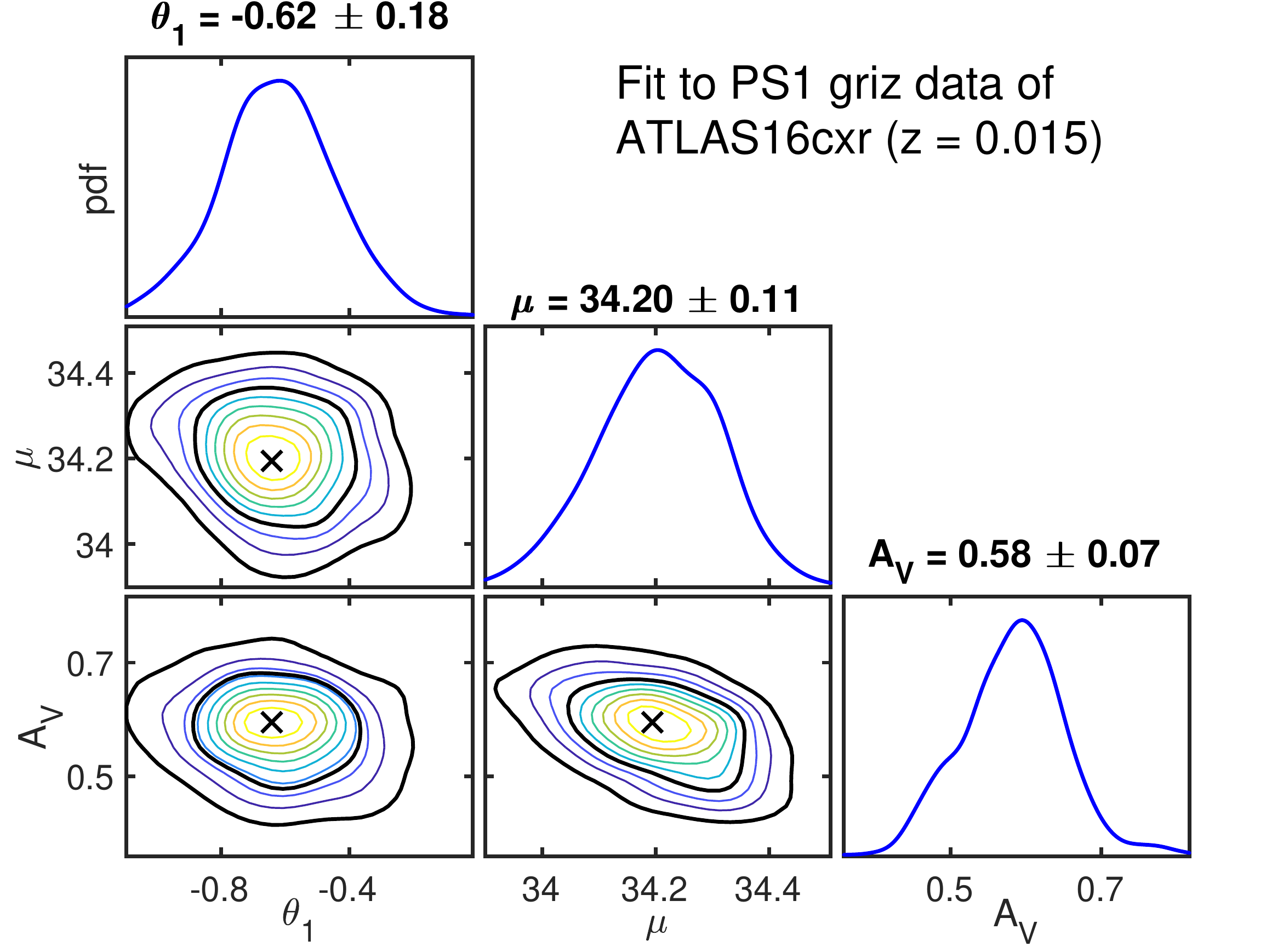}
	\caption{(top) \textsc{BayeSN} light curve fit of Foundation DR1 $griz$ observations of ATLAS16cxr. (bottom) Posterior distribution of \textsc{BayeSN} parameters from the light curve fit.}
	\label{fig:foundation_lc}
\end{figure}

Fig. \ref{fig:foundation_lc} demonstrates a \textsc{BayeSN} fit to Foundation observations of the Type Ia SN 2016gou / ATLAS16cxr.  It shows the well-constrained joint posterior distribution of the parameters obtained from the MCMC fit: the $\theta_1$ coefficient of the 1st FPC, the dust extinction $A_V$, and the photometric distance $\mu$. Because \textsc{BayeSN} is a model for the continuous SED spanning 0.35 to 1.8 $\mu$m, we are able to integrate the model SED under the $griz$ PS1 passbands to fit this data, even though these exact passbands were not used in the training phase.  Our SED model does not require $K$-corrections to be computed as preprocessing step to map observer-frame to rest-frame passbands.  Notably, the SALT2.4 model cannot properly fit rest-frame $z$-band due to the wavelength limits of its SED template, and SNooPy lacks a rest-frame $z$-band light curve template.  However, proper modelling of the rest-frame $z$-band is important for fully utilising $griz$ data from low-$z$ surveys such as Foundation \citep{foley18} and the Young Supernova Experiment.  In future work, we will present a full analysis of the Foundation DR1 dataset using our new \textsc{BayeSN} model.

\section{Conclusion}\label{sec:conclusion}

\subsection{Improvements over current models}

We have constructed a new hierarchical Bayesian model, \textsc{BayeSN}, for SN Ia spectral energy distributions (SEDs) from the optical through NIR.  This is the first statistical model for continuous SN Ia SEDs designed for fitting observed optical and NIR light curve data, and is crucial for properly analysing NIR observations from current and future SN Ia surveys. Our model is capable of statistically leveraging the powerful properties of SN Ia in the NIR, in particular the narrow dispersion in NIR luminosities at peak, and the much diminished effect of dust in the SN Ia host galaxies.  \textsc{BayeSN} jointly leverages the optical and NIR data to constrain the dust extinction $A_V$ and the reddening law $R_V$ more stringently, thereby controlling systematic errors due to the dust correction.    \textsc{BayeSN} coherently estimates the covariance of the residual SED functions across time and wavelength, and incorporates them into the dust and distance estimates in a principled, probabilistic manner.

By generalising the previous hierarchical Bayesian framework of \citet{mandel09,mandel11} from modelling light curves in fixed discrete rest-frame filters to modelling a continuous SED function in phase and wavelength, we obviate the need for ad-hoc $K$-correction pre-processing procedures to compute 1-to-1 mappings between observer-frame and rest-frame filters, which is required by SNooPy.  Instead, observed data are compared directly against the model fluxes implied by the redshifted SED model integrated under the observer's passbands.  Redshifting effects are thereby incorporated directly into the statistical model.

Furthermore, \textsc{BayeSN} has a number of advantages over the SALT2.4 model conventionally used in cosmological analyses.  Whereas the SALT2.4 spectral template has coverage only up to rest-frame 0.9 $\mu$m (inclusive of rest-frame $i$-band), our \textsc{BayeSN} SED model extends to 1.8 $\mu$m (i.e. through rest-frame $H$-band).  The SALT2 model cannot internally discern distinct SED components separately describing the effects of SN Ia intrinsic variation versus host galaxy dust extinction.  Instead, it uses a single colour law to fit a single apparent colour parameter, effectively confounding the two physically-distinct sources of spectral variation.  

In contrast, our \textsc{BayeSN} SED model internally encodes the continuous wavelength-dependent host galaxy dust reddening and extinction on the SN Ia SED, as effects physically distinct from the time-dependent intrinsic components of SED variation.  Our model leverages the photometric constraints on the entire continuous SED to determine the dust properties, fit for the intrinsic modes of variation, and coherently weigh the uncertainties and combine information from across phase and wavelength to compute the probability distribution of the photometric distance modulus.  With the low-$z$ compilation analysed here, \textsc{BayeSN} can determine the distance moduli for SNe Ia with optical and NIR coverage near maximum light to $\approx 0.10$ mag precision (total RMS), compared to $0.14$ mag using SALT2 or SNooPy on the same SNe Ia.  Combining optical and NIR data across the entire phase and wavelength range, we used \textsc{BayeSN} to derive tight constraints on the host galaxy dust law.  For this sample with colour excess $E(B-V)_\text{host} \lesssim 0.4$, we found $R_V = 2.9 \pm 0.2$, consistent with the Milky Way average.

\subsection{Applications to current and future datasets}

Beyond the data compilation analysed here, our \textsc{BayeSN} SED model will be broadly applicable for analysing optical and NIR SN Ia light curve data from more recent and current surveys.  Forthcoming data from the Carnegie Supernova Project-II \citep{phillips19} will enable us to expand our nearby training set with high-quality optical and NIR light curves of SNe Ia further into the Hubble flow (limiting the impact of peculiar velocity uncertainties). We are using Foundation DR1 $griz$ light curves obtained with the well-calibrated Pan-STARRS telescope for training and analysis with \textsc{BayeSN}.

 \textsc{BayeSN} is essential for fully analysing data from recent and ongoing programs that use the Hubble Space Telescope to observe SNe Ia in the rest-frame NIR at high-$z$ (RAISIN) and low-$z$ (SIRAH), in conjunction with optical data from ground-based surveys.  The ESO VISTA Extragalactic Infrared Legacy Survey (VEILS)\footnote{\url{https://people.ast.cam.ac.uk/~mbanerji/VEILS/index.html}} recently concluded a time-domain survey that observed SNe Ia in the observer-frame $J$-band up to $z \approx 0.6$, in conjunction with the Dark Energy Survey and the ESO VOILETTE survey in $griz$. 

 LSST's observer-frame $y$ filter will probe the rest-frame NIR $z$ or $y$ bands to redshifts $z \lesssim 0.3$. The Nancy Grace Roman Space Telescope (RST)'s wide imaging filters will extend to 2.0 $\mu$m \citep[e.g.][]{hounsell18}, and thus will probe rest-frame $H$ to $z \lesssim 0.4$, $J$ to $z \lesssim 0.7$, and $Y$ to $z \lesssim 1$.  \textsc{BayeSN} will be crucial for properly leveraging the full wavelength range of these surveys both to constrain the host galaxy dust properties and to produce optimal distance estimates. It will also be important for fully analysing any potential simultaneous observations of SNe Ia by LSST and RST \citep[e.g.][]{foley18b} or Euclid \citep{rhodes17}.

\subsection{Future analyses and model extensions}

Our hierarchical Bayesian SED modelling and inference framework is modular and flexible and will enable us to expand upon the SED model presented here to explore in greater depth various aspects of SNe Ia. In future work, we will investigate dust distributions by allowing $R_V^s$ to vary for each SN Ia within a population governed by hyperparameters to be inferred, as was done previously by \citet{mandel11}.  We will also be able to test alternative forms of the dust extinction law \citep[e.g.][]{goobar08,amanullah15}.  We will further probe the statistical properties of the intrinsic SED residuals over phase and wavelength, through the modelling and assessment of additional $K > 2$ functional components and improved estimation of residual covariance.

A further shortcoming of current SN Ia models is the lack of incorporation of astrophysical correlations at the fundamental level of the SED.  A ``host mass step'' captures an apparent correlation between host galaxy stellar masses and SN Ia optical luminosities controlling for light curve shape and colour \citep{pkelly10, sullivan10, smith20}.  While the astrophysical nature of this correlation is still under active investigation \citep{jones18,rigault18}, it is typically addressed simplistically by correcting derived distances, or equivalently splitting the scalar absolute magnitude constant in Eq \ref{eqn:tripp}, according to the host mass.  The correlation of SN Ia NIR absolute magnitudes with host mass has been investigated recently by \citet{burns18}, \citet{ponder20}, and \citet{uddin20}.  Our current low-$z$ training set has roughly an average log host mass $\log_{10} (M_* / M_{\odot}) \approx 10.3$ and approximately $80\%$ lie in the ``high-mass'' category $\log_{10} (M_* / M_{\odot}) > 10$.  In future work, we will apply \textsc{BayeSN} to a broader set of SNe Ia to conduct a Bayesian statistical analysis of this effect.  

Similarly, SN Ia ejecta velocities, measured from spectral lines, are correlated with SN Ia intrinsic colour, and can be used to gain leverage on dust estimation and improve the accuracy of distances \citep{foleykasen11, foley12a, mandel14}. Recently, \cite{siebert20} found correlations between ejecta velocity and SALT2 Hubble residuals.  However, these astrophysical correlations should be accounted for at the fundamental physical level of the SN Ia SED functions, rather than by correcting derived distances.  In future work, we will expand our \textsc{BayeSN} framework to explore, estimate, and incorporate the impact of these astrophysical effects on the full SED function $S(t, \lambda_r)$ in a coherent statistical model.  We will do this by adding functional regression terms proportional to $f(\theta_{M_*})\, W_{M_*}(t,\lambda_r)$ or $f(\theta_v)\, W_v(t,\lambda_r)$ to our SED model (Eq. \ref{eqn:model_logsed}), and by modelling potential correlations with host dust population parameters.

In the present work, we have leveraged joint optical and NIR broadband photometry of SNe Ia to learn the statistical properties of the latent intrinsic and dust components of SN Ia SEDs, while using the \citet{hsiaothesis} template as a baseline ``skeleton'' to model spectral features at finer resolutions than the typical passband.  Some of the residual SED covariance and scatter in the Hubble residuals indeed may be caused by per-SN variation in spectral features on wavelength scales much smaller than the typical filter. In future development, we will increase the wavelength resolution of our model, so that we can train simultaneously on spectroscopic sequences and photometric light curves of SNe Ia to improve the latent SED model. We will be able to leverage databases of optical spectra \citep{blondin12,silverman12,folatelli13,siebert19}, as well as forthcoming ground-based NIR spectra from the Magellan FIRE instrument obtained by the CSP-II and CfA Supernova Group \citep{hsiao19}, and space-based NIR spectra from the ongoing Hubble Space Telescope SIRAH program (GO-15889).

In future work, our \textsc{BayeSN} SED model will serve as the centrepiece of a fully hierarchical Bayesian statistical model for principled supernova cosmology analysis.


\section*{acknowledgements}

We dedicate this paper to our friend and colleague, Dr. Andrew S. Friedman (1979--2020), who passed away during preparation of this manuscript. This work builds upon ASF's observations and research on SNe Ia in the NIR. KSM, GN, and AA had the privilege to work with ASF as members of the CfA Supernova Group. Andy's absence is keenly felt.

We thank David Jones, Robert Kirshner, and members of the Nancy Grace Roman Space Telescope Science Investigation Team for supernova cosmology led by Ryan Foley for useful discussions.  ST was supported by the Cambridge CDT for Data-Intensive Science funded by the UK Science and Technology Facilities Council (STFC). This work was performed in part at Aspen Center for Physics, which is supported by National Science Foundation grant PHY-1607611. The participation of KSM at the Aspen Center for Physics was supported by the Simons Foundation. GN was supported by the University of Illinois at Urbana-Champaign and the Center for Astrophysical Surveys at the National Center for Supercomputing Applications.

This work made use of the Illinois Campus Cluster, a computing resource that is operated by the Illinois Campus Cluster Program (ICCP) in conjunction with the National Center for Supercomputing Applications (NCSA) and which is supported by funds from the University of Illinois at Urbana-Champaign.



\bibliographystyle{mnras}
\bibliography{sn.bib, stat.bib}


\appendix

\section{2D Spline Surfaces}\label{sec:2dspline}

We can model a generic function $g(x)$ using a spline with a (generally irregularly-spaced) grid of knot locations $\bm{\xi}$ and a vector of knots $\bm{g}$ through which the function must pass: $g_i = g(\xi_i)$.  Using a natural cubic spline to ensure continuity up to two derivatives (with zero second derivatives at the first and last knots), we can specify a linear smoother (column) vector $\bm{s}(x, \bm{\xi})$ such that $g(x) =   \bm{s}(x, \bm{\xi}) \cdot \bm{g}$ \citep{numrec,givens12}.  Then $g(x)$ at an arbitrary value $x$ is simply a linear combination of the knot values.

We extend this to model a generic, smooth function $g(t, \lambda_r)$ of phase and wavelength in terms of a $\text{dim}(\bm{l}) \times \text{dim}(\bm{\tau})$ matrix $\bm{G}$ of spline knots defined on the phase grid $\bm{\tau}$ and wavelength grid $\bm{l}$.  At any wavelength on the grid, $\lambda_r = l_i$, the function at any phase $t$ can be interpolated as
\begin{equation}
g(t,\lambda_r = l_i) = \bm{G}_{(i,\cdot)} \, \bm{s}(t; \bm{\tau})
\end{equation}
where $\bm{G}_{(i,\cdot)}$ is the $i$th row of $\bm{G}$.  We can compute the above equation for every wavelength knot in $\bm{l}$.  To evaluate the function at an arbitrary wavelength $\lambda_r$, we perform a second spline interpolation in the wavelength dimension.  The composition of these matrix operations is
\begin{equation}
g(t, \lambda_r) = \bm{s}^T(\lambda_r; \bm{l}) \, \bm{G} \, \bm{s}(t; \bm{\tau}).
\end{equation}
The surface $g(t, \lambda_r)$ is linear in the knots matrix $\bm{G}$.

\section{The $W_2$ SED Component}\label{sec:w2}

We describe the second intrinsic functional principal component $W_2(t, \lambda)$ that is learned when we train the model with $K = 2$.  This component can be viewed as the first functional PC of the intrinsic covariance of the residual functions $\epsilon_s(t,\lambda)$ under the $K = 1$ model.  In the $K = 2$ model, we pull out this secondary mode of variation and parametrize its effect through the coefficient $\theta_2$.

\begin{figure}
\includegraphics[scale=0.37]{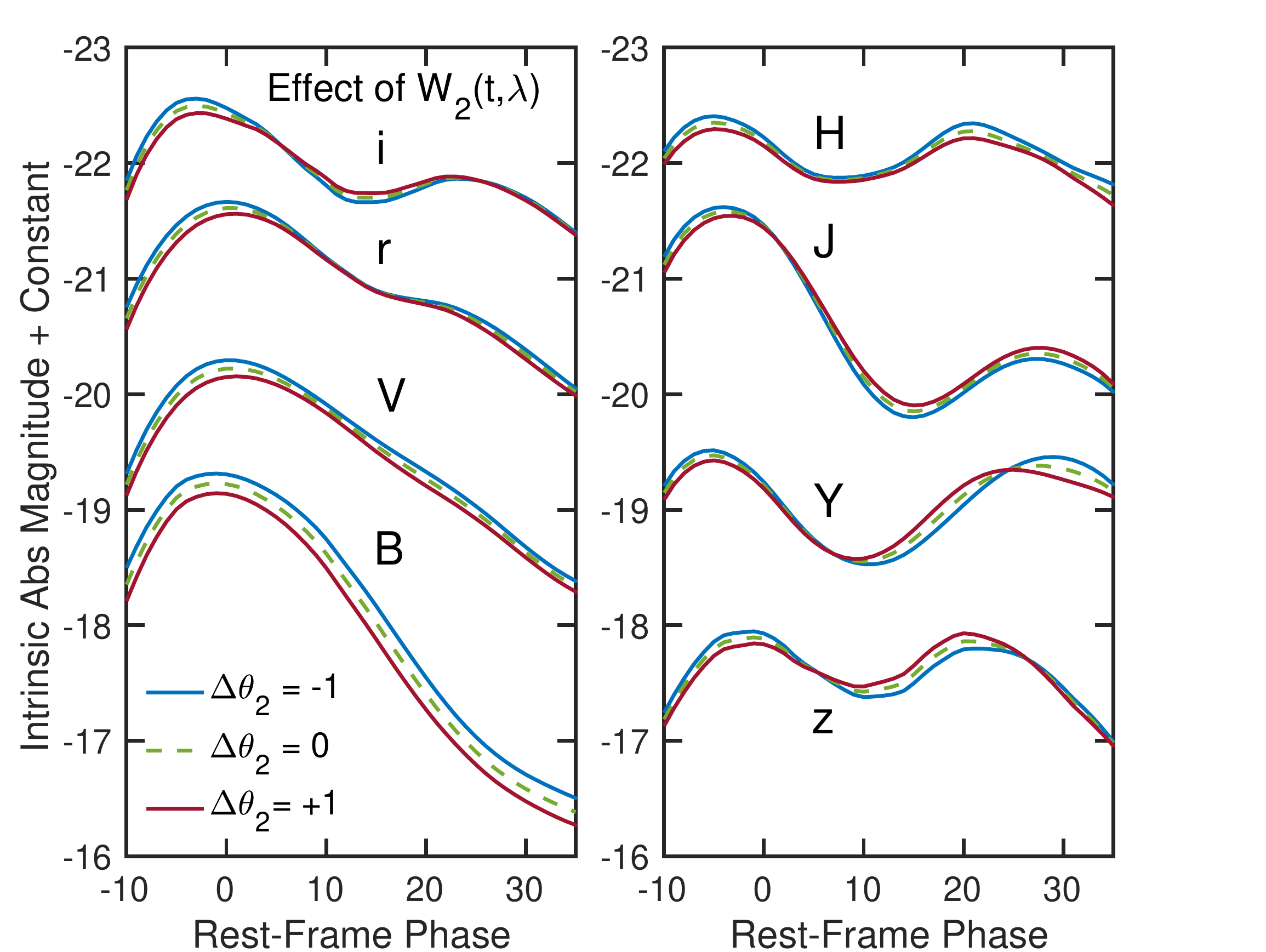}
	\caption{Intrinsic variation in optical and NIR light curves captured by the second functional component $W_2(t, \lambda)$.  We fix $\theta_1 = A_V = 0$ and vary the value of $\theta_2$ by $\bar{\theta}_2 \pm 1\sigma$.   This component captures luminosity variation in the optical that appears to be correlated with the relative amplitudes of the NIR trough and second peak.}
	\label{fig:W2_lcs}
\end{figure}

\begin{figure}
\includegraphics[scale=0.37]{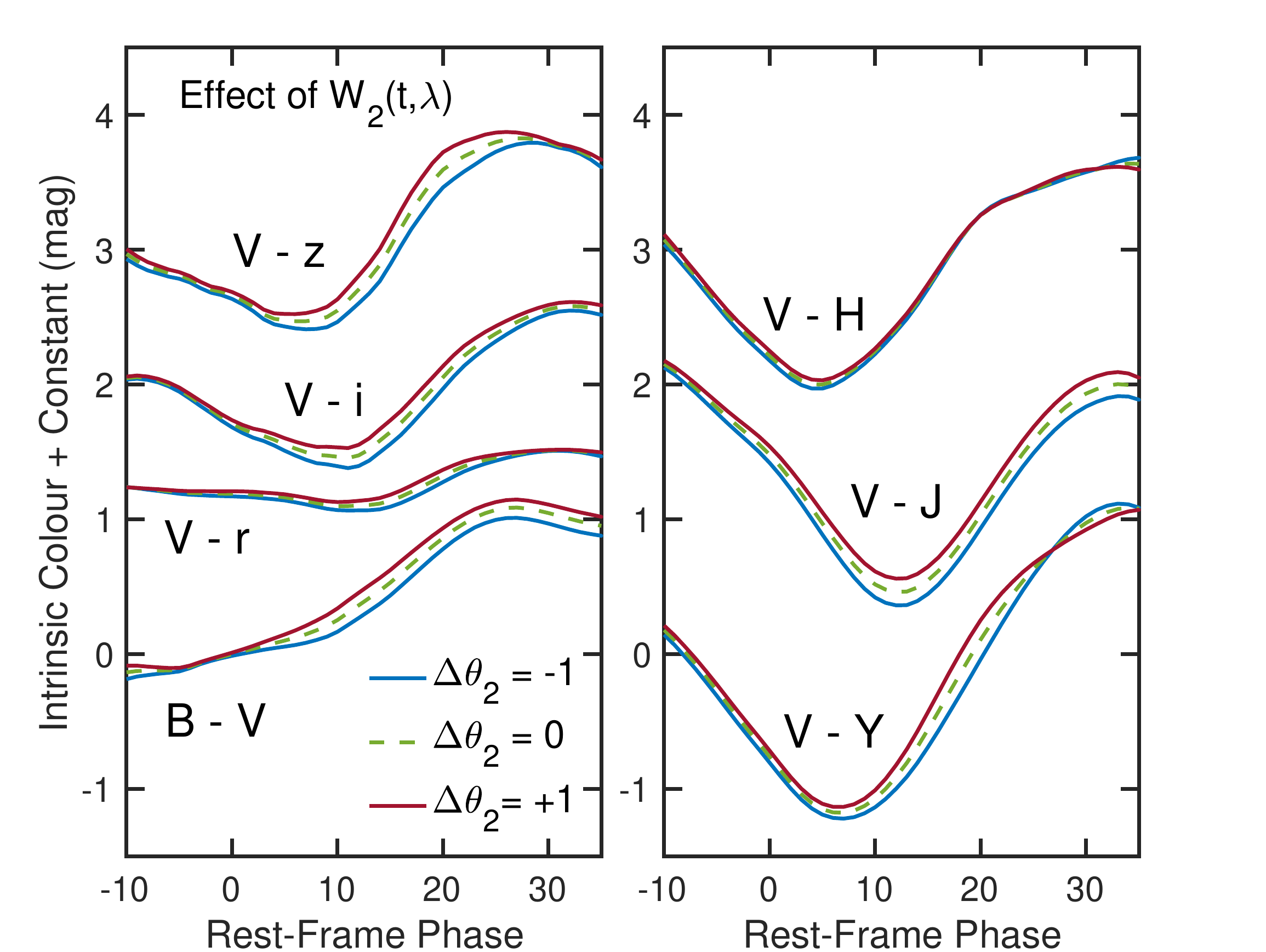}
	\caption{ Variation in optical and NIR intrinsic colour curves captured by the second functional component $W_2(t, \lambda)$.  We fix $\theta_1 = A_V = 0$ and vary the value of $\theta_2$ by $\bar{\theta}_2 \pm 1\sigma$.  This component captures intrinsic colour variation in the post-maximum phases at $t \approx 10$ to $30$ days. 
	}
	\label{fig:W2_cols}
\end{figure}

In Fig. \ref{fig:W2_lcs}, we show the effect of $W_2(t, \lambda)$ on the intrinsic ($A_V = 0$) absolute light curves obtained via integration of the SED model with $\theta_2$ varying between the mean value and $\pm 1 \sigma$.  This component captures some overall luminosity variation in the optical $B$ and $V$ bands, while modulating the relative amplitudes of the first peak, trough, and second peak in the NIR bands.  Unlike $W_1(t,\lambda)$, the second FPC does not significantly change the timing of the second NIR peak, except slightly in $Y$-band.  Fig. \ref{fig:W2_cols} illustrates the effect of $W_2(t, \lambda)$ on the intrinsic optical and optical-NIR colours curves.  This component captures variation in the post-peak colours from 5 to 25 rest-frame days in phase.

In future work with larger datasets, we will further investigate higher-order principal components.

\bsp	
\label{lastpage}
\end{document}